\documentclass[preprint,12pt]{elsarticle}

\usepackage{graphicx}
\usepackage{amssymb}
\usepackage{listings}
\usepackage{hyperref} % CYC

\newcounter{bla}

\newcommand{\genasis}{{\sc GenASiS}}

\journal{Computer Physics Communications}

\begin{document}

\begin{frontmatter}

%% Title, authors and addresses

%% use the tnoteref command within \title for footnotes;
%% use the tnotetext command for the associated footnote;
%% use the fnref command within \author or \address for footnotes;
%% use the fntext command for the associated footnote;
%% use the corref command within \author for corresponding author footnotes;
%% use the cortext command for the associated footnote;
%% use the ead command for the email address,
%% and the form \ead[url] for the home page:
%%
%% \title{Title\tnoteref{label1}}
%% \tnotetext[label1]{}
%% \author{Name\corref{cor1}\fnref{label2}}
%% \ead{email address}
%% \ead[url]{home page}
%% \fntext[label2]{}
%% \cortext[cor1]{}
%% \address{Address\fnref{label3}}
%% \fntext[label3]{}

\title{\textsc{GenASiS }\texttt{Mathematics}: Object-oriented manifolds, operations, and solvers for large-scale physics simulations}

%% use optional labels to link authors explicitly to addresses:
%% \author[label1,label2]{<author name>}
%% \address[label1]{<address>}
%% \address[label2]{<address>}

%\author[a]{First Author\corref{author}}
%\author[a,b]{Second Author}
%\author[b]{Third Author}
\author[a,b]{Christian Y. Cardall\corref{author}\fnref{copyright}}
\author[b,c]{Reuben D. Budiardja\fnref{copyright}}

%\cortext[author] {Corresponding author.\\\textit{E-mail address:} firstAuthor@somewhere.edu}
%\address[a]{First Address}
%\address[b]{Second Address}
\cortext[author] {Corresponding author.\\\textit{E-mail addresses:} cardallcy@ornl.gov (C.Y. Cardall), reubendb@ornl.gov (R.D. Budiardja).}
\address[a]{Physics Division, Oak Ridge National Laboratory, Oak Ridge, TN 37831-6354, USA}
\address[b]{Department of Physics and Astronomy, University of Tennessee, Knoxville, TN 37996-1200, USA}
\address[c]{National Center for Computational Sciences, Oak Ridge National Laboratory, Oak Ridge, TN 37831-6354, USA}

\fntext[copyright]{This manuscript has been authored by UT-Battelle, LLC under Contract No. DE-AC05-00OR22725 with the U.S. Department of Energy. The United States Government retains and the publisher, by accepting the article for publication, acknowledges that the United States Government retains a non-exclusive, paid-up, irrevocable, worldwide license to publish or reproduce the published form of this manuscript, or allow others to do so, for United States Government purposes. The Department of Energy will provide public access to these results of federally sponsored research in accordance with the DOE Public Access Plan (http://energy.gov/downloads/doe-public-access-plan).}

\begin{abstract}
The large-scale computer simulation of a system of physical fields governed by partial differential equations requires some means of approximating the mathematical limit of continuity.
For example, conservation laws are often treated with a `finite-volume' approach in which space is partitioned into a large number of small `cells,' with fluxes through cell faces providing an intuitive discretization modeled on the mathematical definition of the divergence operator.
Here we describe and make available Fortran 2003 classes furnishing extensible object-oriented implementations of simple meshes and the evolution of generic conserved currents thereon, along with individual `unit test' programs and larger example problems demonstrating their use.
These classes inaugurate the \texttt{Mathematics} division of our developing astrophysics simulation code \textsc{GenASiS} (\textit{Gen}eral \textit{A}strophysical \textit{Si}mulation \textit{S}ystem), which will be expanded over time to include additional meshing options, mathematical operations, solver types, and solver variations appropriate for many multiphysics applications.  
\end{abstract}


\end{frontmatter}

%%
%% Start line numbering here if you want
%%
% \linenumbers

% Computer program descriptions should contain the following
% PROGRAM SUMMARY.

%{\bf PROGRAM SUMMARY/NEW VERSION PROGRAM SUMMARY}
{\bf PROGRAM SUMMARY}
  %Delete as appropriate.

\begin{small}
\noindent
{\em Program Title:} \\
SineWave, SawtoothWave, RiemannProblem, RayleighTaylor, SedovTaylor, and FishboneMoncrief (fluid dynamics example problems illustrating \textsc{GenASiS} \texttt{Mathematics})                                          \\
%{\em Licensing provisions(please choose one): CC0 1.0/CC By 4.0/MIT/Apache-2.0/BSD 3-clause/BSD 2-clause/GPLv3/CC BY NC 3.0 }                                   \\
{\em Licensing provisions:} \\ GPLv3                                   \\
{\em Programming language:}                                   \\
Fortran 2003 (tested with gfortran 6.2.0, Intel Fortran 16.0.3, Cray Compiler 8.5.3)  \\                                  
{\em External routines/libraries:} \\
MPI [1] and Silo [2]                          \\
%{\em Supplementary material:}                                 \\
  % Fill in if necessary, otherwise leave out.
%{\em Journal reference of previous version:}                  \\
  %Only required for a New Version summary, otherwise leave out.
%{\em Does the new version supersede the previous version?:}   \\
  %Only required for a New Version summary, otherwise leave out.
%{\em Reasons for the new version:}\\
  %Only required for a New Version summary, otherwise leave out.
%{\em Summary of revisions:}*\\
  %Only required for a New Version summary, otherwise leave out.
%{\em Nature of problem(approx. 50-250 words):}\\
  %Describe the nature of the problem here. \\
{\em Nature of problem:} \\
By way of illustrating \textsc{GenASiS} \texttt{Mathematics} functionality, solve example fluid dynamics problems.\\
%{\em Solution method(approx. 50-250 words):}\\
  %Describe the method solution here.
{\em Solution method:} \\
Finite-volume discretization; second-order slope-limited reconstruction; HLL Riemann Solver; Runge-Kutta time integration. \\
%{\em Additional comments including Restrictions and Unusual features (approx. 50-250 words):}\\
  %Provide any additional comments here.
{\em Additional comments including Restrictions and Unusual features:}\\
The example problems named above are not ends in themselves, but serve to illustrate the functionality available though \textsc{GenASiS} \texttt{Mathematics}. 
In addition to these more substantial examples, we provide individual unit test programs for the classes comprised by \textsc{GenASiS} \texttt{Mathematics}.

\textsc{GenASiS} \texttt{Mathematics} is available in the CPC Program Library and also at \\
https://github.com/GenASiS.  \\

%* Items marked with an asterisk are only required for new versions
%of programs previously published in the CPC Program Library.\\
\end{small}

% For alleviating margin overruns due to monospace type
% The \allowbreak command within \texttt{} may also be useful
\sloppy 

%% main text
\section{Introduction}
\label{sec:Introduction}

Increasing computational power is both boon and bane to scientists in many fields.
Many problems can only be addressed by computer simulation, and the fidelity of those simulations---in terms of, for example, the physics approximations employed, and the dimensionality and resolution of position (and perhaps momentum) space---can be improved with larger machines furnishing greater computing resources.
Unfortunately, such increases in processing power, memory, and storage seem only to come with increased hardware complexity.
In an era of distributed memory and multicore (and even heterogeneous) processing capacity, working scientists may feel that the science itself recedes into the background as they are forced to devote attention and energy developing codes tailored to these increasingly sophisticated machines.

Computational physicists are better able to focus on problem setup and testing, shepherding production runs to completion, and data analysis and interpretation---the tasks they would like to spend time on---to the extent generic but relevant mathematics capabilities are made available in codes that can be readily tailored or extended to their particular problems.
Very often these needs center on solvers for various classes of partial differential equations.
One very important class of equations is conservation laws (continuity equations).
In their broadest form, and including sources, these cover for example both fluid dynamics and Boltzmann kinetic theory or radiation transport; the former involves fluid fields with position space dependence, and the latter treats particle distributions in phase space (position space plus momentum space).
%For purposes of computer simulation, one common means of approximating the mathematical limit of continuity is to model position space (and momentum space, where relevant) with a discrete mesh. 

Because fluid dynamics and radiation transport are central to our particular target application of core-collapse supernovae, we have implemented a solver for conservation laws in our developing code \textsc{GenASiS} ({\em Gen}eral {\em A}strophysical {\em Si}mulation {\em S}ystem).
Initial capabilities of \textsc{GenASiS} for fluid dynamics on a refinable mesh, with emphasis on test results rather than code features, have been reported from an astrophysics perspective elsewhere \cite{Cardall2014GENASIS:-Genera}.
\textsc{GenASiS} has recently been deployed in studies of the core-collapse supernova environment \cite{Budiardja2015Accelerating-Ou,Cardall2015Stochasticity-a,Endeve2016Convection--and}.
`General' in the code title denotes the capacity of the code to include and refer to multiple algorithms, solvers, and physics and numerics choices with the same abstracted names and/or interfaces. 
In \textsc{GenASiS} this is accomplished with features of Fortran 2003 that support the object-oriented programming paradigm (e.g. \cite{Reid2007The-New-Feature}).
`Astrophysical' roughly suggests---over-broadly, at least initially---the types of systems at which the code is aimed, and the kinds of physics and solvers it will include. 
`Simulation System' indicates that the code is not a single program, but a collection of modules, organized as classes, that can be invoked by a suitable driver program set up to characterize and initialize a particular problem.
While we are initially and primarily developing and using \textsc{GenASiS} for astrophysics problems, our goal---facilitated by our object-oriented approach---is to develop solvers with enough care and generality to make them useful for physics simulations in other fields.

\begin{figure}
\centering
\includegraphics[height=3.0in]{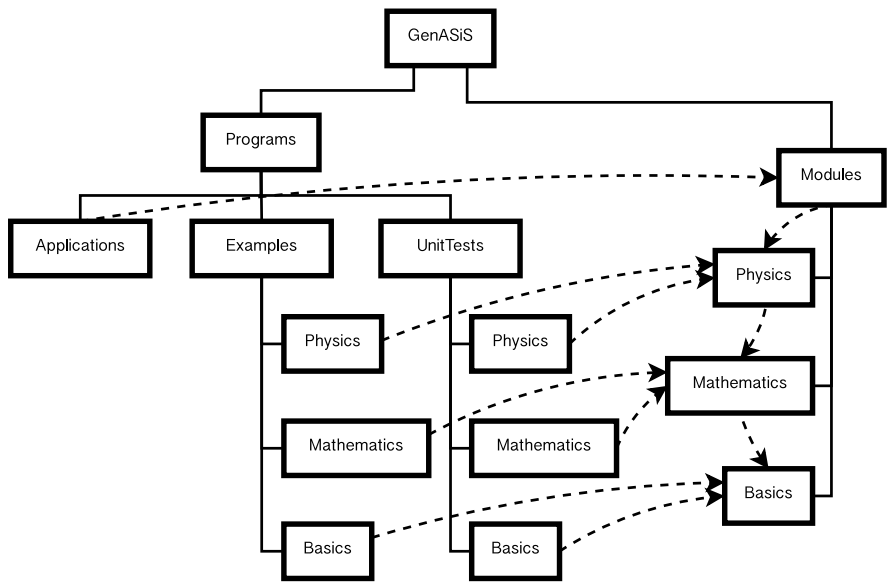}
\caption{High-level structure of the core of \textsc{GenASiS}. Solid lines outline the directory hierarchy, and dashed arrows indicate compilation dependencies.}
\label{fig:GenASiS_Structure}
\end{figure}

The high-level structure of the core of \textsc{GenASiS} is sketched in Fig.~\ref{fig:GenASiS_Structure}. 
The \texttt{Basics} division of \textsc{GenASiS}---which includes classes under \texttt{Modules}, as well as individual \texttt{UnitTests} and larger integrative \texttt{Examples} under \texttt{Programs}---has been publicly released \cite{Cardall2015GenASiS-Basics:,Cardall2017GenASiS-Basics:}.
Included in that release are two categories of nontrivial example problems for which solutions are built upon \texttt{Basics} functionality: fluid dynamics and molecular dynamics.
These are fundamentally different models, requiring the solution of different equations, using different techniques and different parallelization strategies. 
Nevertheless, \textsc{GenASiS} \texttt{Basics} serves as an excellent basis for coding solutions in both cases. 
These examples foreshadow the \texttt{Mathematics} and \texttt{Physics} portions of \textsc{GenASiS} by illustrating in a simple way the object-oriented mechanisms of inheritance and polymorphism that we use to separate lower-level coding of generic, reusable solvers from higher-level coding for specific physical systems (see Ref.~\cite{Cardall2015GenASiS-Basics:} for detailed discussion of a specific example).

\begin{figure}
\centering
\includegraphics[width=5.5in]{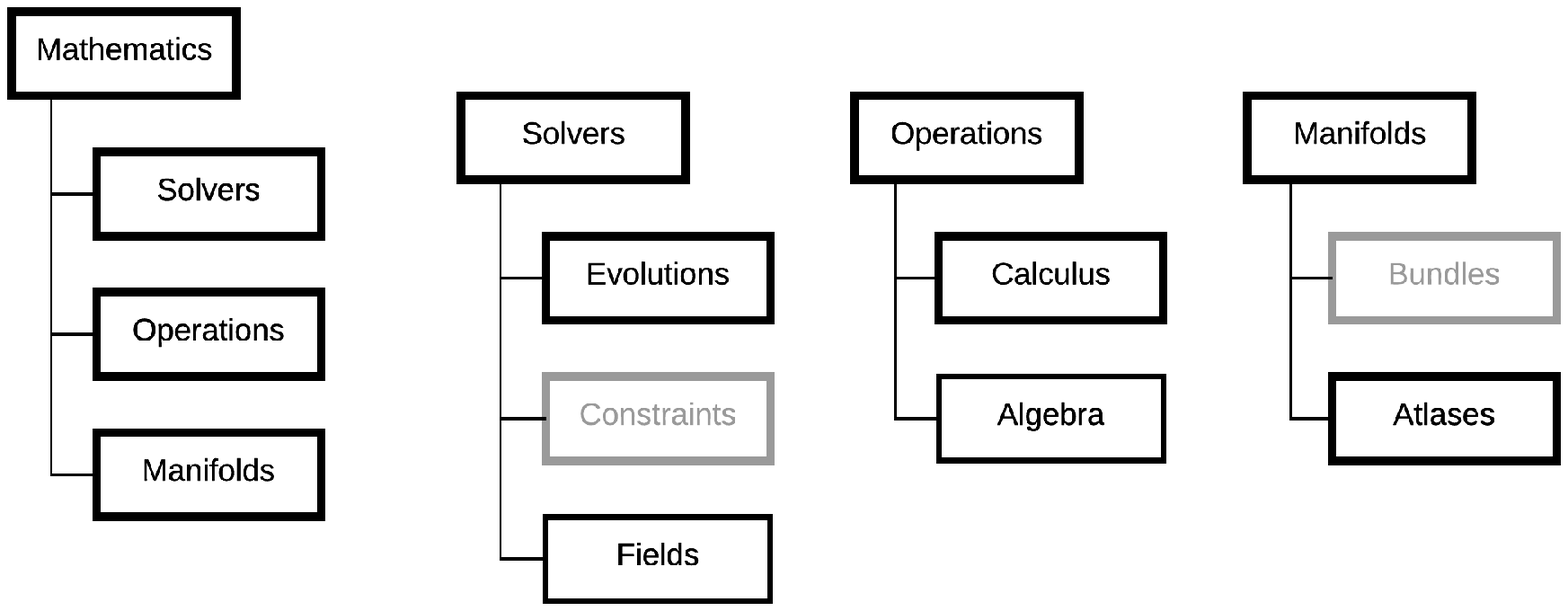}
\caption{{\em Left:} Structure of \texttt{Mathematics}. 
{\em Middle Left:} Substructure of \texttt{Solvers}.
{\em Middle Right:} Substructure of \texttt{Operations}. 
{\em Right:} Substructure of \texttt{Manifolds}.
{\em All:} Solid lines outline the directory hierarchy. 
Boxes framed with thinner linewidths (\texttt{Algebra}, \texttt{Fields}) denote `leaf' divisions of the code with no additional subdirectories.
Boxes in light grey (\texttt{Bundles}, \texttt{Constraints}) denote code divisions not included in this inaugural release of \texttt{Mathematics}.
The compilation order is from bottom to top; thus dependencies essentially flow in reverse, from top to bottom.}
\label{fig:Mathematics_Structure}
\end{figure}

In this paper and code release we take the next step suggested by that foreshadowing, by inaugurating the \texttt{Mathematics} division of \textsc{GenASiS}.
In accordance with our own scientific focus, this release of \texttt{Mathematics} contemplates systems of physical fields governed by partial differential equations,
as reflected in its contents sketched in Fig.~\ref{fig:Mathematics_Structure}.
The main divisions are displayed in the left diagram; in compilation order (from bottom to top) these are \texttt{Manifolds}, \texttt{Operations}, and \texttt{Solvers}.
The right diagram shows the structure within \texttt{Manifolds}, which contains classes used to represent spaces.
The classes in \texttt{Atlases} are sufficient for representing position space, while \texttt{Bundles}---not included in this release, as denoted by the light grey coloring---will in future releases provide functionality for the representation of phase space (position space plus momentum space).
The middle right diagram of Fig.~\ref{fig:Mathematics_Structure} shows the structure within \texttt{Operations}, which includes \texttt{Algebra} and \texttt{Calculus}.
The middle left diagram displays the structure within \texttt{Solvers}: \texttt{Fields} contains templates of physical fields addressed by particular classes of equations; \texttt{Constraints} (not included in this release) contains solvers for equations expressing constraints on the position space dependence of fields within a given time slice; and \texttt{Evolutions} contains solvers that integrate systems of fields forward in time.

We describe the functionality available in this version of \texttt{Mathematics} in both top-down and bottom-up fashion.
A top-down introduction is provided in the context of example problems in Section~\ref{sec:ExampleProblems}.
\texttt{Mathematics} functionality is described more particularly and systematically in the bottom-up discussion of Section~\ref{sec:MathematicsFunctionality}.
Section~\ref{sec:Building} provides instructions for compiling and building the examples and unit test programs. 
We conclude in Section~\ref{sec:Conclusion}.

%%%%%%%%%%%%%%%%%%%%%%%%%%%%%%%%%%%%%%%%%%%%%%%%%%
\section{Example Problems}
\label{sec:ExampleProblems}

In order to illustrate the functionality available in this initial release of \textsc{GenASiS} \texttt{Mathematics} we present several fluid dynamics problems.
The first is the periodic advection of a plane wave in mass density in one, two, or three position space dimensions (1D, 2D, 3D).
The second is a Riemann problem---also 1D, 2D, 3D---which ventures beyond periodic advection to full fluid evolution with shocks and reflecting boundary conditions.
The third is a demonstration of the Rayleigh-Taylor instability in 2D, which illustrates the specification of non-default values for the dimensionality and extent of the domain, and the number of cells; the introduction of source terms into the conservation laws; and a custom tally of global diagnostics.
A fourth problem, the Sedov-Taylor blast wave, introduces the use of different coordinate systems---in this case, spherical coordinates in 1D, cylindrical coordinates in 2D, and Cartesian coordinates in 3D.
Finally, a 2D Fishbone-Moncrief equilibrium torus exemplifies the use of physical units and non-uniform grid spacing, providing also an additional example and test of source terms and spherical coordinates.

\begin{figure}
\centering
\includegraphics[width=5.0in]{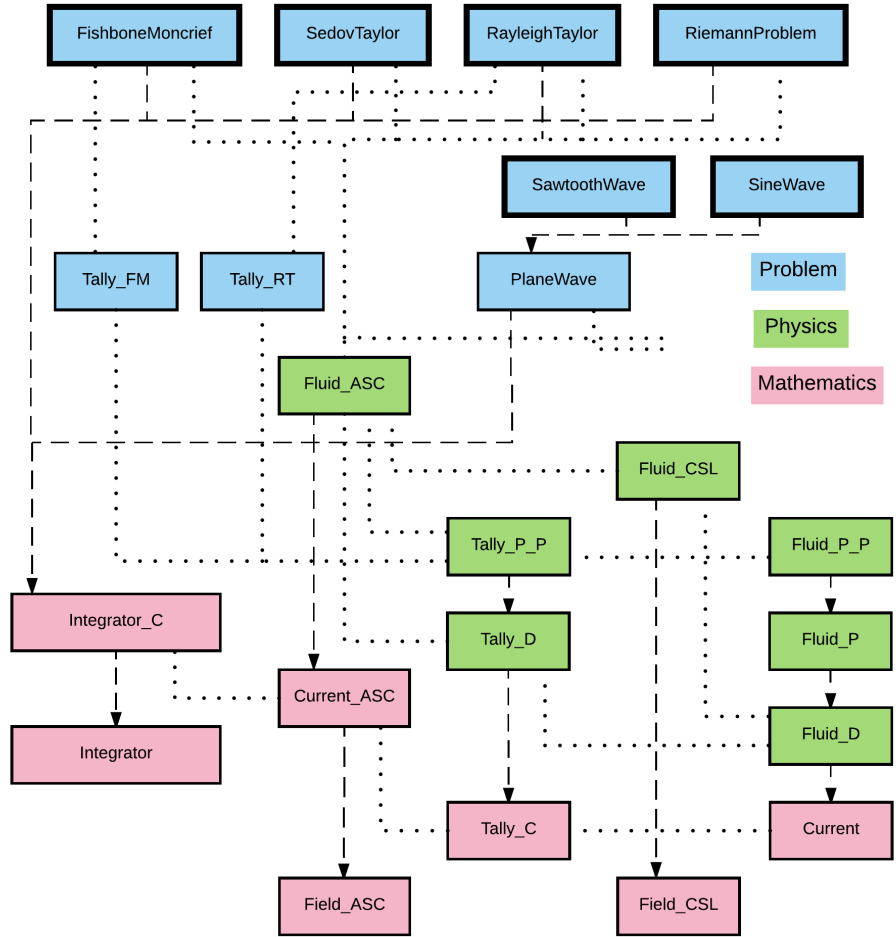}
\caption{Classes built on \textsc{GenASiS} \texttt{Mathematics} classes for fluid dynamics example problems, categorized as `Problem' definition (blue) and `Physics' (green).
Their ancestor classes in `Mathematics' (red) are also shown, ripped from the organizing context of Fig.~\ref{fig:Mathematics_Structure}, which will be discussed more systematically in Section~\ref{sec:MathematicsFunctionality}. 
Light dotted lines indicate compilation dependencies.
Heavy dashed lines with arrows denote class inheritance by type extension.
The boxes with heavy outlines---\texttt{SineWave}, \texttt{SawtoothWave}, \texttt{RiemannProblem}, \texttt{RayleighTaylor}, \texttt{SedovTaylor}, and \texttt{FishboneMoncrief}---represent fully defined problems ready for execution by a very short and simple driver program, such as \texttt{SineWave} in Listing~\ref{lst:Program_SW_Outline}.
The generic mathematical structure of conservation laws is separated from specific physics and particular problems.}
\label{fig:FluidDynamics}
\end{figure}

Classes initializing these problems appear across the top of Fig.~\ref{fig:FluidDynamics}, which diagrams the relationships among the example `Problem' definition (blue) and `Physics' (green) classes we discuss below.
These examples foreshadow the future release of the \texttt{Physics} portion of \textsc{GenASiS} by illustrating how the generic, reusable classes in \texttt{Mathematics} (red in Fig.~\ref{fig:FluidDynamics}) can be extended with relative ease to the specification of particular physical systems and initial conditions.
In discussing these examples we present selected snippets of code; the programs are included in full in the accompanying submission to the CPC Program Library.

%%%%%%%%%%%%%%%%%%%%%%%%
\subsection{Plane Wave Advection}
\label{sec:PlaneWave}

Consider first the periodic advection of a plane wave in mass density for two different waveforms, beginning with the \texttt{program} in Listing~\ref{lst:Program_SW_Outline}, which is specialized to a sine wave.
\begin{lstlisting}[float,frame=tb,numbers=left,numbersep=5pt,xleftmargin=10pt,label=lst:Program_SW_Outline,caption={\texttt{program SineWave}.}]
program SineWave

  use Basics
  use SineWave_Form  !-- See Listing \ref{lst:Module_SW_Outline}
  implicit none

  type ( SineWaveForm ), allocatable :: SW  !-- See Listing \ref{lst:Module_SW_Outline}
  
  allocate ( PROGRAM_HEADER )
  call PROGRAM_HEADER % Initialize ( 'SineWave' )

  allocate ( SW )
  call SW % Initialize ( PROGRAM_HEADER % Name )
  call SW % Evolve ( )
  call SW % ComputeError ( )
  deallocate ( SW )

  deallocate ( PROGRAM_HEADER )

end program SineWave
\end{lstlisting}
The statement \texttt{use Basics} in line~3 gives access to all the classes in the \texttt{Basics} division of \textsc{GenASiS} \cite{Cardall2015GenASiS-Basics:}; among these is the \texttt{PROGRAM\_HEADER} singleton allocated and initialized in lines~9-10, an object used by all \textsc{GenASiS} programs.
The physical system is represented by the instance \texttt{SW} of the class \texttt{SineWaveForm}, declared in line~7 and allocated in line~12, whose methods called in lines~13-15 constitute the main computational tasks: set up the system and its initial conditions (\texttt{Initialize}); evolve the system to some final time (\texttt{Evolve}); and, an analytic solution at arbitrary time of a given waveform being readily available, compute the L1 norm of the difference between the final state and the initial state (\texttt{ComputeError}).
The argument \texttt{PROGRAM\_HEADER \% Name} in line~13 is the base name \texttt{'SineWave'} that is an argument in line~10, but augmented by a dimensionality suffix (\texttt{'3D'} by default, but overridable by a command line argument or parameter file entry detected by \texttt{PROGRAM\_HEADER}, for instance \texttt{Dimensionality=2D}).

Looking for insight into how the highest-level tasks---initializing and evolving the problem, and evaluating the results---are implemented,  we turn to Listing~\ref{lst:Module_SW_Outline}, which outlines the class \texttt{SineWaveForm}.
\begin{lstlisting}[float,frame=tb,numbers=left,numbersep=5pt,xleftmargin=10pt,label=lst:Module_SW_Outline,caption={Outline of \texttt{module SineWave\_Form}. Used at line~4 of Listing~\ref{lst:Program_SW_Outline}.}]
module SineWave_Form

  use Basics
  use PlaneWave_Template  !-- See Listing \ref{lst:Module_PW_Outline}
  implicit none
  private

  type, public, extends ( PlaneWaveTemplate ) :: SineWaveForm  
      real ( KDR ) :: Offset, Amplitude
  contains
    procedure, public, pass :: Initialize
    procedure, private, pass :: Waveform
    final :: Finalize
  end type SineWaveForm

contains

  !-- Definitions of {\tt subroutine}s {\tt Initialize} and {\tt Finalize}, and {\tt function Waveform}, omitted

end module SineWave_Form
\end{lstlisting}
Two members of \texttt{SineWaveForm} are declared in line~9: the \texttt{Offset} and \texttt{Amplitude} of the sine wave.
(An offset is included because the plane wave is in the density, and our fluid dynamics methods expect the density to be positive.) 
The \texttt{kind} of the real numbers \texttt{Offset} and \texttt{Amplitude} is \texttt{KDR} or `kind default real,' normally double precision, as defined in \textsc{GenASiS} \texttt{Basics} \cite{Cardall2015GenASiS-Basics:}.
The method \texttt{Initialize} called in the main program is declared as a type-bound procedure in line~11, but we look in vain for the other methods \texttt{Evolve} and \texttt{ComputeError}.
 
The reason the methods \texttt{Evolve} and \texttt{ComputeError} do not appear in \texttt{SineWaveForm} is that they are inherited from ancestor classes.
This illustrates the philosophy of pushing more generic code down into lower-level classes, enhancing reusability, so that the highest-level classes need specify only what defines a particular problem.
In Fig.~\ref{fig:FluidDynamics}, we see that \texttt{SineWaveForm} extends \texttt{PlaneWaveTemplate} (see also line~8 of Listing~\ref{lst:Module_SW_Outline});\footnote{We use the suffix \texttt{Template} to denote \texttt{abstract} classes with \texttt{deferred} methods, as opposed to the suffix \texttt{Form} for non-\texttt{abstract} classes.} the latter implements the method \texttt{ComputeError} with code suitable for any waveform.
Figure~\ref{fig:FluidDynamics} also shows that \texttt{PlaneWaveTemplate} extends the \texttt{Mathematics} class \texttt{Integrator\_C\_Template}, which evolves a generic conserved current; this in turn extends \texttt{Integrator\_Template}, for generic time-evolved systems, in which we come finally to the \texttt{Evolve} method called in Listing~\ref{lst:Program_SW_Outline}.
The class \texttt{PlaneWaveTemplate} is outlined in Listing~\ref{lst:Module_PW_Outline}, showing its status as an extension of \texttt{Integrator\_C\_Template} in line~10, and the declaration of its \texttt{ComputeError} method in line~15.
\begin{lstlisting}[float,frame=tb,numbers=left,numbersep=5pt,xleftmargin=10pt,label=lst:Module_PW_Outline,caption={Outline of {\tt module PlaneWave\_Template}. Used at line~4 of Listing~\ref{lst:Module_SW_Outline}.}]
module PlaneWave_Template

  use Basics
  use Mathematics
  use Fluid_D__Form
  use Fluid_ASC__Form
  implicit none
  private

  type, public, extends ( Integrator_C_Template ), abstract :: &
    PlaneWaveTemplate
      !-- Members omitted
  contains
    procedure, public, pass :: InitializeTemplate_PW  !-- See Listing \ref{lst:Subroutine_Initialize_PW}
    procedure, public, pass :: ComputeError
    procedure, public, pass :: FinalizeTemplate_PW
    procedure ( Waveform ), private, pass, deferred :: Waveform
  end type PlaneWaveTemplate

  !-- Declaration of the {\tt interface} for {\tt function Waveform} omitted
  
    private :: &
      SetFluid, &  !-- See Listing \ref{lst:Subroutine_SetFluid}
      SetReference

contains

  !-- Definitions of member {\tt subroutine}s {\tt InitializeTemplate\_PW} and {\tt ComputeError} omitted
  !-- Definitions of private local {\tt subroutine}s {\tt SetFluid} and {\tt SetReference} omitted

end module PlaneWave_Template
\end{lstlisting}

Having located all the methods called in the main program, we turn to initialization, which consists of five tasks to define the problem: 
(1)~Specify a physical space and its geometry. 
(2)~Specify the forms of stress-energy in that space. 
(3)~Specify the solver that advances the system forward in time. 
(4)~Set the initial conditions.
(5)~Initialize the parent time integrator. 
In the case of \texttt{SineWaveForm}, the code unique to its short \texttt{Initialize} method merely sets the sine wave's \texttt{Offset} and \texttt{Amplitude}, and then calls the \texttt{InitializeTemplate\_PW} method of its parent \texttt{PlaneWaveTemplate} to accomplish tasks (1)-(5).
This method is declared in line~14 of Listing~\ref{lst:Module_PW_Outline}, and sketched in Listing~\ref{lst:Subroutine_Initialize_PW}.
\begin{lstlisting}[float,frame=tb,numbers=left,numbersep=5pt,xleftmargin=10pt,label=lst:Subroutine_Initialize_PW,caption={{\tt subroutine InitializeTemplate\_PW}. Fits in line~28 of Listing~\ref{lst:Module_PW_Outline}.}]
subroutine InitializeTemplate_PW ( PW, Name )

  class ( PlaneWaveTemplate ), intent ( inout ) :: PW
  character ( * ), intent ( in ) :: Name
 
  integer ( KDI ) :: nPeriods
  integer ( KDI ), dimension ( 3 ) :: nWavelengths
  real ( KDR ) :: Period
  class ( Fluid_D_Form ), pointer :: F

  !-- PositionSpace
  allocate ( Atlas_SC_Form :: PW % PositionSpace )
  select type ( PS => PW % PositionSpace )
  class is ( Atlas_SC_Form )
  call PS % Initialize ( Name, PROGRAM_HEADER % Communicator )
  call PS % CreateChart ( )

  !-- Geometry of PositionSpace
  allocate ( PW % Geometry_ASC )
  associate ( GA => PW % Geometry_ASC )  !-- GeometryAtlas
  call GA % Initialize ( PS )
  call PS % SetGeometry ( GA )
  end associate !-- GA

  !-- Fluid (Dust, i.e. pressureless fluid)
  allocate ( Fluid_ASC_Form :: PW % Current_ASC )
  select type ( FA => PW % Current_ASC )  !-- FluidAtlas
  class is ( Fluid_ASC_Form )
  call FA % Initialize ( PS, 'DUST' )

  !-- Step
  allocate ( Step_RK2_C_Form :: PW % Step )
  select type ( S => PW % Step )
  class is ( Step_RK2_C_Form )
  call S % Initialize ( Name )
  end select !-- S
  
  !-- Initialization related to diagnostics omitted

  !-- Initial conditions
  !-- Initialization of plane wave parameters omitted
  F => FA % Fluid_D ( )
  call SetFluid ( PW, F, Time = 0.0_KDR )
  
  !-- Initialize template
  call PW % InitializeTemplate_C &
        ( Name, FinishTimeOption = nPeriods * Period )

  !-- Cleanup
  end select !-- FA
  end select !-- PS
  
end subroutine InitializeTemplate_PW
\end{lstlisting}

Specification of the physical space begins in lines~11-16 of Listing~\ref{lst:Subroutine_Initialize_PW}.
The object \texttt{PW} of class \texttt{PlaneWaveTemplate} inherits a polymorphic member \texttt{PositionSpace} from its ancestors, which here is allocated as type \texttt{Atlas\_SC\_Form} in line~12.
Position space is a manifold, covered by one or more coordinate patches; in mathematical terms each coordinate patch is called a `chart,' and the collection of charts is called an `atlas.'
In the simplest case a manifold is covered by single chart; this is the case embodied in the class \texttt{Atlas\_SC\_Form}.
Because \texttt{PositionSpace} is polymorphic, the members and methods particular to the chosen type can only be accessed via the \texttt{select type} construct in lines~13-14.
Initialization of the manifold---including the dimensionality as determined by \texttt{PROGRAM\_HEADER}---occurs in line~15.
The coordinate chart comes into existence in line~16; in this call to \texttt{CreateChart} with no options, default values for the parameters characterizing the chart are used in the absence of command line options and/or parameter file entries.

Specification of the geometry of physical space is intertwined with that of the \texttt{PositionSpace} member, as seen in lines~18-23 of Listing~\ref{lst:Subroutine_Initialize_PW}.
\texttt{PW} inherits a member \texttt{Geometry\_ASC} representing the geometry of an instance \texttt{Atlas\_SC\_Form}; its allocation without any type specification in line~19 means it will be the default geometry of flat space.
Initialization of the geometry in line~21 requires as an argument an initialized atlas with which the geometry is to be associated.
Then, specification of the physical space is completed by giving the initialized geometry object (aliased as \texttt{GA}) as an argument to the \texttt{SetGeometry} method of the \texttt{PositionSpace} object (aliased above as \texttt{PS}) in line~22.

In this example, the only form of stress-energy in the system is a pressureless fluid---`dust,' as known in astrophysics---specified in lines~25-29 of Listing~\ref{lst:Subroutine_Initialize_PW}.
The inherited polymorphic member \texttt{Current\_ASC} of \texttt{PW} represents a generic conserved current on an instance of \texttt{Atlas\_SC\_Form}.
Here it is allocated as an instance of \texttt{Fluid\_ASC\_Form} in line~26, which, as seen towards the middle of Fig.~\ref{fig:FluidDynamics}, is a child of the \texttt{Mathematics} class \texttt{Current\_ASC\_Template}.
Initialization in line~29 includes the string \texttt{'DUST'} as an argument in order to identify the type of fluid desired.

Selection of the means of evolving forward in time is accomplished in lines~31-36 of Listing~\ref{lst:Subroutine_Initialize_PW}.
Once again an inherited polymorphic member is allocated as a chosen type; here, the type \texttt{Step\_RK2\_C\_Form} used in line~32 denotes second-order Runge-Kutta (\texttt{RK2}) evolution of a conserved current (\texttt{C}).

Before addressing the next task in the above list---setting the initial conditions, which takes place in lines~40-43 of Listing~\ref{lst:Subroutine_Initialize_PW}---we should say more about the data structures containing the fluid variables.
The fluid object we have mentioned so far---an instance of \texttt{Fluid\_ASC\_Form}, representing a fluid on an atlas with a single coordinate chart---is a couple of steps removed from the actual fluid data.
Note that in Fig.~\ref{fig:FluidDynamics}, \texttt{Fluid\_ASC\_Form} depends on a class called \texttt{Fluid\_CSL\_Form} (middle right); in fact one of its members is an instance of this class.
Here \texttt{CSL} stands for `chart single level', denoting the fact that an instance of this class is associated with a coordinate chart that is a simple unigrid, as opposed to a multi-level refinable mesh. 
\texttt{Fluid\_CSL\_Form} is a child of the \texttt{Mathematics} class \texttt{Field\_CSL\_Template}.
The latter has a polymorphic member \texttt{Field} of class \texttt{VariableGroupForm} \cite{Cardall2015GenASiS-Basics:} to represent a collection of related physical fields on one single-level coordinate chart.
In \texttt{Fluid\_CSL\_Form}, the polymorphic \texttt{Field} member is allocated to one of the fluid types descending from \texttt{CurrentTemplate} (lower right of Fig.~\ref{fig:FluidDynamics}).
The latter is an \texttt{abstract} class representing a generic conserved current---a 4-vector (or set of related 4-vectors, and associated fields) whose 4-divergence vanishes in the absence of sources. 

\texttt{CurrentTemplate} is \texttt{abstract} because some of its methods---such as those for computing conserved variables from primitive variables and vice-versa, computing fluxes, etc.---are \texttt{deferred} at the level of generality aimed for in \texttt{Mathematics} classes.
Provision of these deferred methods allows generic solver code to be implemented, culminating here in the \texttt{Mathematics} class \texttt{Integrator\_C\_Template}, which as we have seen is the generic ancestor extended to the example problems along the top of Fig.~\ref{fig:FluidDynamics}.
As seen towards the lower right of Fig.~\ref{fig:FluidDynamics}, \texttt{CurrentTemplate} is extended to \texttt{Fluid\_D\_Form} (`dust') and beyond in the example physics classes, in which primitive and conserved (and other) fields are spelled out and the deferred methods given concrete implementations.
In line~42 of Listing~\ref{lst:Subroutine_Initialize_PW}, the method \texttt{Fluid\_D} of \texttt{Fluid\_ASC\_Form} is a function returning an appropriately typed pointer to the fluid data living on the single level of the single chart of the instance of \texttt{Fluid\_ASC\_Form}; it provides a shortcut to the fluid data by providing a target for the pointer \texttt{F} declared in line~9.

The local subroutine \texttt{SetFluid}---noted at lines~23 and 29 of Listing~\ref{lst:Module_PW_Outline}, and called at line~43 of Listing~\ref{lst:Subroutine_Initialize_PW}---is where the initial fluid data are actually set, as shown in Listing~\ref{lst:Subroutine_SetFluid}.
\begin{lstlisting}[float,frame=tb,numbers=left,numbersep=5pt,xleftmargin=10pt,label=lst:Subroutine_SetFluid,caption={{\tt subroutine SetFluid}. Fits in line~29 of Listing~\ref{lst:Module_PW_Outline}. Called at line~43 of Listing~\ref{lst:Subroutine_Initialize_PW}.}]
subroutine SetFluid ( PW, F, Time )

  class ( PlaneWaveTemplate ), intent ( in ) :: PW
  class ( Fluid_D_Form ), intent ( inout ) :: F
  real ( KDR ), intent ( in ) :: Time
    
  class ( GeometryFlatForm ), pointer :: G

  select type ( PS => PW % PositionSpace )
  class is ( Atlas_SC_Form )
  G => PS % Geometry ( )

  associate &
    ( K     => PW % Wavenumber, &
      Abs_K => sqrt ( dot_product ( PW % Wavenumber, PW % Wavenumber ) ), &
      V     => PW % Speed, &
      T     => Time, &
      X     => G % Value ( :, G % CENTER ( 1 ) ), &
      Y     => G % Value ( :, G % CENTER ( 2 ) ), &
      Z     => G % Value ( :, G % CENTER ( 3 ) ), &
      VX    => F % Value ( :, F % VELOCITY_U ( 1 ) ), &
      VY    => F % Value ( :, F % VELOCITY_U ( 2 ) ), &
      VZ    => F % Value ( :, F % VELOCITY_U ( 3 ) ), &
      N     => F % Value ( :, F % COMOVING_DENSITY ) )

  VX = V * K ( 1 ) / Abs_K
  VY = V * K ( 2 ) / Abs_K
  VZ = V * K ( 3 ) / Abs_K

  N = PW % Waveform &
        (    K ( 1 ) * ( X  -  VX * T ) &
          +  K ( 2 ) * ( Y  -  VY * T ) &
          +  K ( 3 ) * ( Z  -  VZ * T ) )

  call F % ComputeFromPrimitive ( G )

  end associate !-- K, etc.
  end select    !-- PS

end subroutine SetFluid
\end{lstlisting}
This routine has a \texttt{Time} argument because it is also used to generate a reference solution to compare against the computed solution; for purposes of initial conditions, this argument is set to \texttt{0.0\_KDR}.%\footnote{The suffix \texttt{KDR}, appearing also in line~5 of Listing~\ref{lst:Subroutine_SetFluid}, is a Fortran \texttt{kind} specifier defined in \textsc{GenASiS} \texttt{Basics} \cite{Cardall2015GenASiS-Basics:}. It is shorthand for \texttt{KIND\_DEFAULT \% REAL} and normally corresponds to double precision.} 
Line~11 provides a shortcut to the geometry data by setting a pointer, declared in line~7, through a mechanism analogous to that discussed in the previous paragraph for accessing the fluid data.
Reference to variables is further simplified by the \texttt{associate} statement in lines~13-24. 
Some of the aliased variables are plane wave parameters omitted from Listings~\ref{lst:Module_PW_Outline} and \ref{lst:Subroutine_Initialize_PW} (see their lines~12 and 41 respectively).
The others refer to fields stored in the geometry and fluid data structures; these are extensions of the very important class \texttt{VariableGroupForm} defined in \textsc{GenASiS} \texttt{Basics} \cite{Cardall2015GenASiS-Basics:}.
These classes contain both field data and metadata regarding those fields (field names, units, etc.). 
The numerical data itself is stored in the rank-2 \texttt{real} member \texttt{Value}, whose first index corresponds to data points (cells in the coordinate chart, in this case) and whose second index corresponds to different variables (geometry or fluid fields, in this case).

With position, velocity, and density aliased, the primitive fluid variables are set in lines~26-28 and 30-33 of Listing~\ref{lst:Subroutine_SetFluid} as functions of the plane wave parameters and position.
From the point of view of \texttt{PlaneWaveTemplate}, its deferred method \texttt{Waveform} (declared at line~17 of Listing~\ref{lst:Module_PW_Outline}; see also line~20) could be any periodic function on the interval $[ 0, 1]$;
the extensions \texttt{SineWaveForm} and \texttt{SawtoothWaveForm} provide concrete overriding functions for a sine wave and sawtooth wave respectively (see lines~12 and 18 of Listing~\ref{lst:Module_SW_Outline}).
The fluid is then fleshed out as conserved and auxiliary fields are set by the call to the \texttt{ComputeFromPrimitive} method of \texttt{Fluid\_D\_Form} in line~35 of Listing~\ref{lst:Subroutine_SetFluid}. 

Note the degree of abstraction afforded application scientists in this example of setting initial conditions.
Having chosen a representation of position space and a fluid associated with it, the user need not worry about array indexing or other details of how the data are associated with the mesh;
all that is needed is the desired functional form, in this case a traveling wave of the form $W\left[ \mathbf{k} \cdot ( \mathbf{x} - \mathbf{v} t ) \right]$.
Indeed an important reason more generally for separating \texttt{Mathematics} from \texttt{Physics} in \textsc{GenASiS} is to allow problem definition and setup at a relatively high level of abstraction.
This both frees the user from the need to know the details of data structures of the mesh and solvers, and also allows problem specification that can readily apply to \textit{different} mesh and solver variations under the hood.

The final task in setting up a problem is initializing the parent time integrator.
This is accomplished in lines~45-47 of Listing~\ref{lst:Subroutine_Initialize_PW}, in which the call includes an optional argument specifying the time to which the simulation should evolve.

\begin{figure}
\begin{centering}
  \includegraphics[width=0.49\textwidth]{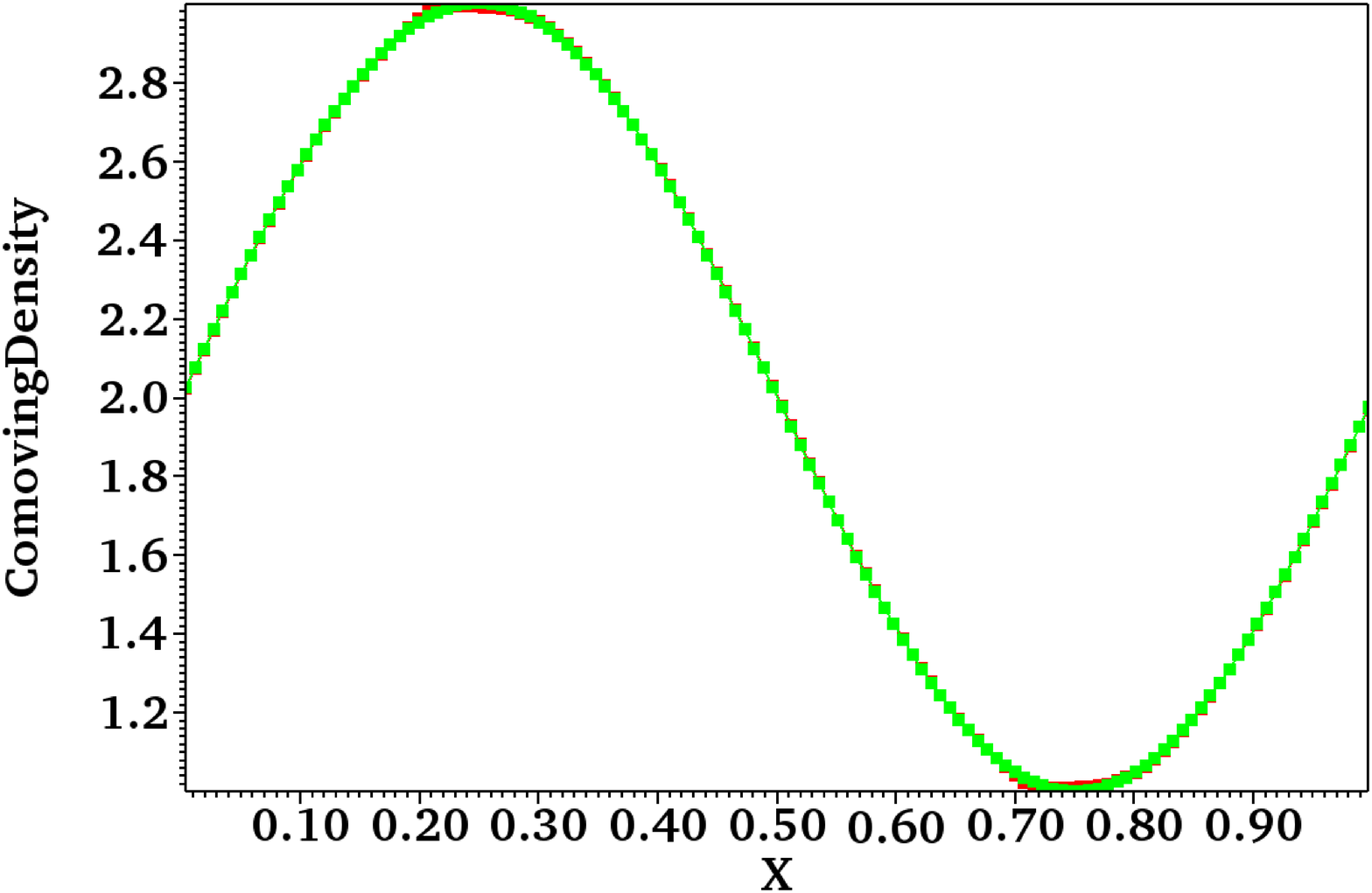}
  \includegraphics[width=0.49\textwidth]{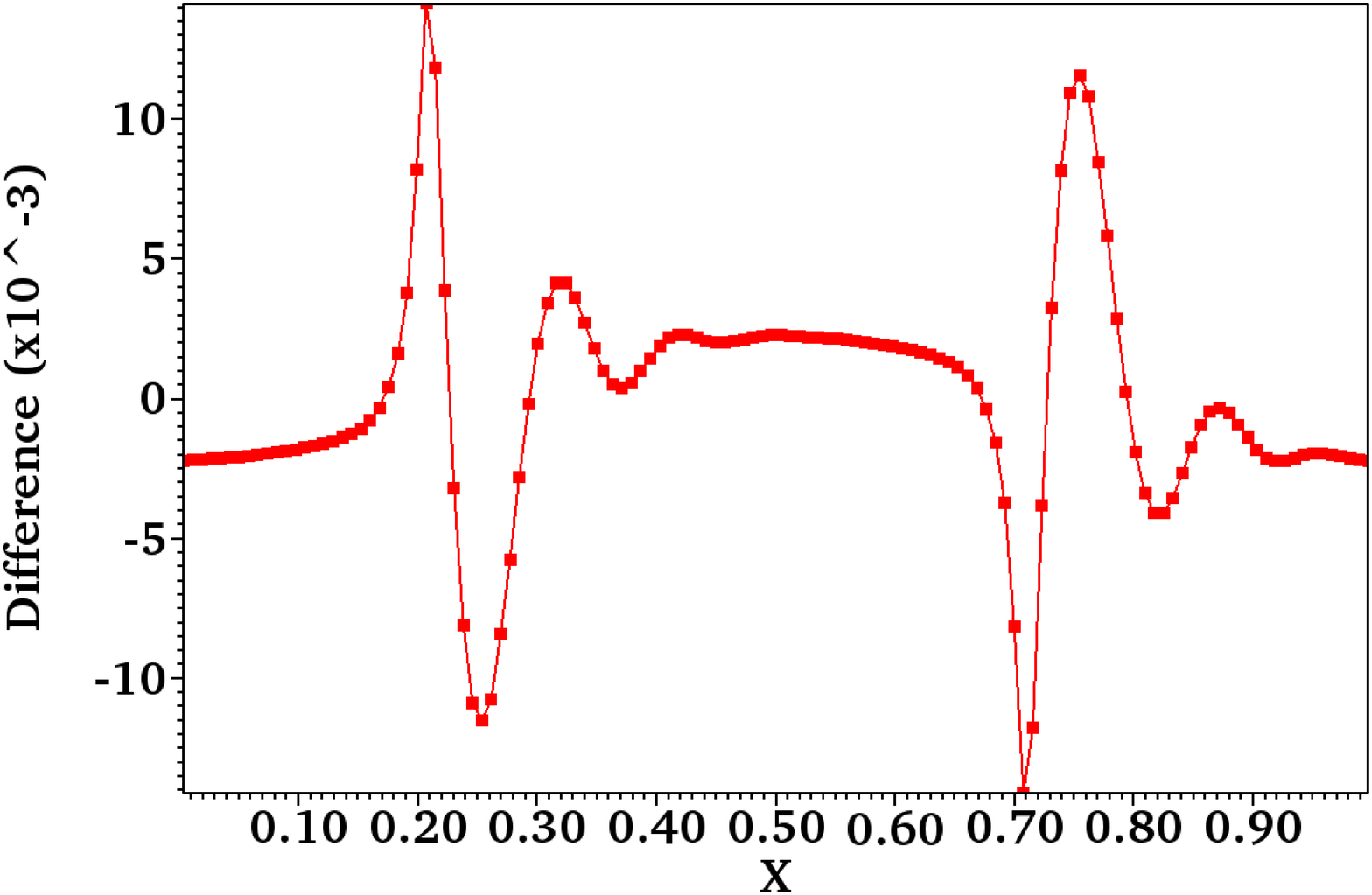}
  \includegraphics[width=0.49\textwidth]{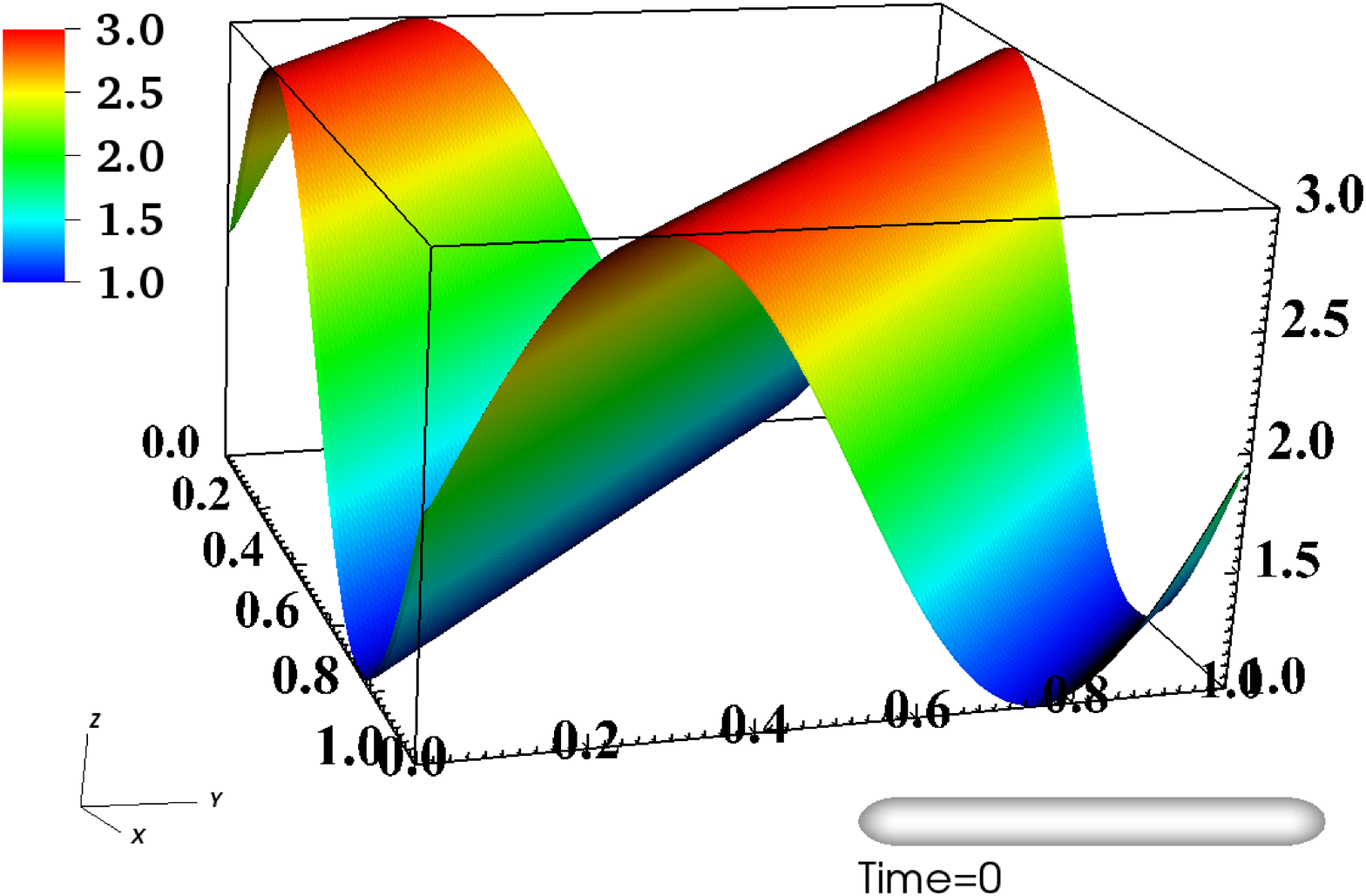}
  \includegraphics[width=0.49\textwidth]{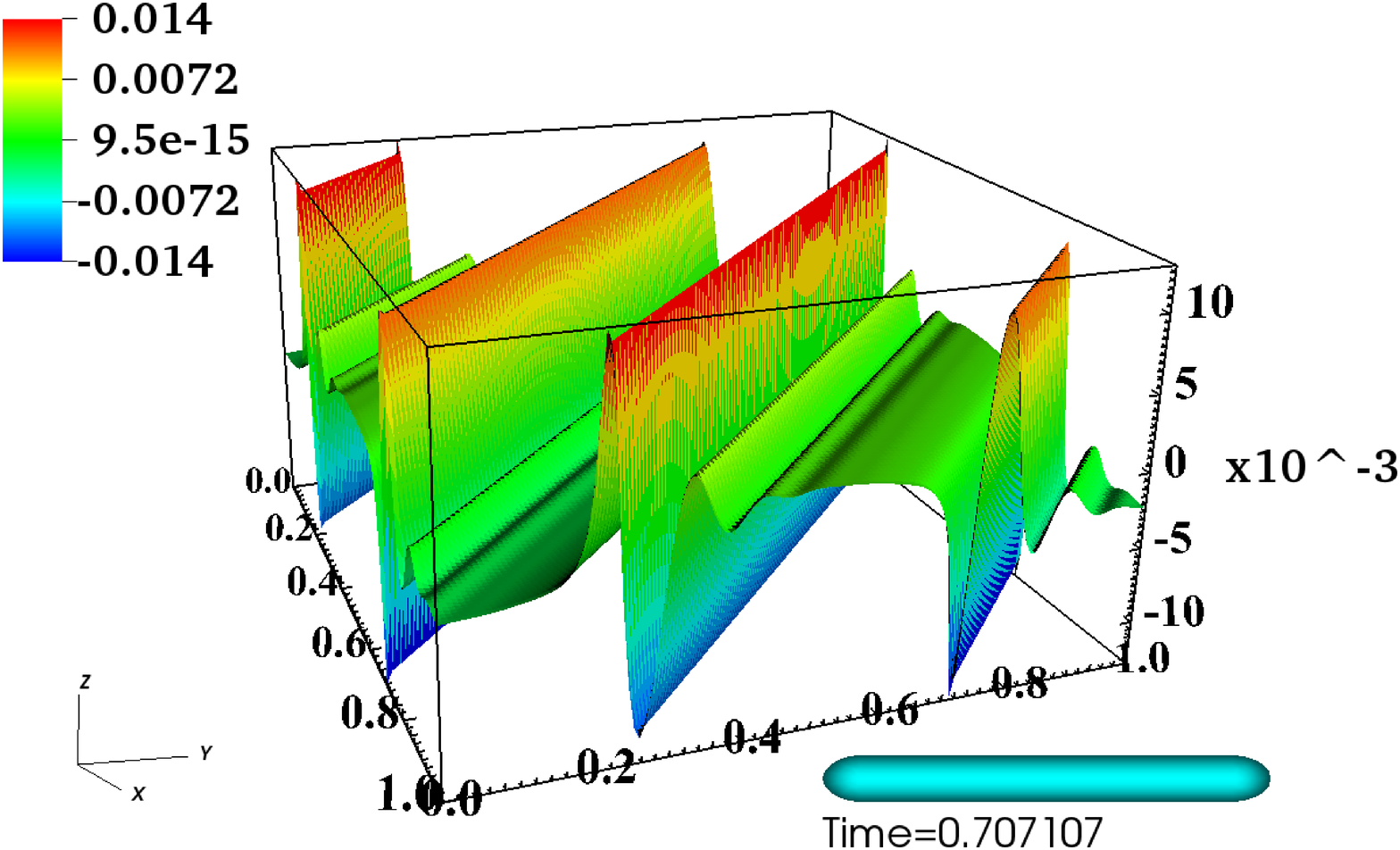}
  \includegraphics[width=0.49\textwidth]{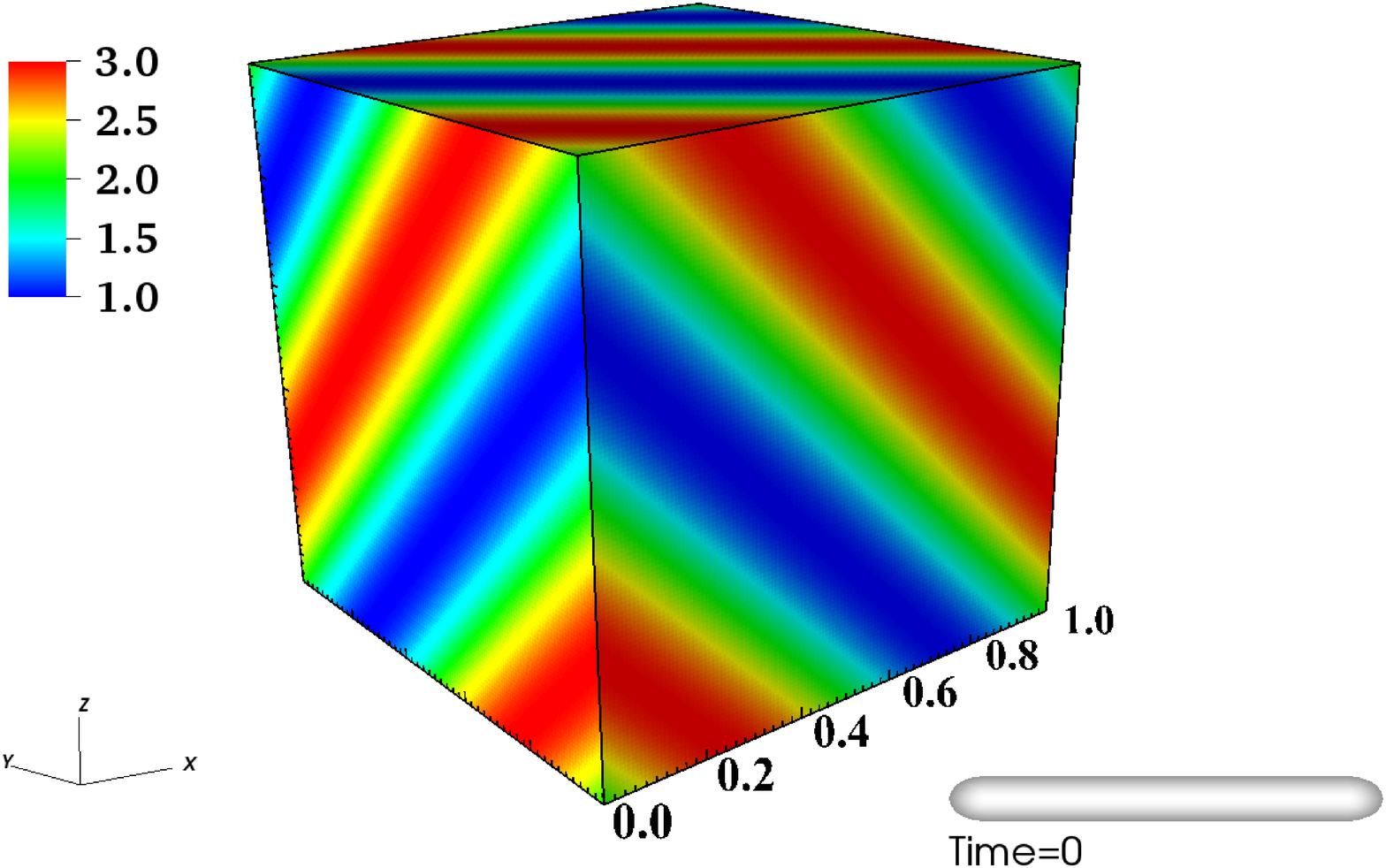}
  \includegraphics[width=0.49\textwidth]{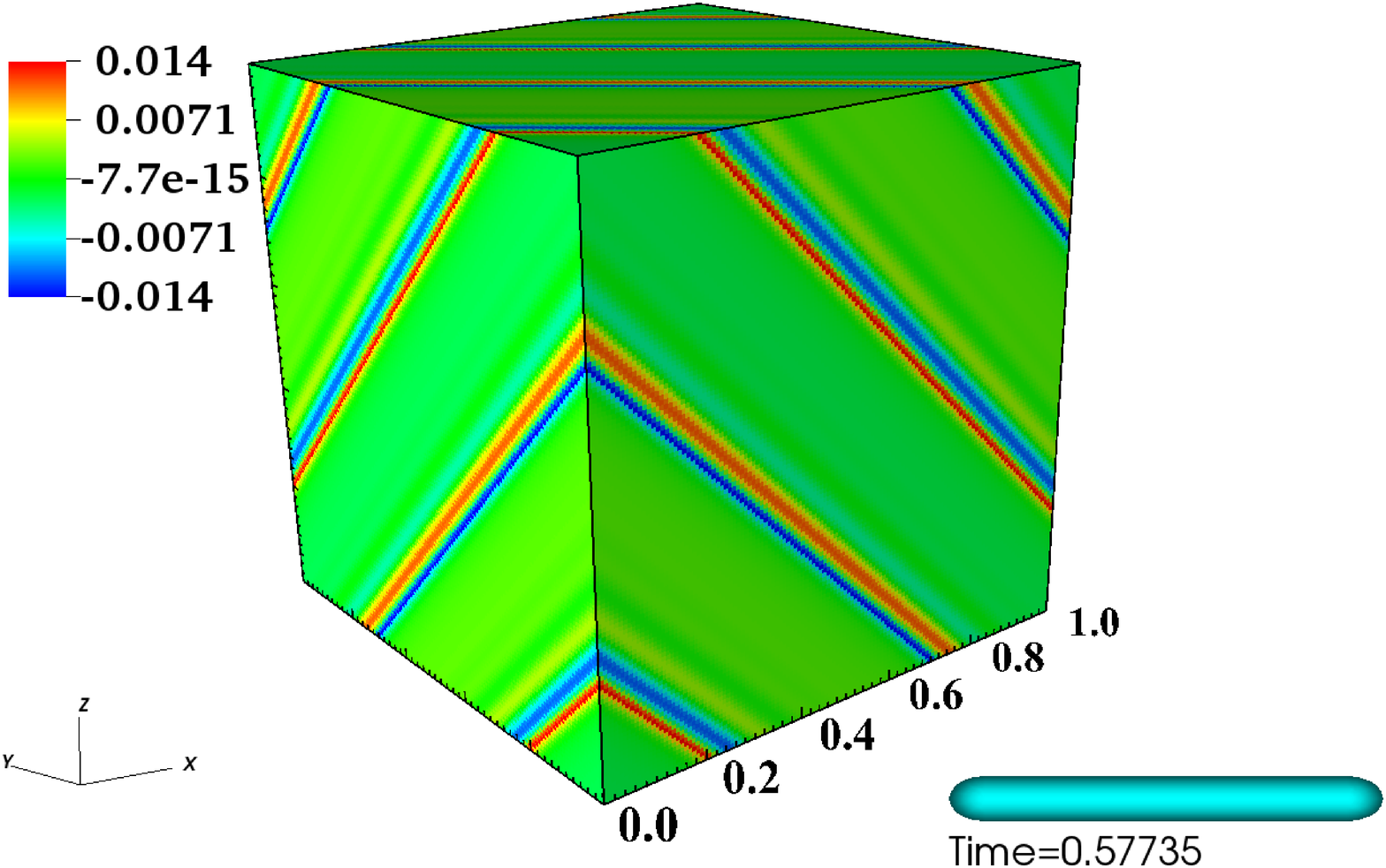}
  \caption{Plots of density for the 1D (upper), 2D (middle), and 3D (lower) versions of the \texttt{SineWave} initial conditions (one wavelength, left) and their difference from the initial conditions after one period of evolution (right) at a resolution of 128 cells in each dimension.
The 1D plot of density (upper left) shows both the final (red) and initial (green) states, which visually overlap almost perfectly.}
  \label{fig:SineWave}
\end{centering}
\end{figure}

\begin{figure}
\begin{centering}
  \includegraphics[width=0.49\textwidth]{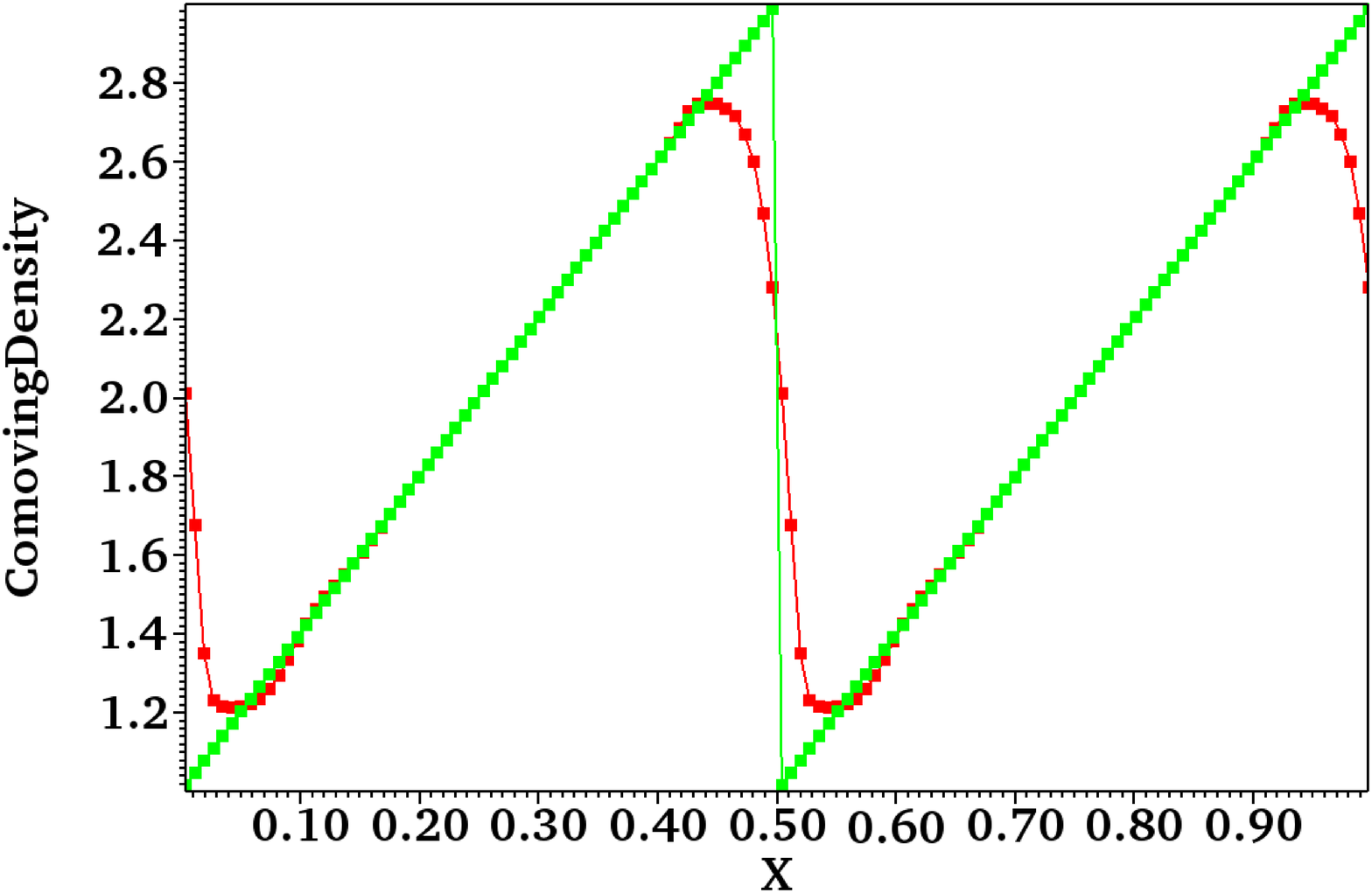}
  \includegraphics[width=0.49\textwidth]{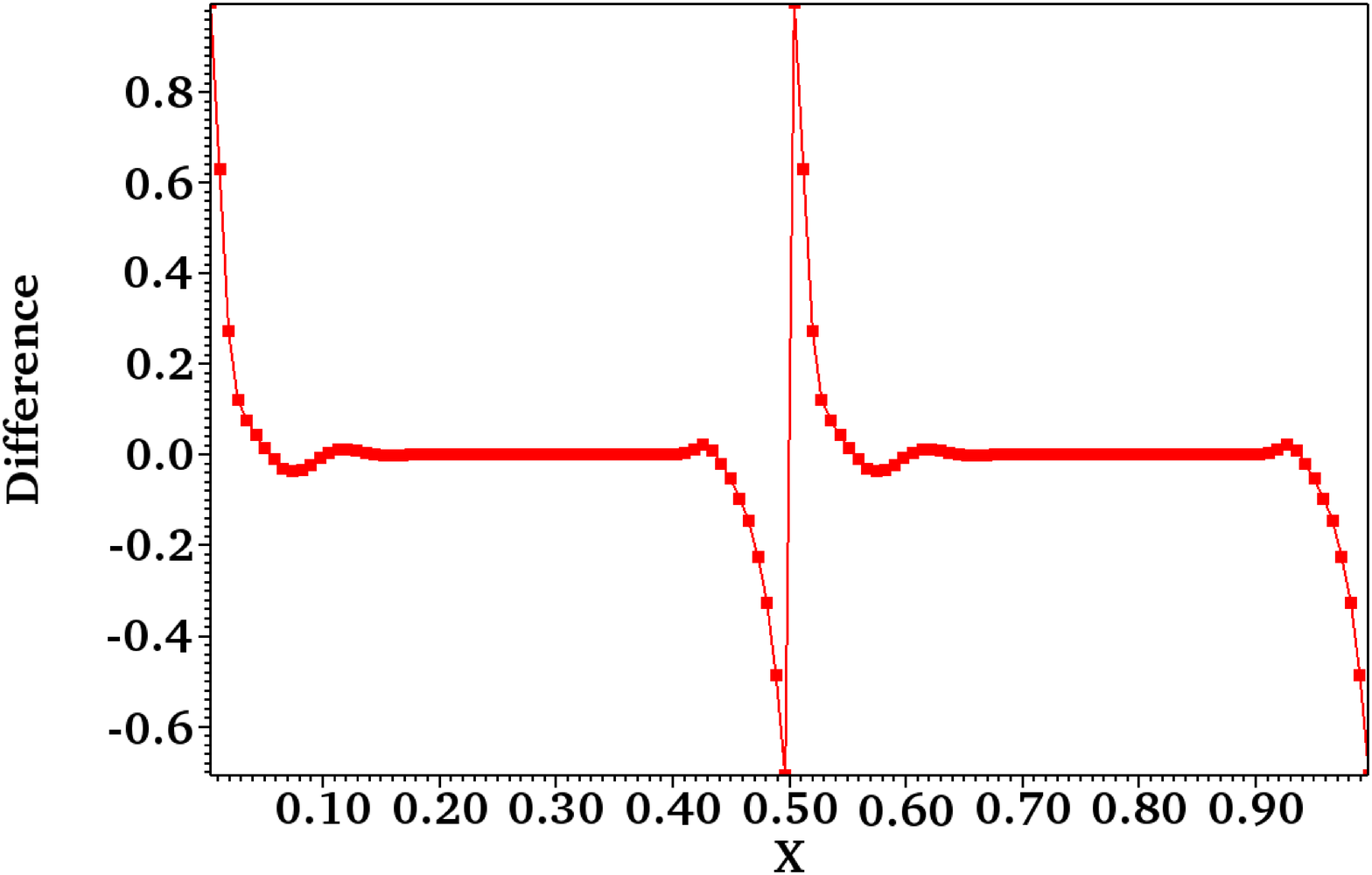}
  \includegraphics[width=0.49\textwidth]{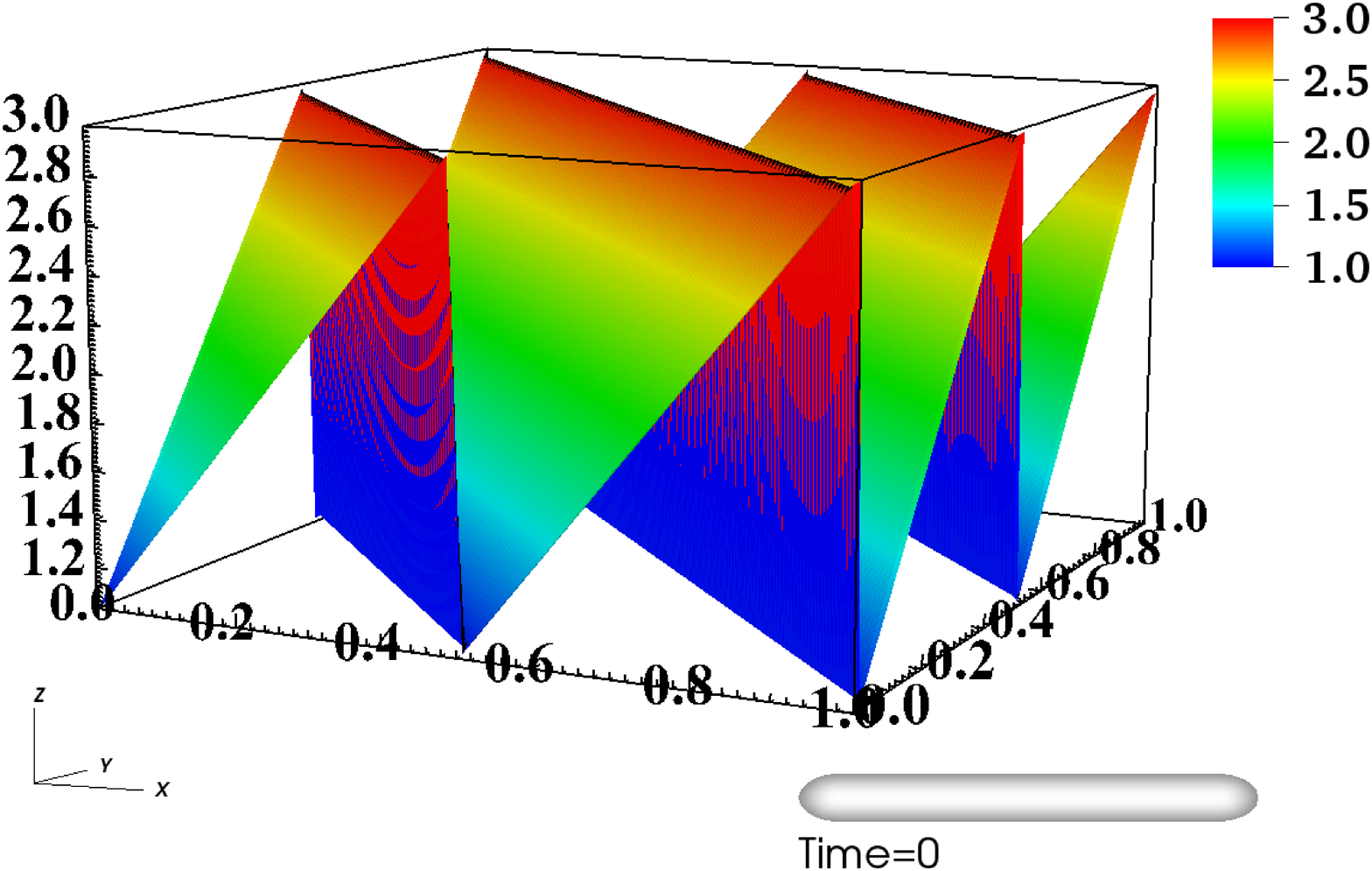}
  \includegraphics[width=0.49\textwidth]{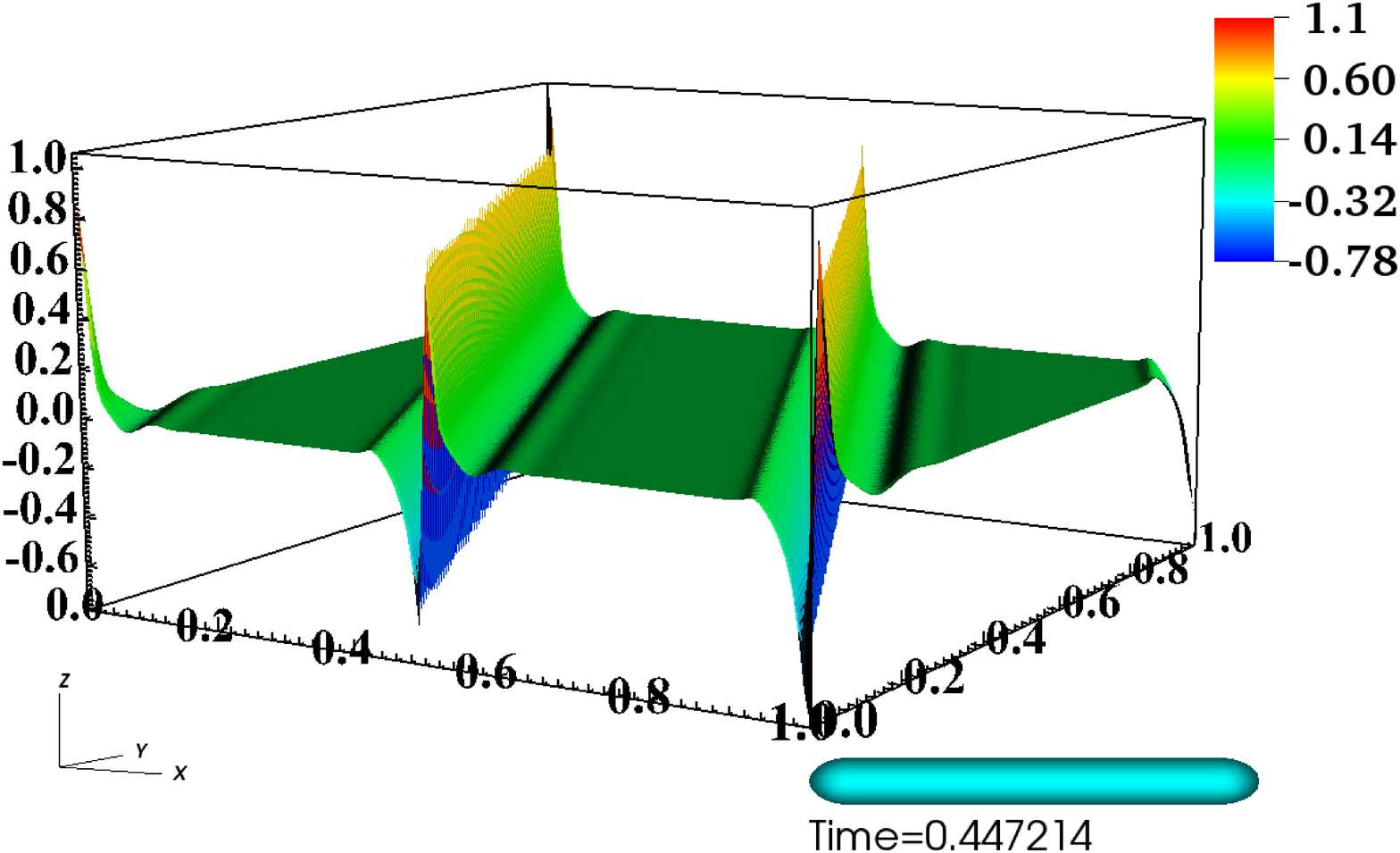}
  \includegraphics[width=0.49\textwidth]{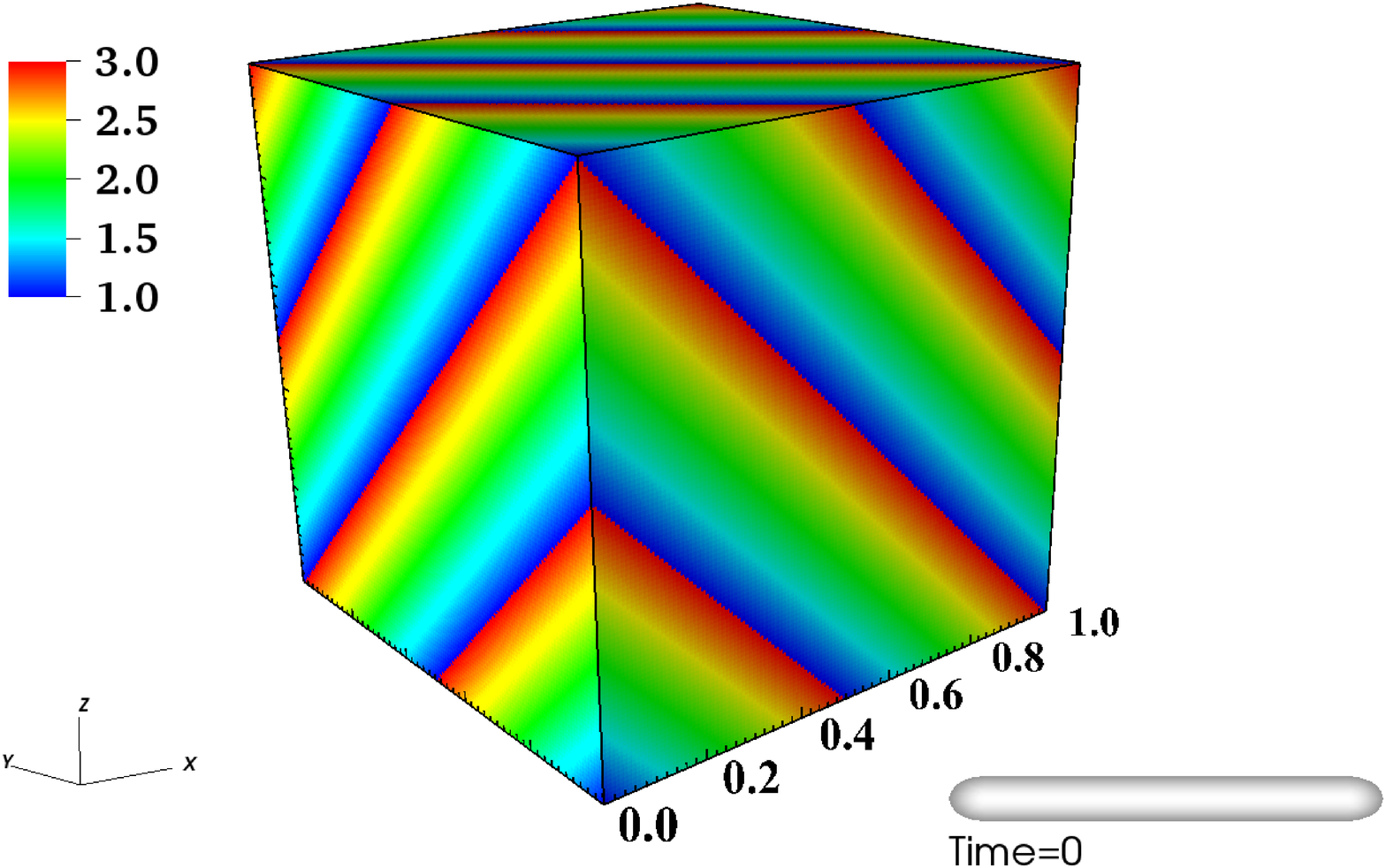}
  \includegraphics[width=0.49\textwidth]{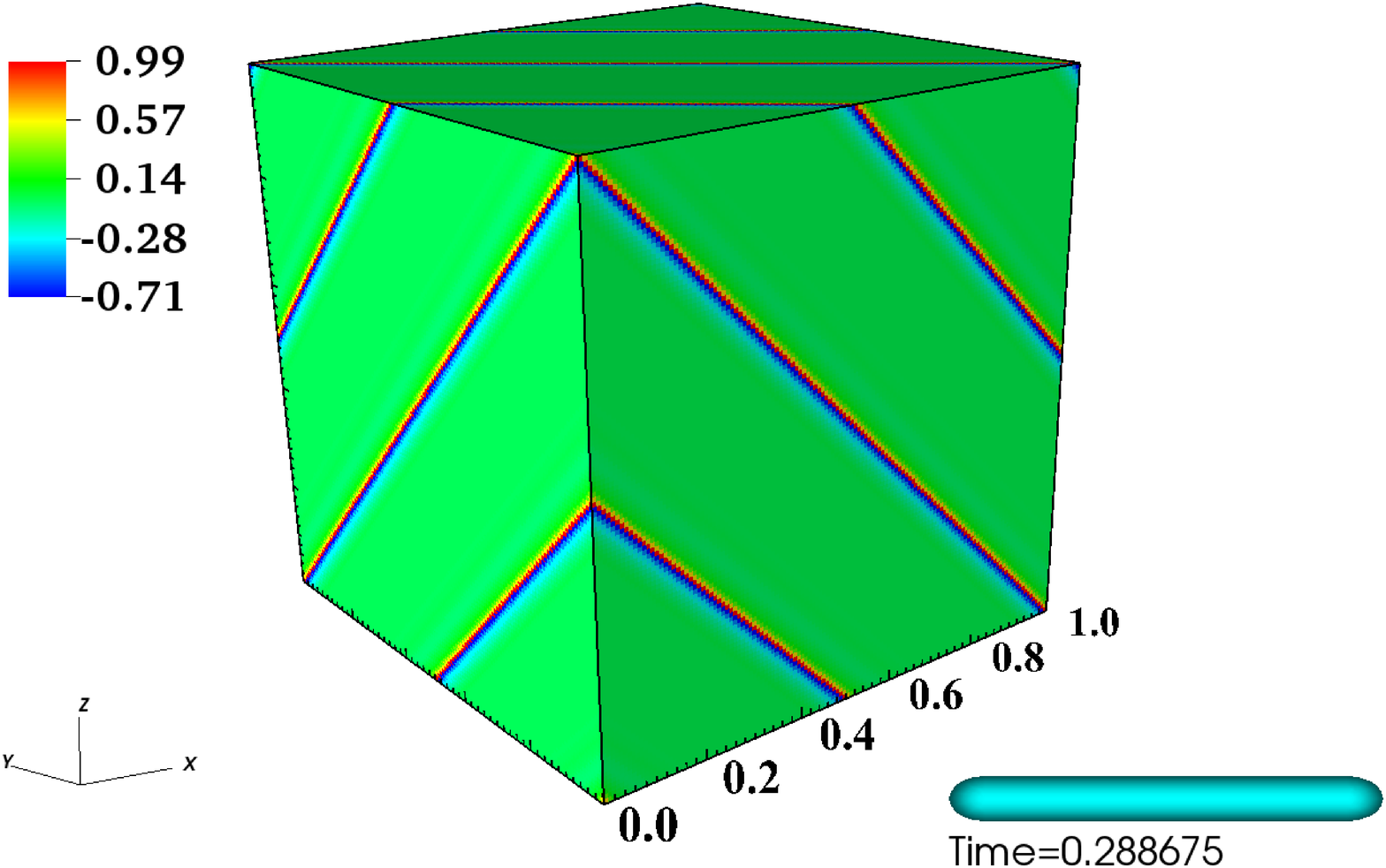}
  \caption{Plots of density for the 1D (upper), 2D (middle), and 3D (lower) versions of the \texttt{SawtoothWave} initial conditions (two wavelengths, left) and their difference from the initial conditions after one period of evolution (right) at a resolution of 128 cells in each dimension.
The 1D plot of density (upper left) shows both the final (red) and initial (green) states, revealing the degradation in accuracy resulting from the drop to first order at discontinuities.}
  \label{fig:SawtoothWave}
\end{centering}
\end{figure}

\begin{figure}
\begin{centering}
  \includegraphics[width=0.49\textwidth]{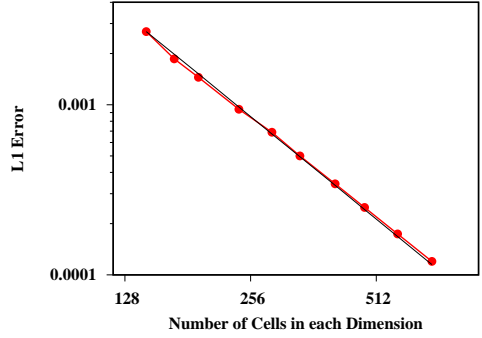}
  \caption{The L1 error (red) of the 3D \texttt{SineWave} evolved for one period converges at second order (black) as expected for a smooth problem.}
  \label{fig:SineWaveConvergence}
\end{centering}
\end{figure}

\begin{figure}
\begin{centering}
  \includegraphics[width=0.49\textwidth]{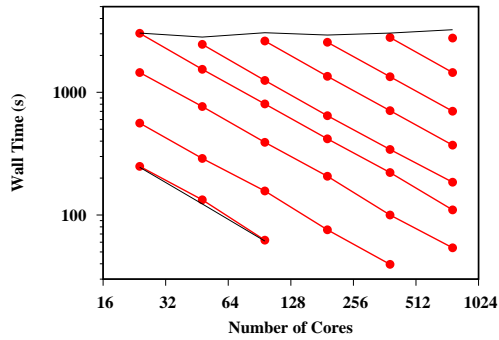}
  \caption{Scaling of the 3D \texttt{SineWave} problem with a single thread.
The red curves connect runs of the same problem size with different numbers of processes, thus demonstrating `strong scaling,' with the black curve in the lower left showing the ideal.
The problem sizes from lower left to upper right are 144, 168, 192, 240, 288, 336, 408, 480, 576, and 696 cells in each dimension.
The black curve across the top shows ideal `weak scaling' (rate of increase of wall time with increase in both problem size and number of cores) for the problem sizes considered.
}
  \label{fig:SineWaveScaling}
\end{centering}
\end{figure}

\begin{figure}
\begin{centering}
  \includegraphics[width=0.51\textwidth]{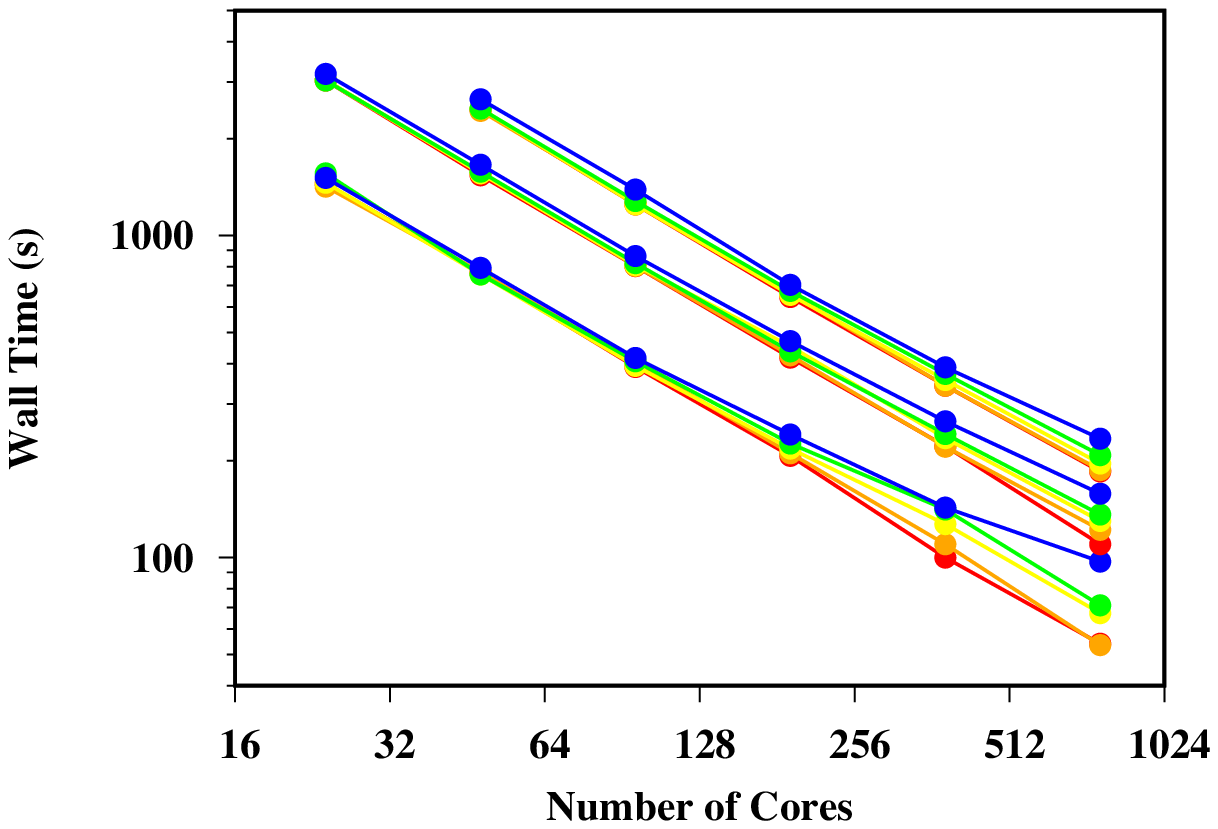}
  \caption{Scaling of the 3D \texttt{SineWave} problem with multiple threads.
Five strong scaling curves are shown for each of the selected problem sizes of 240, 288, and 336 cells in each dimension; colors correspond to thread counts 1 (red), 2 (orange), 4 (yellow), 6 (green), and 12 (blue) per process (the number of MPI processes is equal to the number of cores divided by the number of threads per process).}
  \label{fig:SineWaveThreading}
\end{centering}
\end{figure}

Figures~\ref{fig:SineWave}-\ref{fig:SineWaveThreading} show both computational and performance results from the plane wave test problems.
The initial density field and its error after one period of evolution in 1D, 2D, and 3D versions of the
\texttt{SineWave} and \texttt{SawtoothWave} problems are shown in Figures~\ref{fig:SineWave} and \ref{fig:SawtoothWave} respectively.
The second order convergence we expect from a smooth problem is demonstrated for the 3D \texttt{SineWave} in Fig.~\ref{fig:SineWaveConvergence}.
The scaling results in Figs.~\ref{fig:SineWaveScaling} and \ref{fig:SineWaveThreading}---obtained on the machine Comet at the San Diego Supercomputing Center---are also shown for the 3D \texttt{SineWave} problem.
Both strong and weak scaling with pure MPI parallelism are excellent (Fig.~\ref{fig:SineWaveScaling}). The scaling of our initial draft of hybrid parallelism using OpenMP threading is not as good as pure MPI at higher core counts (Fig.~\ref{fig:SineWaveThreading}) for modest problem sizes.
Because the spread becomes less severe with increasing problem size, we attribute this to `work starvation': the smaller amount of work per thread as the number of threads increases corresponds to increased relative overhead of frequent thread launching.
Nevertheless these initial results with threading are encouraging in terms of possible future exploitation of accelerators on heterogeneous architectures.

%%%%%%%%%%%%%%%%%%%%%%%%
\subsection{Riemann Problem}
\label{sec:RiemannProblem}

A Riemann problem features two piecewise constant initial states separated by an initial discontinuity. 
Tests with a few different sets of initial conditions are shown in Ref.~\cite{Cardall2014GENASIS:-Genera}; here we focus on the classic Sod shock tube values
\begin{eqnarray}
  \left[\rho,\, v,\, p \right]_{L} &=& \left[1.0,\, 0.0,\, 1.0 \right], \nonumber \\
  \left[\rho,\, v,\, p \right]_{R} &=& \left[0.1,\, 0.0,\, 0.125\right]  
\end{eqnarray}
for density $\rho$, speed $v$, and pressure $p$ on the `left' ($L$) and `right' ($R$) sides of the discontinuity.
The fluid is governed by a polytropic equation of state with adiabatic index $1.4$.
The classic 1D problem locates the initial discontinuity at position $x = 0.5$, and by default we extend this to 2D and 3D by defining the plane of discontinuity to also intersect the $y$ and $z$ axes at $y = 0.5$ and $z = 0.5$ respectively.

Setting up the Riemann problem involves the same five tasks required for the advection of a plane wave discussed in Section~\ref{sec:PlaneWave}, the difference in terms of deployment of \textsc{GenASiS} \texttt{Mathematics} classes being the use of reflecting boundary conditions.
Like \texttt{PlaneWaveTemplate}, the class \texttt{RiemannProblemForm} is an extension of the \texttt{Mathematics} class \texttt{Integrator\_C\_Template}, which evolves a generic conserved current (see Fig.~\ref{fig:FluidDynamics}).
The five setup tasks are accomplished in the \texttt{Initialize} method of \texttt{RiemannProblemForm}, which has a structure similar to Listing~\ref{lst:Subroutine_Initialize_PW}.
The first task---initializing position space---is just as in lines~11-16 of Listing~\ref{lst:Subroutine_Initialize_PW}, except that the lines
\begin{lstlisting}
  do iD = 1, PS % nDimensions
    call PS % SetBoundaryConditionsFace &
           ( [ 'REFLECTING', 'REFLECTING' ], iDimension = iD )
  end do !-- iD
\end{lstlisting}
appear \textit{between} lines~15 and 16, the calls to the \texttt{Initialize} and \texttt{CreateChart} methods of the position space object.
Each iteration of the above loop sets the inner and outer boundary conditions in one of the position space dimensions.

Aside from the different boundary and initial conditions, the other important difference from the plane wave case is the type of fluid selected.
Whereas a pressureless fluid was specified by the string argument \texttt{'DUST'} in line~29 of Listing~\ref{lst:Subroutine_Initialize_PW}, the string \texttt{'POLYTROPIC'} is used in the initialization of the Riemann problem (and all the problems in subsequent subsections).
Consequently the polymorphic \texttt{Field} member of \texttt{Field\_CSL\_Template} will be allocated
as \texttt{Fluid\_P\_P\_Form} (`fluid perfect polytropic') instead of \texttt{Fluid\_D\_Form} (`fluid dust'; see the lower right of Fig.~\ref{fig:FluidDynamics}).

\begin{figure}
\begin{centering}
  \includegraphics[width=0.49\textwidth]{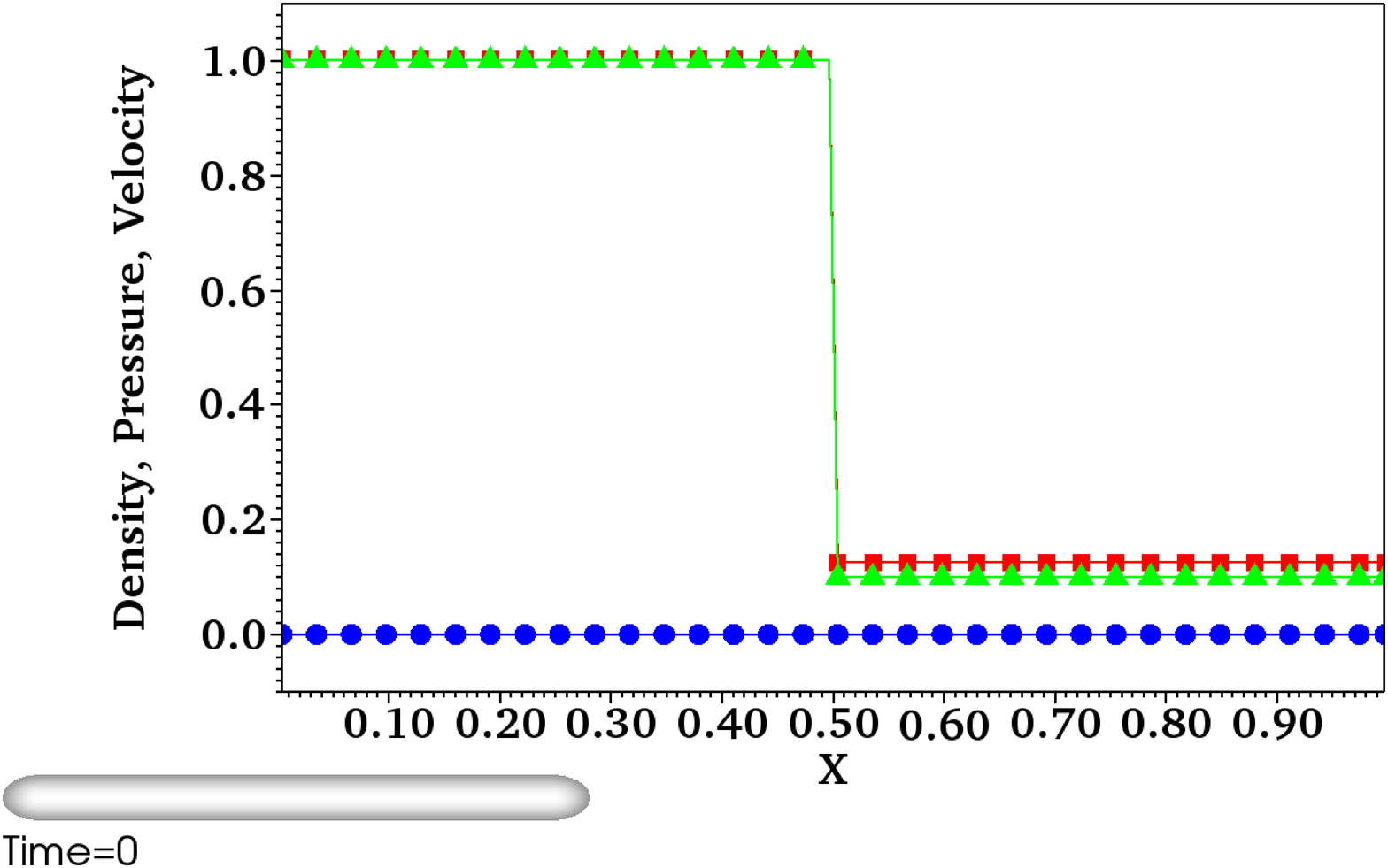}
  \includegraphics[width=0.49\textwidth]{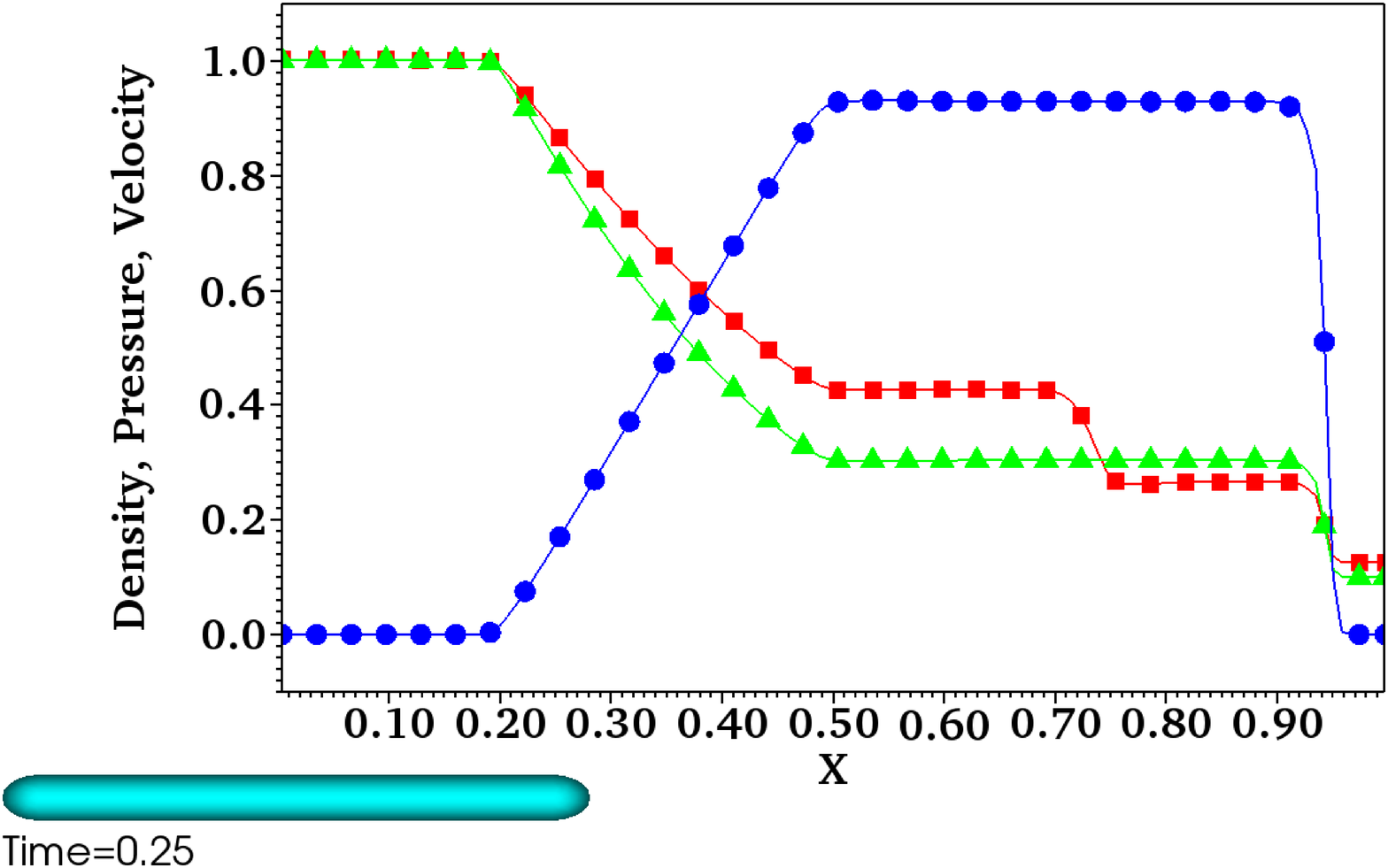}
  \includegraphics[width=0.49\textwidth]{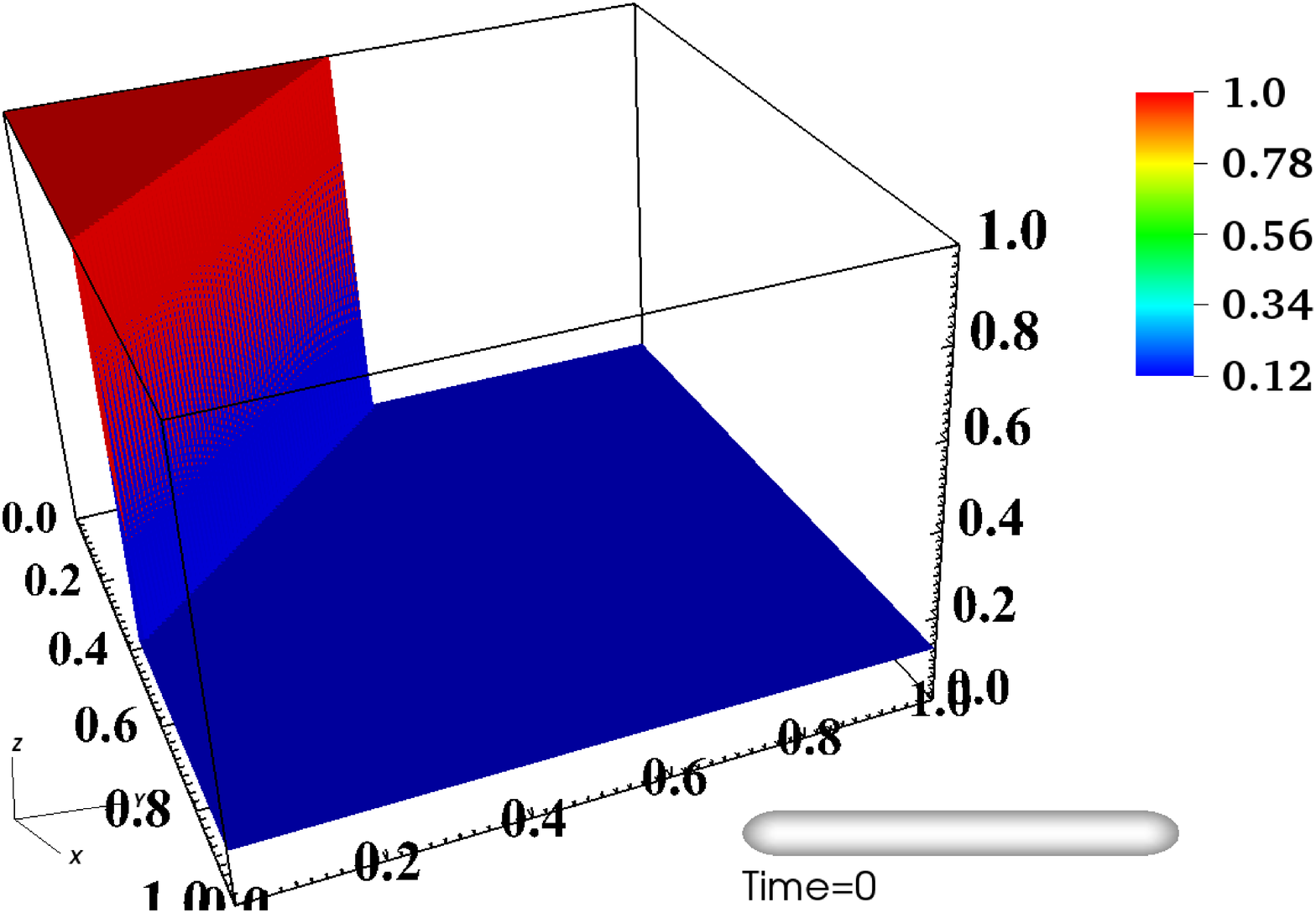}
 \includegraphics[width=0.49\textwidth]{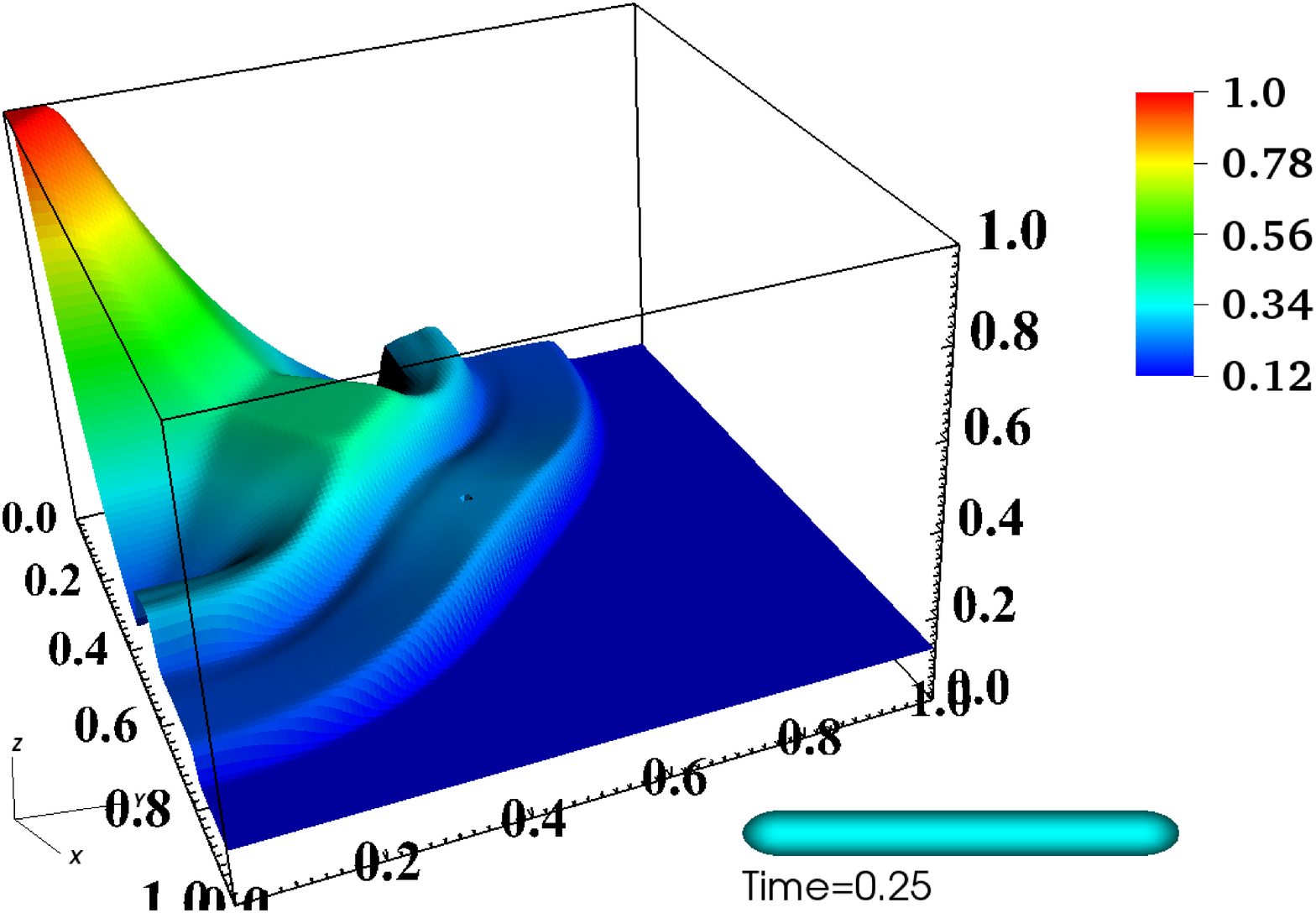}
  \includegraphics[width=0.49\textwidth]{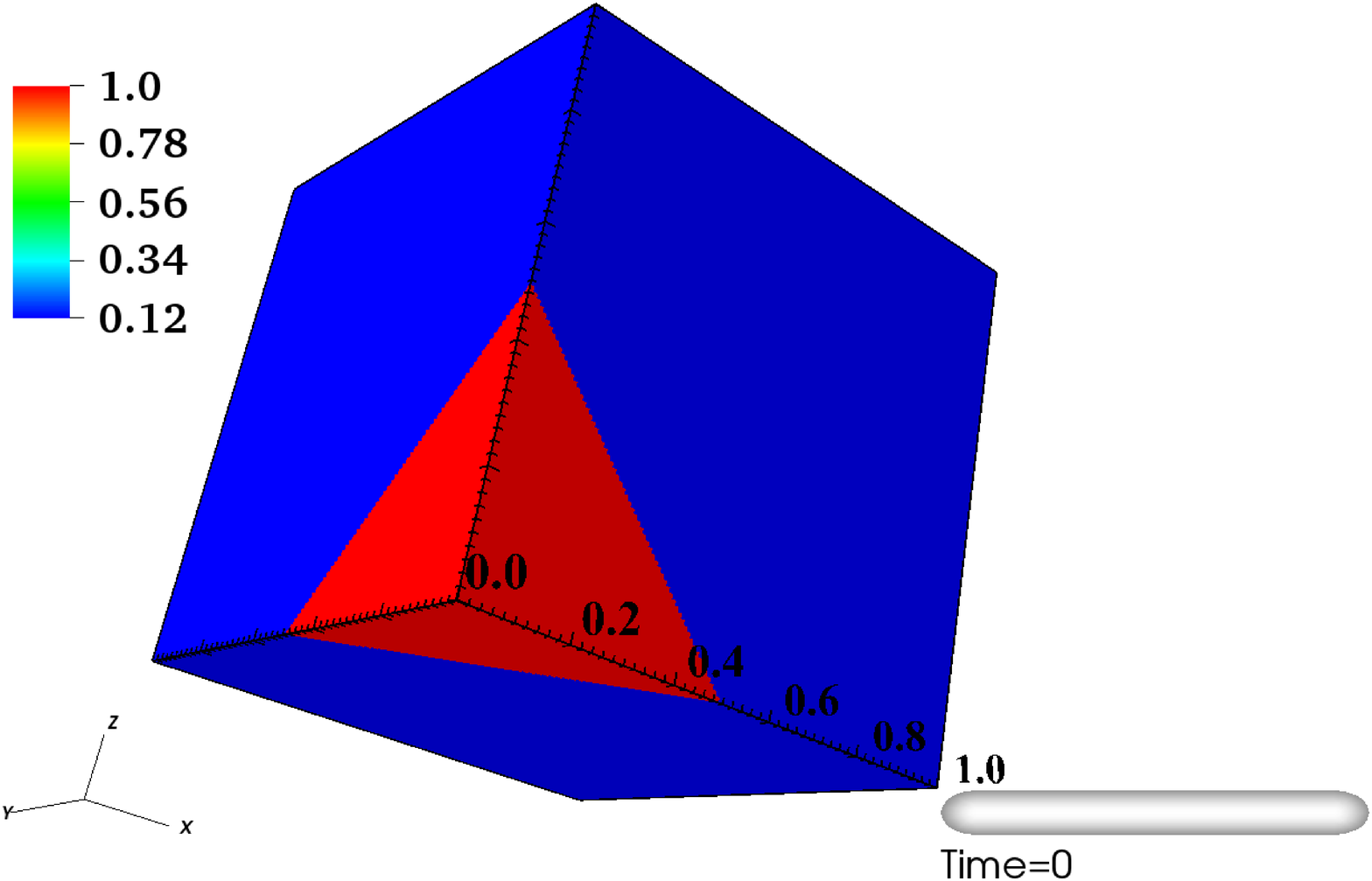}
  \includegraphics[width=0.49\textwidth]{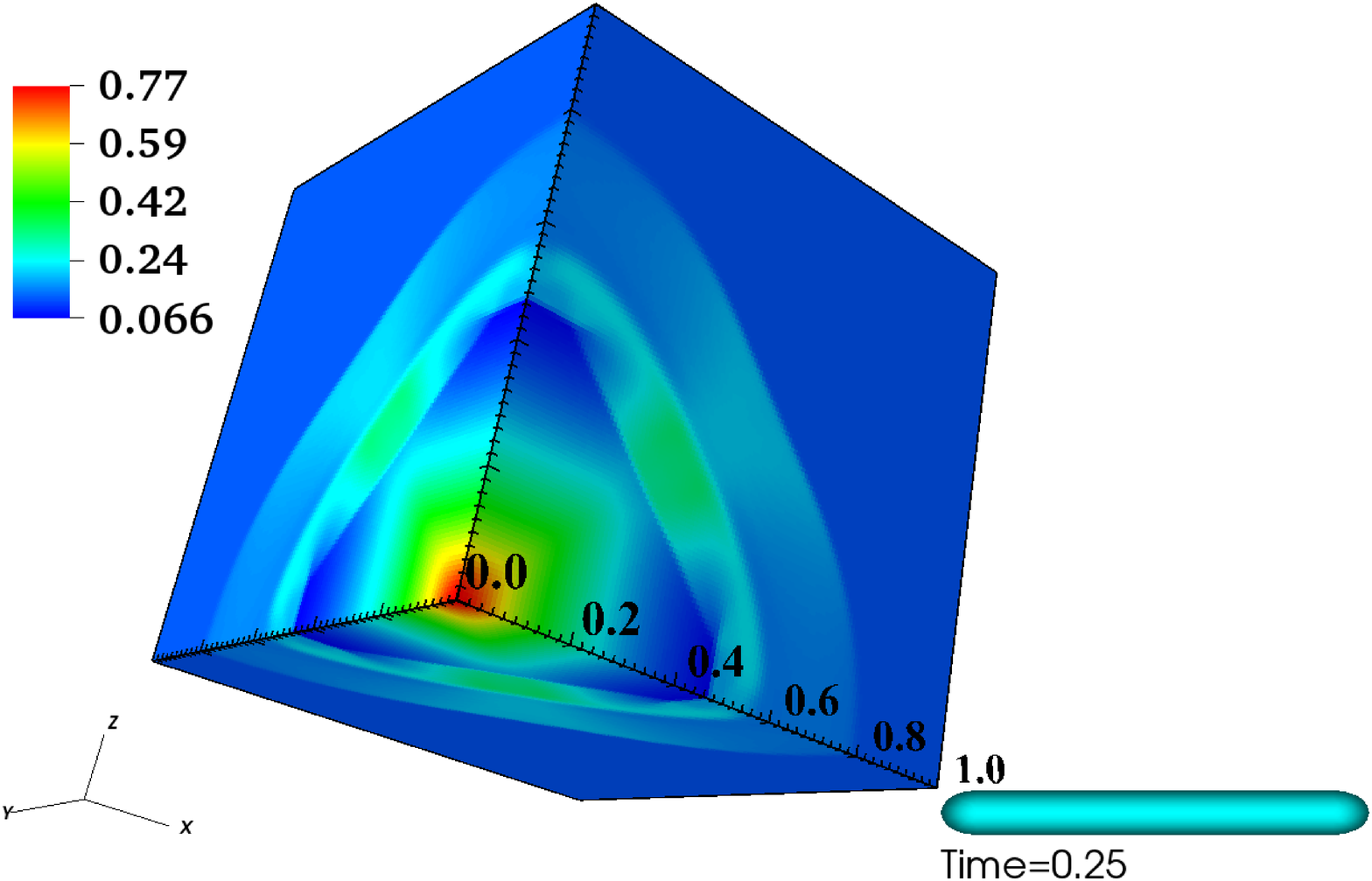}
  \caption{Density in 1D (upper), 2D (middle), and 3D (lower) versions of {\tt RiemannProblem} at $t=0$ (left) and $t=0.25$ (right) at a resolution of 128 cells in each dimension. For 1D, pressure (green triangles) and velocity (blue circles) are also plotted in addition to density (red squares; in all cases, one symbol every four cells).}
  \label{fig:RiemannProblem}
\end{centering}
\end{figure}

Figure~\ref{fig:RiemannProblem} shows results from the \texttt{RiemannProblem} in 1D, 2D, and 3D.
The left panels show the initial condition at time $t = 0.0$.
The familiar shock, contact discontinuity, and rarefaction waves emanating from the initial discontinuity are apparent in the right panels at $t = 0.25$, before the shock arrives at and reflects off the opposite walls.

\begin{figure}
\begin{centering}
  \includegraphics[width=0.49\textwidth]{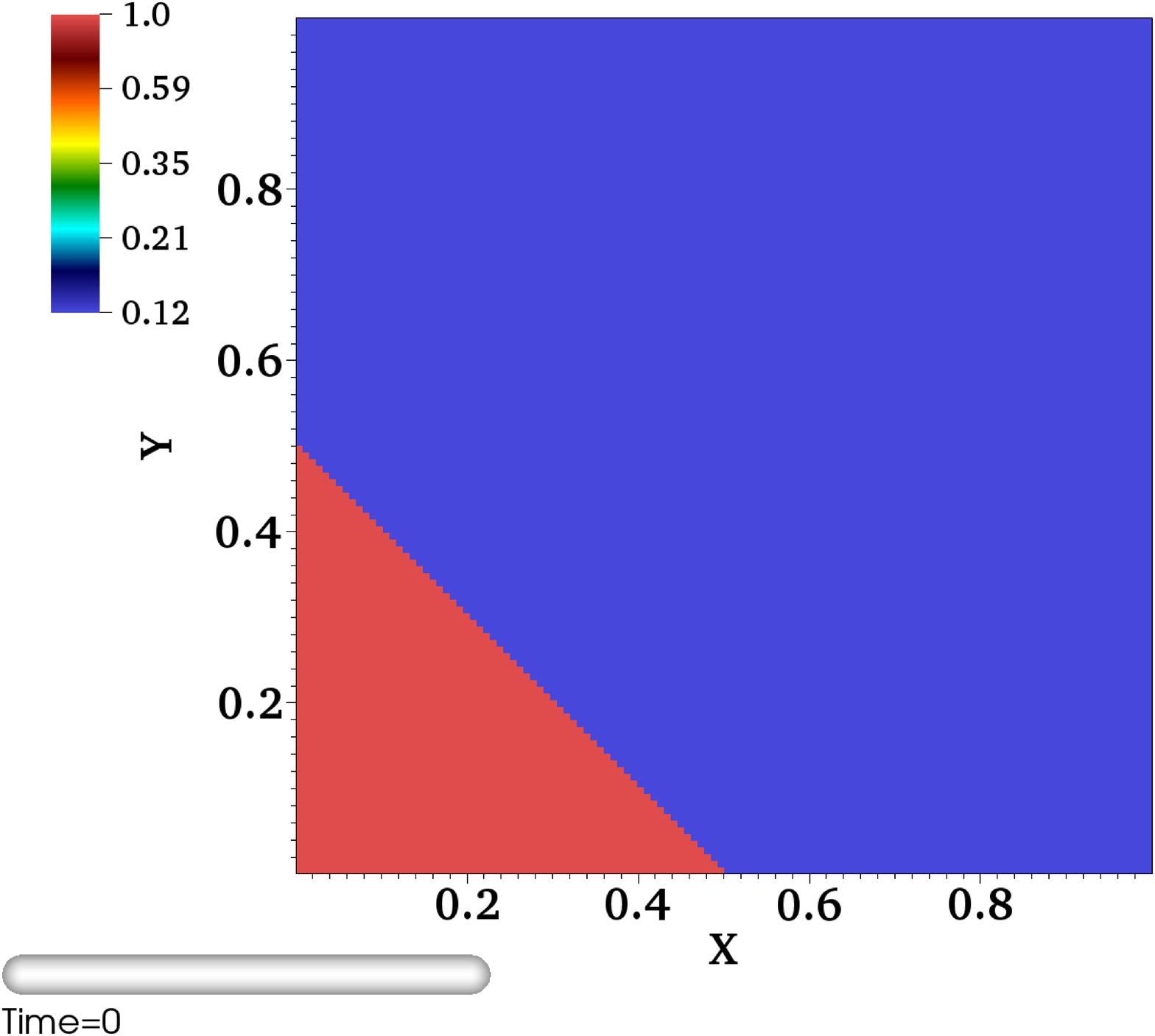}
  \includegraphics[width=0.49\textwidth]{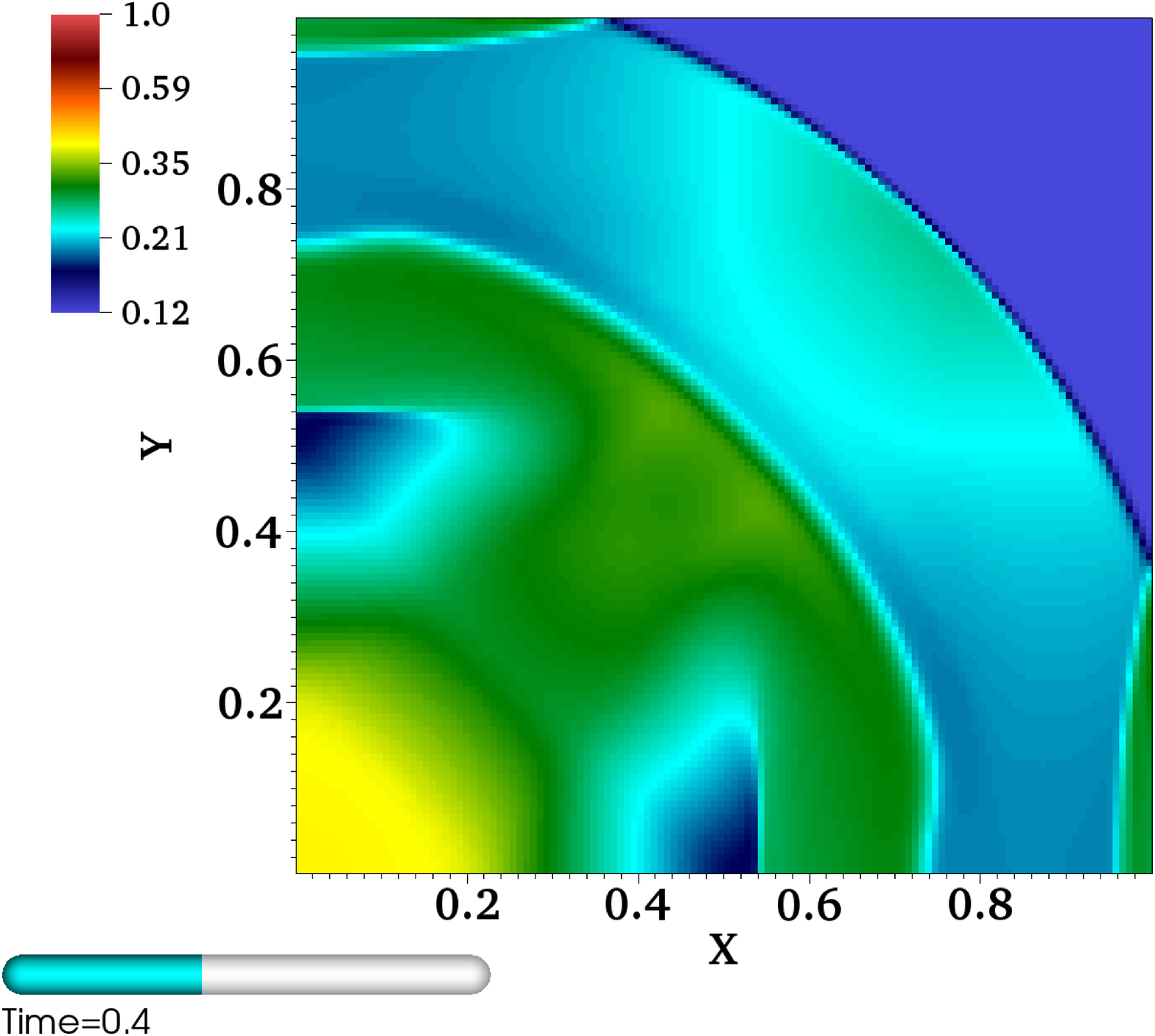}
  \includegraphics[width=0.49\textwidth]{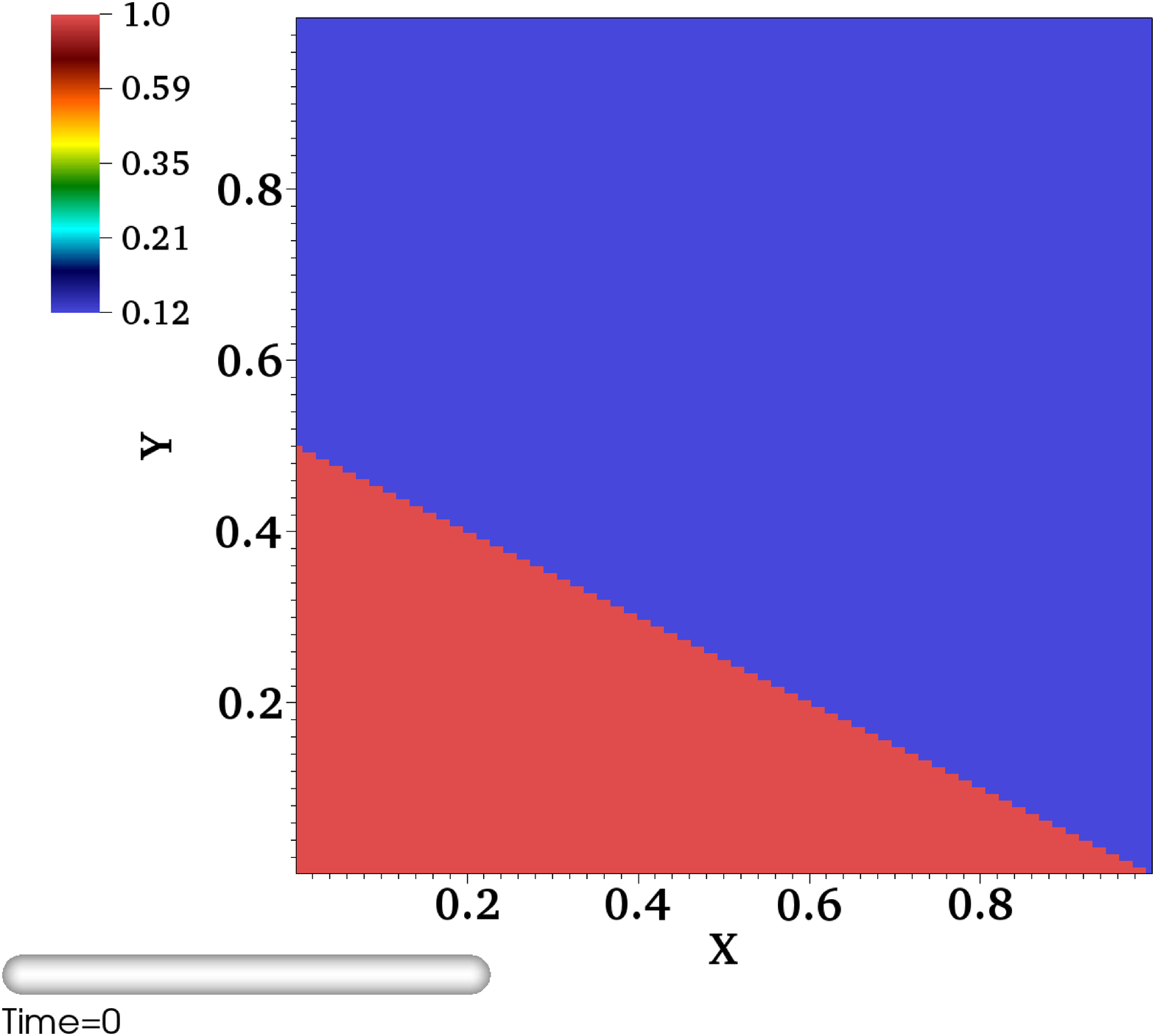}
  \includegraphics[width=0.49\textwidth]{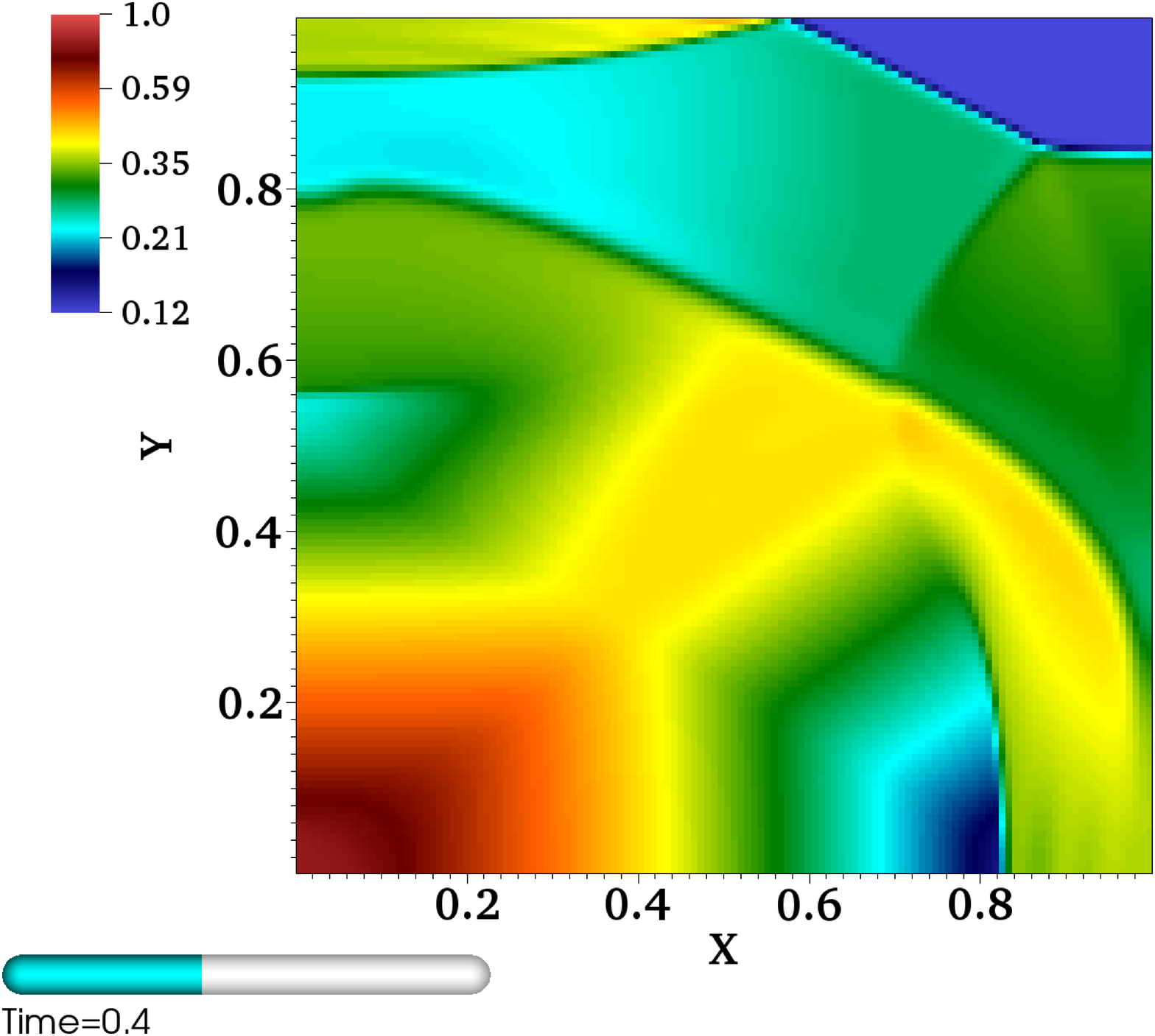}
  \caption{Density in symmetric (upper panels) and asymmetric (lower panels) versions of the 2D \texttt{RiemannProblem}, at time $t = 0.0$ (left panels) and $t = 0.4$ (right panels).}
  \label{fig:RiemannProblemComparison}
\end{centering}
\end{figure}

Figure~\ref{fig:RiemannProblemComparison} introduces an asymmetric 2D \texttt{RiemannProblem}.
Different from the symmetric initial condition in the upper left panel of Fig.~\ref{fig:RiemannProblemComparison}, the initial discontinuity crosses the $x$-axis at $x = 1.0$ in the asymmetric case shown in the lower left panel.
Snapshots at $t = 0.4$ in the right panels show that the shock has just begun reflecting off both opposite walls in the symmetric case, while shock reflection off the right wall and passage through the contact discontinuity began right away and is further advanced in the asymmetric case.

\begin{figure}
\begin{centering}
  \includegraphics[width=0.49\textwidth]{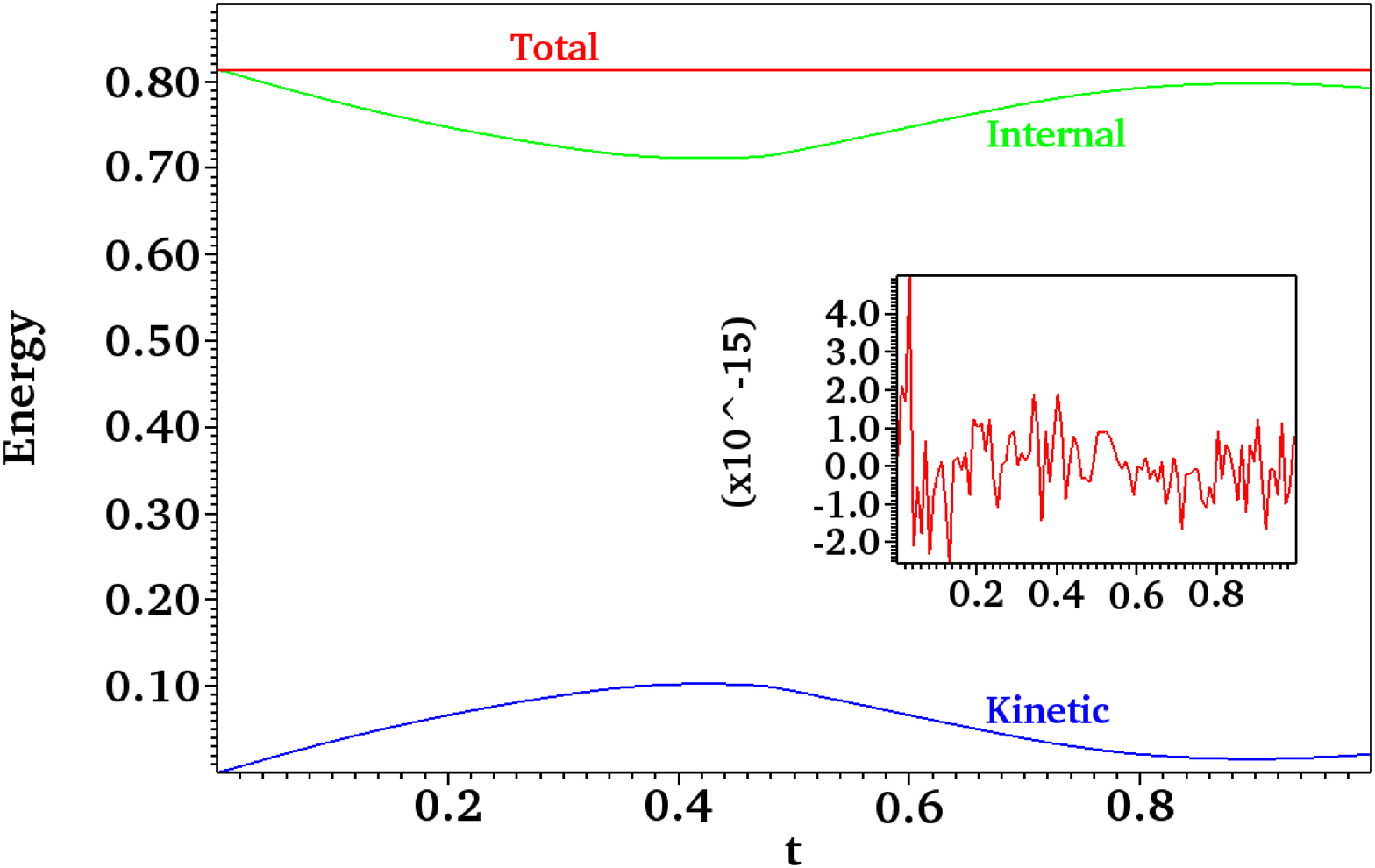}
  \includegraphics[width=0.49\textwidth]{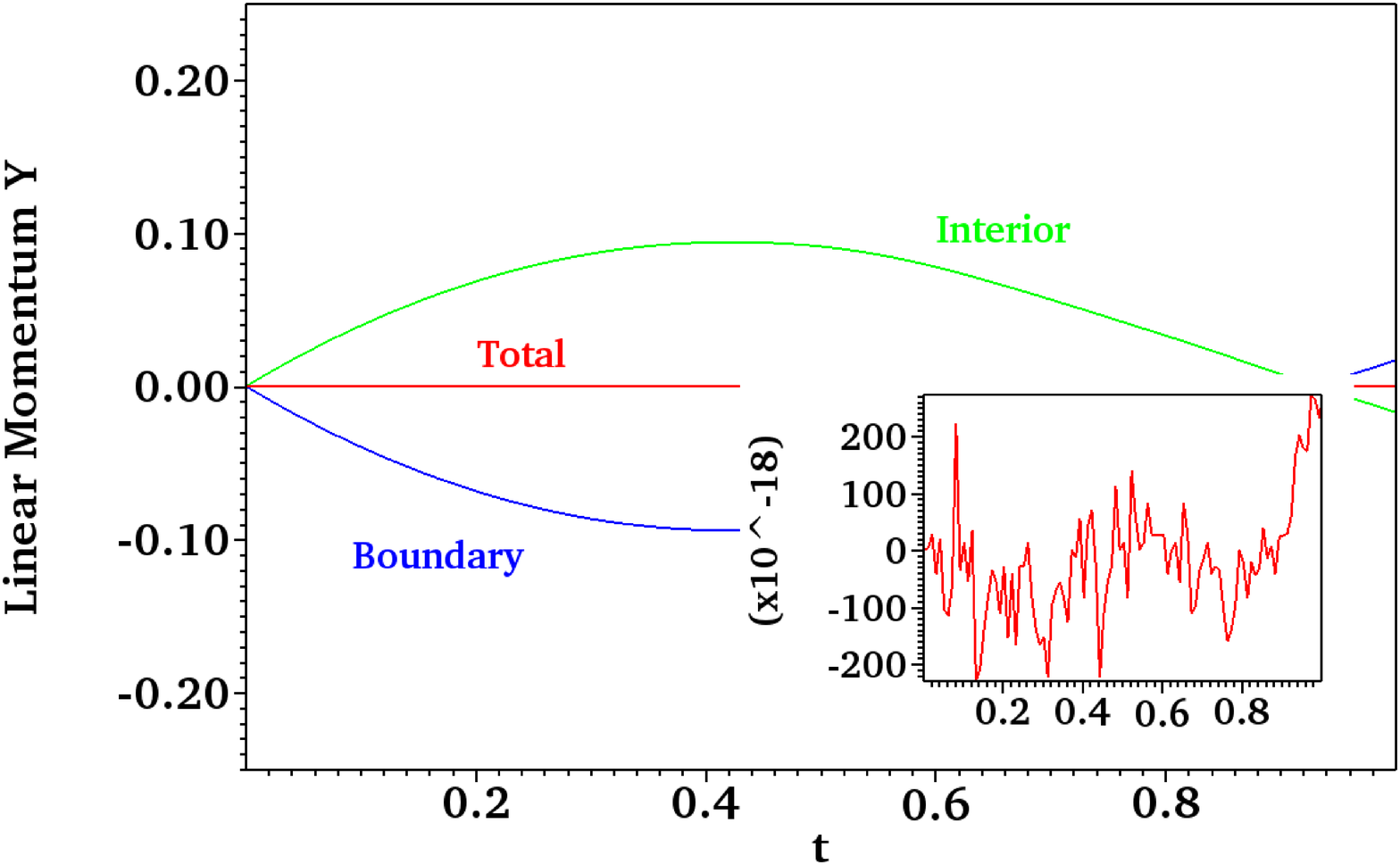}
  \includegraphics[width=0.49\textwidth]{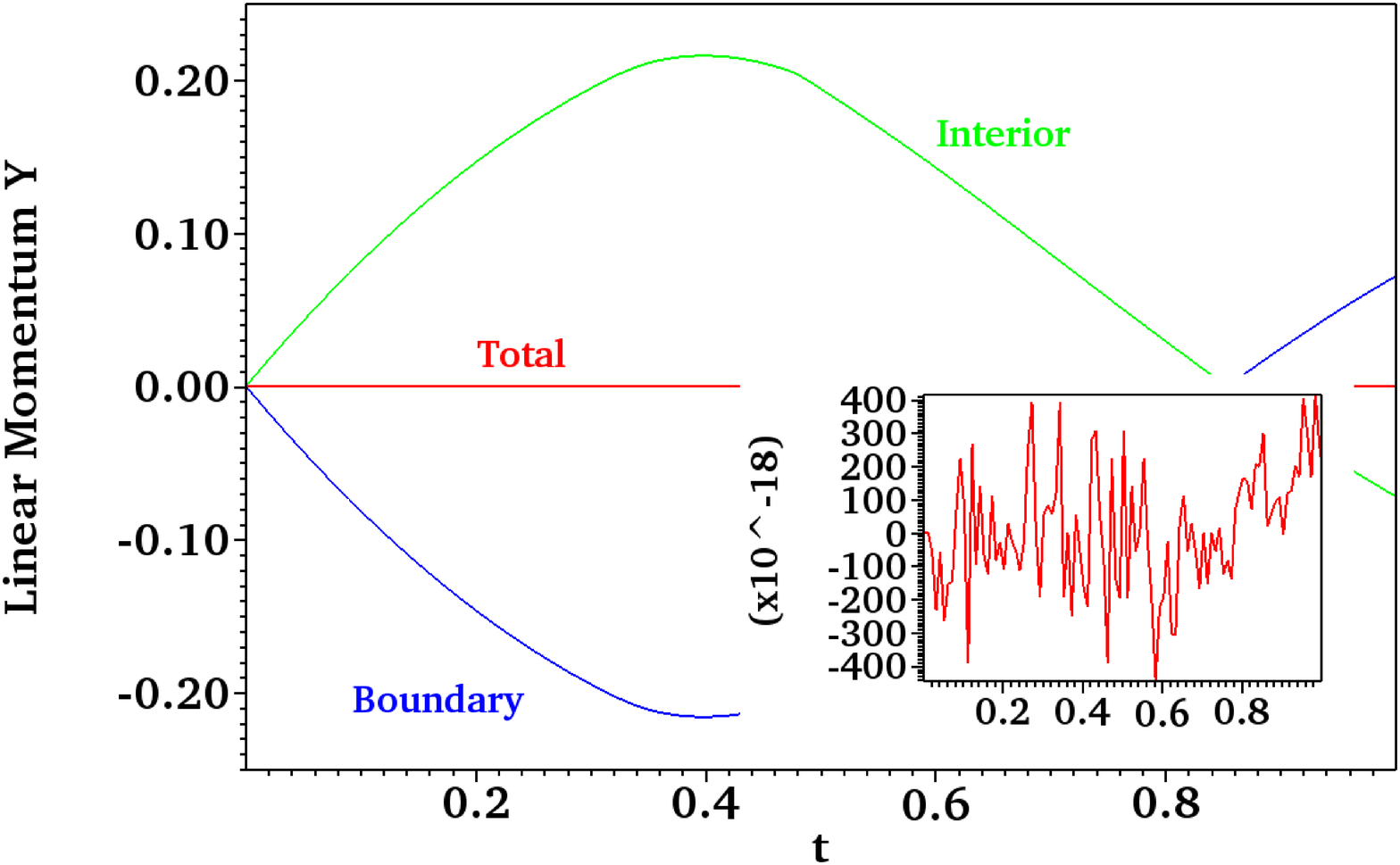}
  \includegraphics[width=0.49\textwidth]{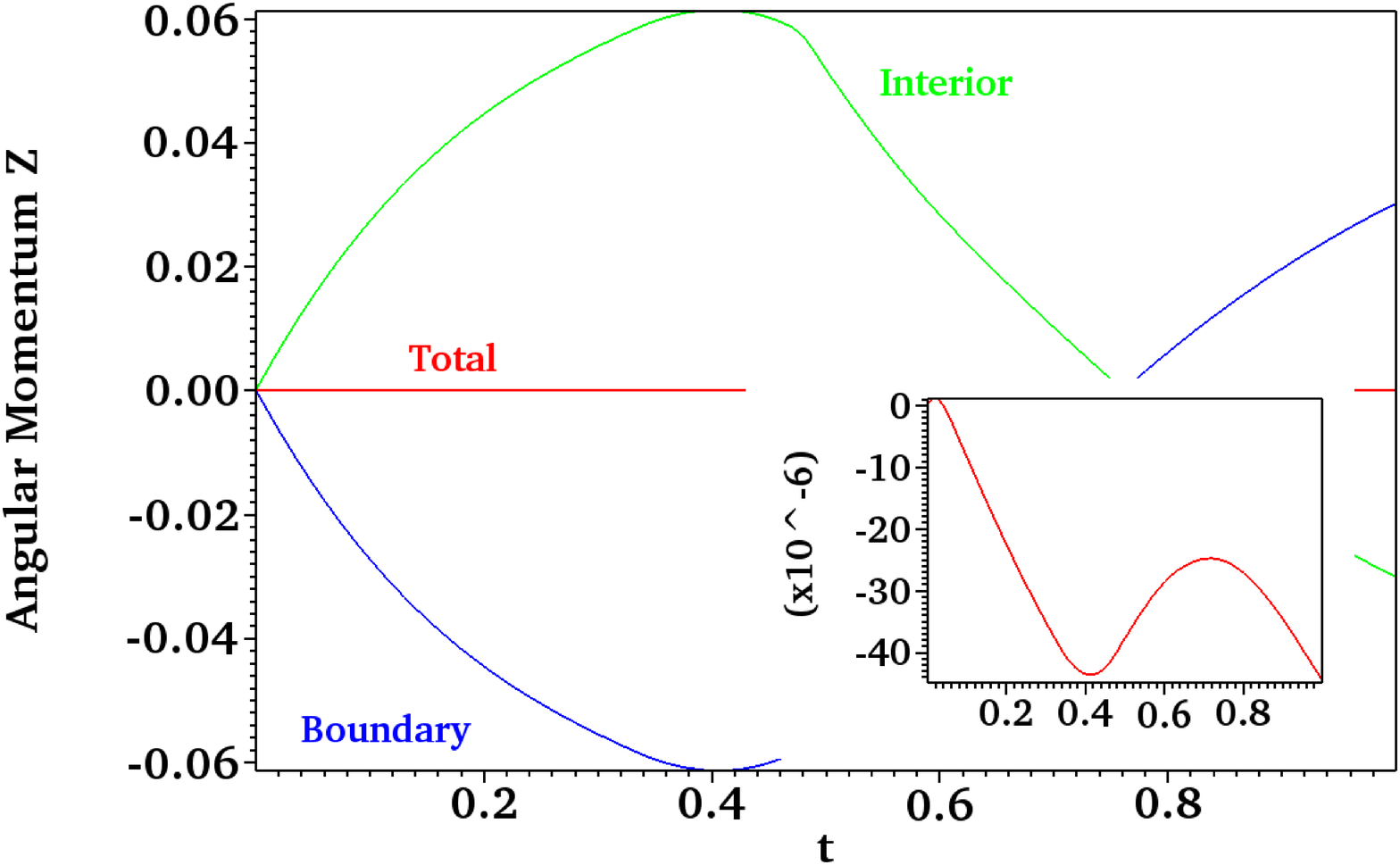}
  \caption{Selected global tallies in the asymmetric 2D \texttt{RiemannProblem}: energy (upper left), $x$ and $y$ components of linear momentum (upper right, lower left), and $z$ component of angular momentum (lower right). 
 Upper left, the internal (green) and kinetic (blue) energies on the domain sum to a total (red) that is conserved to machine precision (inset). 
 Upper right and lower left, the components of linear momentum on the domain (green) balance the momentum imparted to the boundary (blue), again summing to totals (red) that are conserved to machine precision. 
 Lower right, the imbalance in linear momentum components produces a net angular momentum on the domain (green) that is balanced (red) by that imparted to the boundary (blue), but not to machine precision (inset).
}
  \label{fig:RiemannProblemAsymmetric}
\end{centering}
\end{figure}

\begin{figure}
\begin{centering}
  \includegraphics[width=0.49\textwidth]{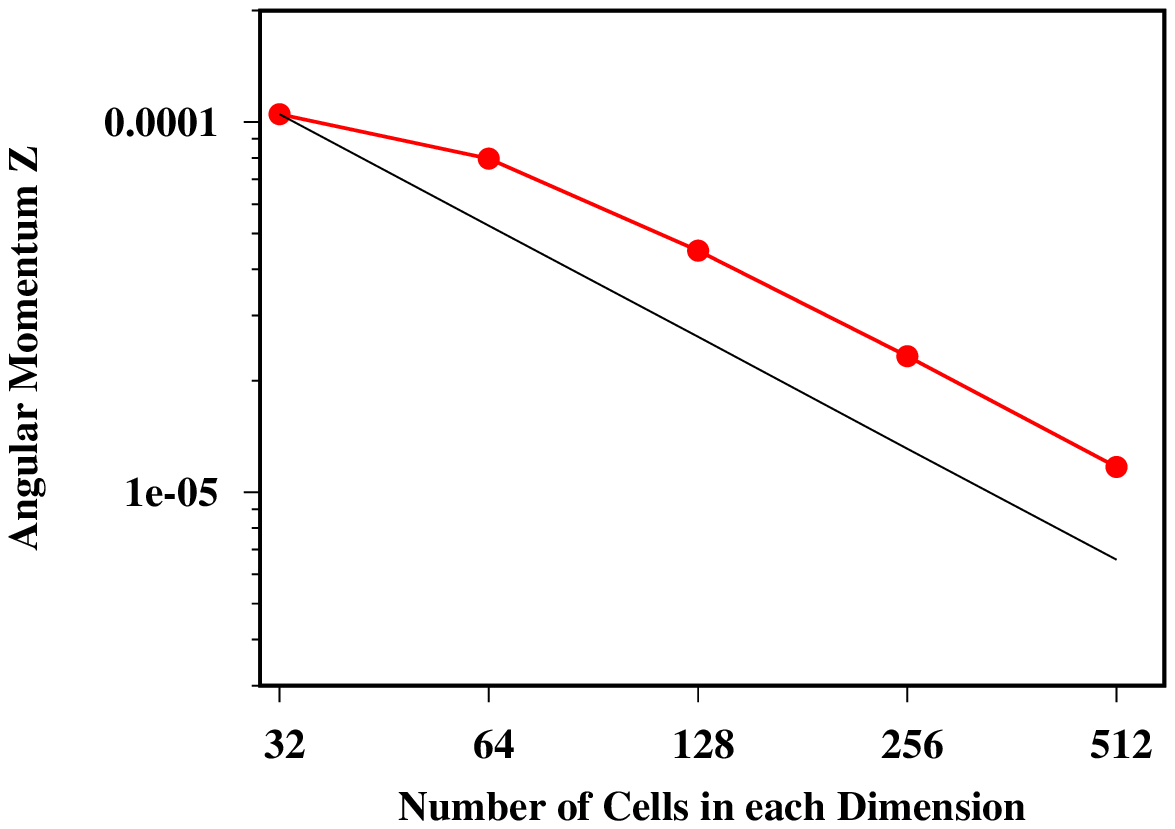}
  \caption{The total angular momentum on the domain and boundary should tally to zero in an asymmetric 2D \texttt{RiemannProblem}; in practice, as resolution increases the numerical solution (red) approaches this ideal at first order (black) consonant with the discontinuous nature of the problem.}
  \label{fig:RiemannProblemConvergence}
\end{centering}
\end{figure}

The asymmetric 2D \texttt{RiemannProblem} provides an interesting illustration in Figs.~\ref{fig:RiemannProblemAsymmetric} and \ref{fig:RiemannProblemConvergence} of diagnostic output from volume and surface integrals performed by tally classes.
(Classes \texttt{Tally\_P\_P\_Form} and \texttt{Tally\_D\_Form} are specialized for polytropic and dust fluids respectively, as extensions from a generic conserved current \texttt{Tally\_C\_Form}, as seen in the lower center of Fig.~\ref{fig:FluidDynamics}.)
The reflecting boundary conditions keep energy inside in the computational domain; the upper left panel of Fig.~\ref{fig:RiemannProblemAsymmetric} and its inset show that the finite volume scheme conserves the total fluid energy to machine precision, even as kinetic energy develops and evolves at the expense of internal energy as a consequence of the initial discontinuity.
The components of linear momentum are also conserved to machine precision in the upper right and lower left panels, but in this case only when the momentum imparted to the walls is taken into account: physically, the reflecting boundary conditions can be interpreted as a containing box of infinite mass that absorbs arbitrary momentum.
The unequal magnitude of the linear momentum components in the domain implies a net interior angular momentum, which should be exactly balanced by an opposing net torque on the bounding box.
To the naked eye this is indeed the case in the lower right panel; but the inset shows that total angular momentum conservation---while of physically reasonable (to the naked eye) magnitude---is actually far from machine precision, as the solver does not achieve this by construction.
Figure~\ref{fig:RiemannProblemConvergence} shows first order convergence for angular momentum conservation consonant with the discontinuous nature of the problem.

%%%%%%%%%%%%%%%%%%%%%%%%
\subsection{Rayleigh-Taylor Instability}
\label{sec:RayleighTaylor}

The Rayleigh-Taylor instability test we implement here follows that presented in Ref.~\cite{Cardall2014GENASIS:-Genera} and references therein, in which a 2D region of higher density $\rho = 2$ initially overlies a region of lower density $\rho = 1$ in the presence of a uniform downward acceleration $g = 0.1$, but in hydrostatic equilibrium with pressure $p = p_b - \rho g y$, with base pressure $p_b = 2.5$ at the interface at $y = 0$.
As before the fluid is governed by a polytropic equation of state with adiabatic index $1.4$.
A 2D computational domain covers the region $[x,y]\in[-0.25,0.25]\times[-0.75,0.75]$, with periodic and reflecting boundaries in the $x$ and $y$ directions respectively.
Vanishing velocity components characterize the initial equilibrium; the instability is initiated with a perturbation of the $y$ velocity component,
\begin{equation}
v_y = \frac{A}{4} \left[ 1 + \cos \left( 4\pi x \right) \right] \left[ 1 + \cos \left( 3\pi y \right) \right],
\end{equation}
with $A = 0.01$.

This problem introduces a few new elements.
\textsc{GenASiS} assumes a 3D problem by default, which can be overridden with a command line argument as discussed at the end of the first paragraph of Section~\ref{sec:PlaneWave}.
But this default can be also be overridden with an optional argument in the initialization of \texttt{PROGRAM\_HEADER} by the driver \texttt{program}.
Specifically, instead of line~10 of Listing~\ref{lst:Program_SW_Outline}, we have
\begin{lstlisting}
  call PROGRAM_HEADER % Initialize &
         ( 'RayleighTaylor', DimensionalityOption = '2D' )
\end{lstlisting}
in order to set up and run a problem we intend specifically for 2D without having to use a command line option to override the 3D default.
In terms of interfacing with \textsc{GenASiS} \texttt{Mathematics} classes, the new elements are specification of non-default values for the extent of the domain and the number of cells; the use of source terms; and a custom tally of global diagnostics.
These differences occur in the \texttt{Initialize} method of \texttt{RayleighTaylorForm}; 
as with previously discussed problems, the latter is an extension of the \texttt{Mathematics} class \texttt{Integrator\_C\_Template}, and the former has a structure similar to Listing~\ref{lst:Subroutine_Initialize_PW}.

In contrast to the previously discussed problems, \texttt{RayleighTaylor} does not accept the default domain size of $[0,1]$ in each dimension, nor the default resolution of $32$ cells in each dimension.
(Figures in the previous subsections with $128$ cells in each dimension were generated with command line arguments.)
While the call to the \texttt{CreateChart} method in line~16 of Listing~\ref{lst:Subroutine_Initialize_PW} contained no arguments (aside from the passed position space object), we have instead 
\begin{lstlisting}
  call PS % CreateChart &
         ( MinCoordinateOption = [ -0.25_KDR, -0.75_KDR ], &
           MaxCoordinateOption = [ +0.25_KDR, +0.75_KDR ], &
           nCellsOption = [ 128, 384 ] )
\end{lstlisting}
for the corresponding call in \texttt{RayleighTaylor}, showing how these particular default chart parameters are overridden with optional arguments.
These parameters can be further overridden at runtime with command line arguments. 

Source terms are introduced via the \texttt{Step} member, which in Listing~\ref{lst:Subroutine_Initialize_PW} is prepared in lines~31-36. 
(Recall from Section~\ref{sec:PlaneWave} that the class \texttt{Step\_RK2\_C\_Form} implements second-order Runge-Kutta (\texttt{RK2}) evolution of a conserved current (\texttt{C}).) 
In the case of \texttt{RayleighTaylor}, the line 
\begin{lstlisting}
  S % ApplySources => ApplySources
\end{lstlisting}
is added after initialization of \texttt{Step} in line~35 of Listing~\ref{lst:Subroutine_Initialize_PW}.
This sets a procedure pointer member of \texttt{Step} to the routine \texttt{ApplySources} in \texttt{RayleighTaylorForm}. 
This subroutine, which must have a specific interface, is given in Listing~\ref{lst:Subroutine_ApplySources}.
\begin{lstlisting}[float,frame=tb,numbers=left,numbersep=5pt,xleftmargin=10pt,label=lst:Subroutine_ApplySources,caption={{\tt subroutine ApplySources} in \texttt{RiemannProblem\_Form}.}]
  subroutine ApplySources ( S, Increment, Fluid, TimeStep )

    class ( Step_RK_C_Template ), intent ( in ) :: &
      S
    type ( VariableGroupForm ), intent ( inout ) :: &
      Increment
    class ( CurrentTemplate ), intent ( in ) :: &
      Fluid
    real ( KDR ), intent ( in ) :: &
      TimeStep

    integer ( KDI ) :: &
      iMomentum_2, &
      iEnergy   

    select type ( F => Fluid )
    class is ( Fluid_P_P_Form )

    call Search ( F % iaConserved, F % MOMENTUM_DENSITY_D ( 2 ), iMomentum_2 )
    call Search ( F % iaConserved, F % CONSERVED_ENERGY, iEnergy )

    associate &
      ( KVM => Increment % Value ( :, iMomentum_2 ), &
        KVE => Increment % Value ( :, iEnergy ), &
        N   => F % Value ( :, F % COMOVING_DENSITY ), &
        VY  => F % Value ( :, F % VELOCITY_U ( 2 ) ), &
        A   => Acceleration, &
        dT  => TimeStep )
    
    KVM  =  KVM  -  dT * N * A
    KVE  =  KVE  -  dT * N * A * VY

    end associate !-- KVM, etc.
    end select !-- F
    
  end subroutine ApplySources
\end{lstlisting}
Because this routine is called by the \texttt{Mathematics} class \texttt{Step\_RK2\_C\_Form}, its argument list is correspondingly generic.
Thus the \texttt{Fluid} argument is of class \texttt{CurrentTemplate}, and the specific type of conserved current for this particular problem is identified as a polytropic fluid in lines~16-17 of Listing~\ref{lst:Subroutine_ApplySources}.
The \texttt{Increment} argument is an instance of \texttt{VariableGroupForm}, our generic data container from \textsc{GenASiS} \texttt{Basics} \cite{Cardall2015GenASiS-Basics:};
it contains the Runge-Kutta increments for the conserved fields being evolved.
Upon entry it contains the increments due to fluxes through cell faces, and it is the job of this routine to add the source terms.
For \texttt{RayleighTaylor} there is the downward force in $y$ direction and the corresponding power. 
The indices corresponding to $y$ momentum and energy are set in lines~19-20 using the \texttt{Basics} command \texttt{Search}, and used in lines~23-24 to alias those increment fields.
Also aliased are the density and $y$-velocity fields (lines~25-26), the uniform acceleration (line~27, a local \texttt{RayleighTaylor} module variable), and the \texttt{TimeStep} argument.
The source terms are added to the relevant increments in lines~30-31.

\begin{figure}
\begin{centering}
  \includegraphics[width=0.24\textwidth]{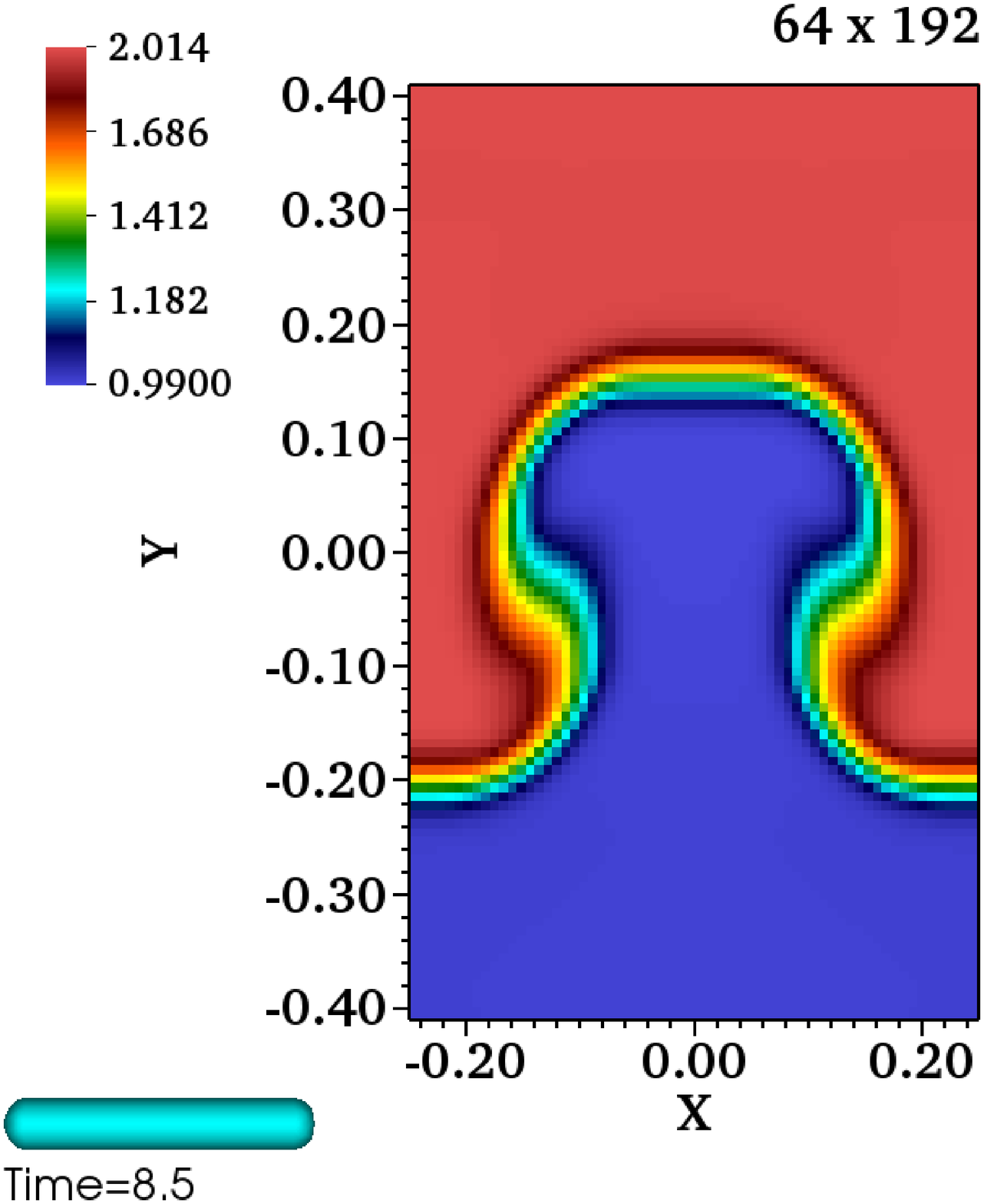}
  \includegraphics[width=0.24\textwidth]{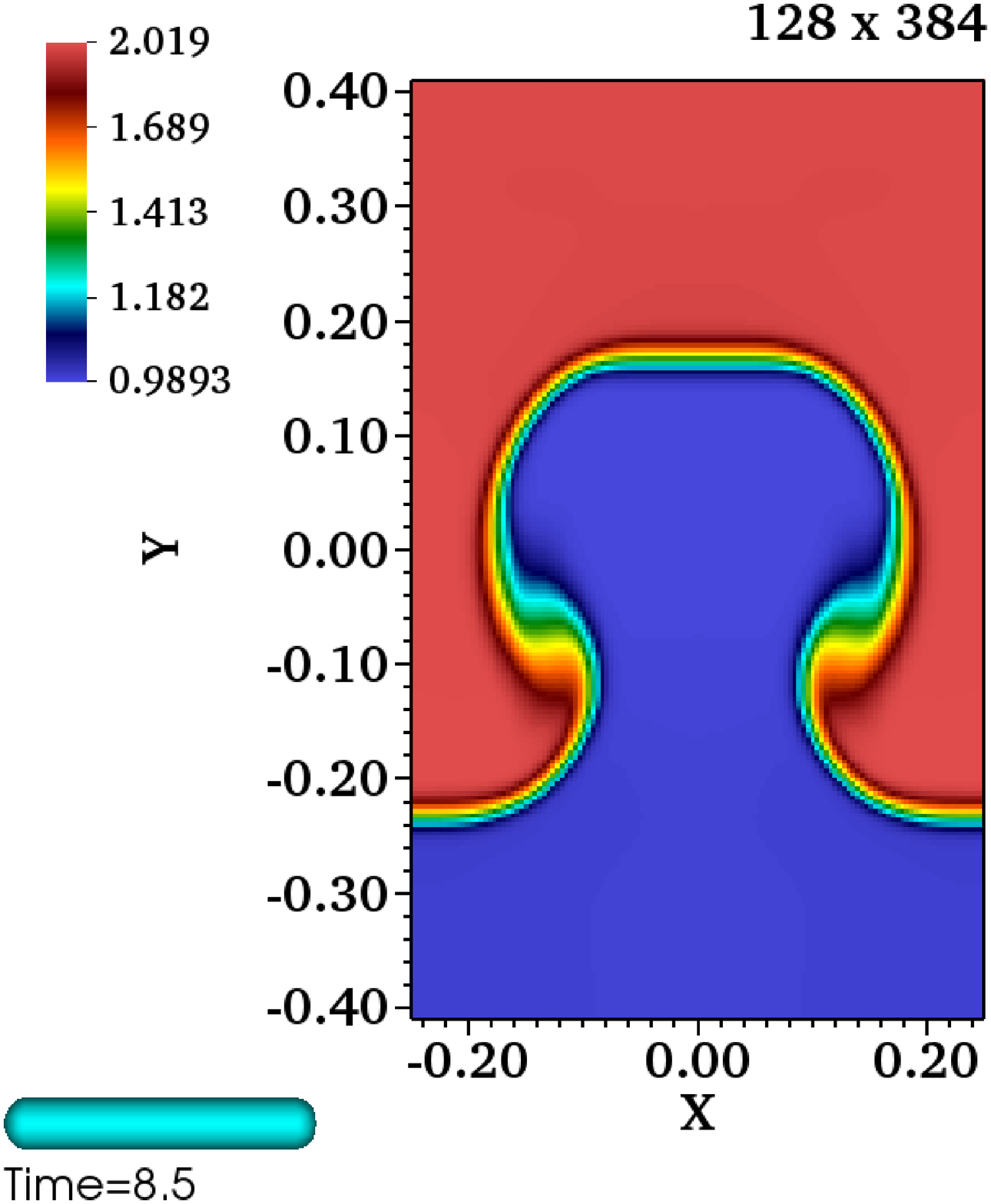}
  \includegraphics[width=0.24\textwidth]{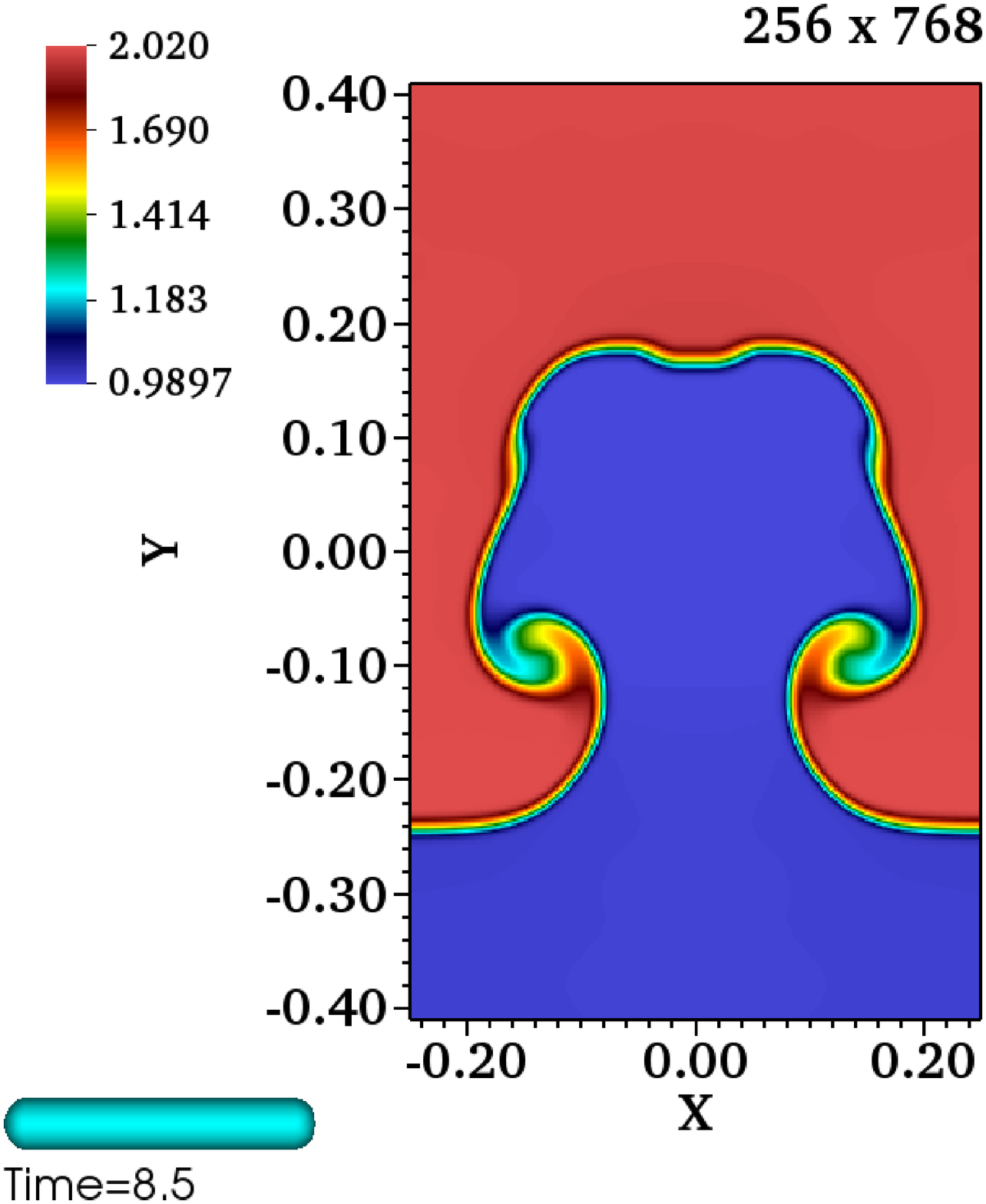}
  \includegraphics[width=0.24\textwidth]{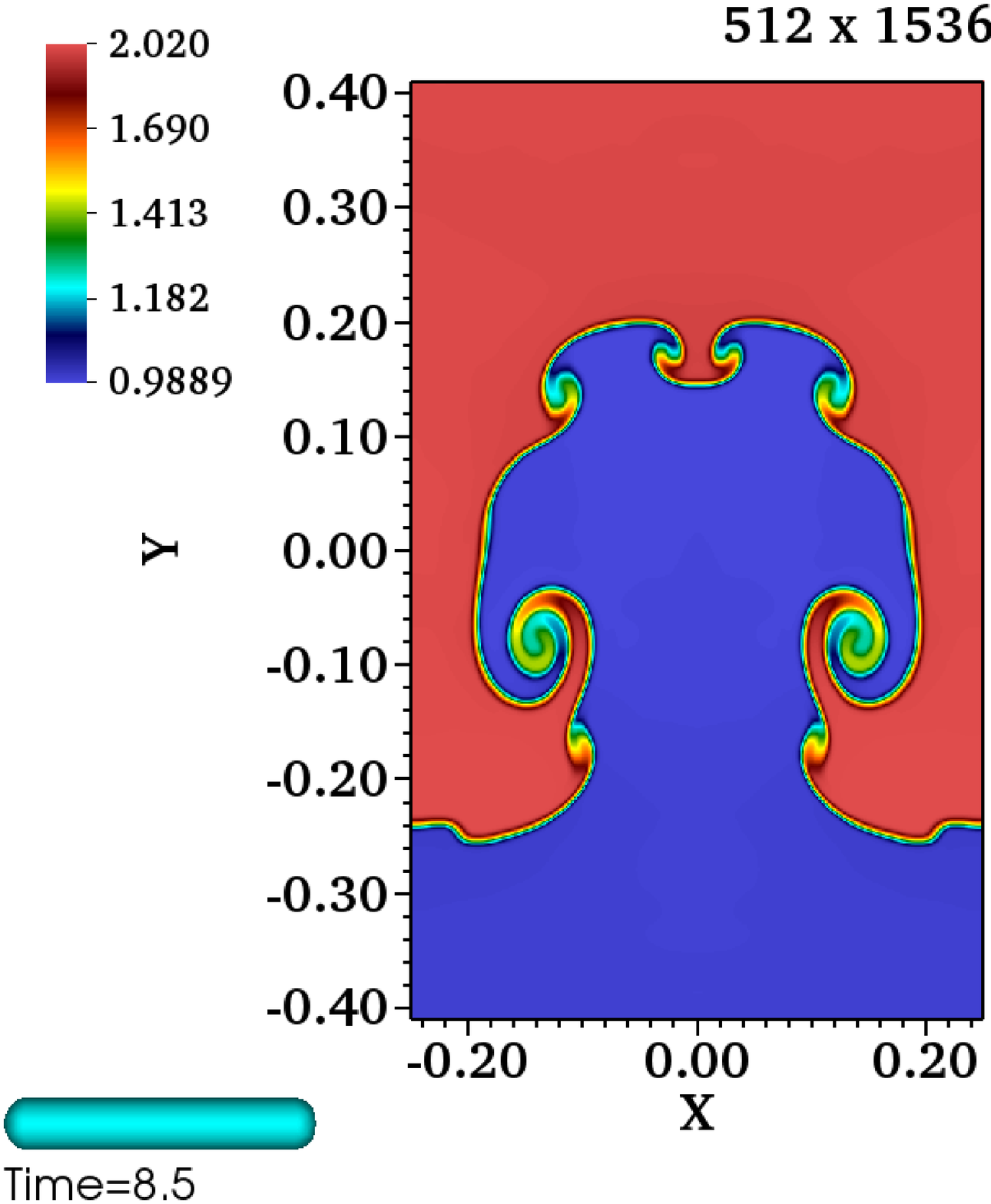}
  \caption{Density at $t = 8.5$ in the \texttt{RayleighTaylor} problem for four different resolutions as labeled in the upper right corner of each panel. 
(The full extent of the domain in $y$ is not shown.)
Increasing resolution yields a thinner interface layer and more structure in the characteristic `mushroom cap.'}
  \label{fig:RayleighTaylor}
\end{centering}
\end{figure}

Figure~\ref{fig:RayleighTaylor} shows the density at $t = 8.5$ in the \texttt{RayleighTaylor} problem for four different resolutions as labeled in the upper right corner of each panel.
Increasing resolution yields a thinner interface layer and more structure in the characteristic `mushroom cap.'

\begin{figure}
\begin{centering}
  \includegraphics[width=0.49\textwidth]{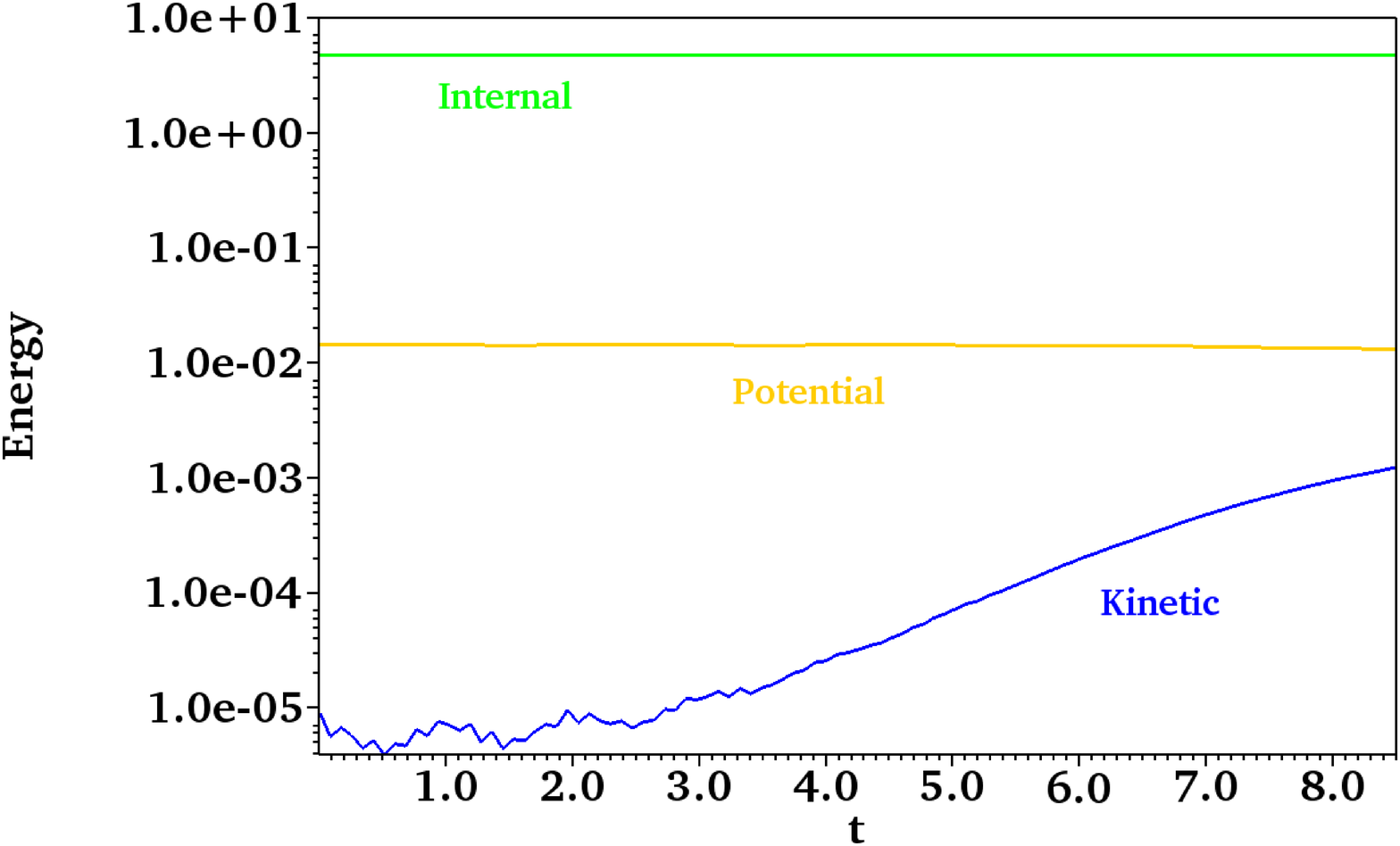}
  \includegraphics[width=0.49\textwidth]{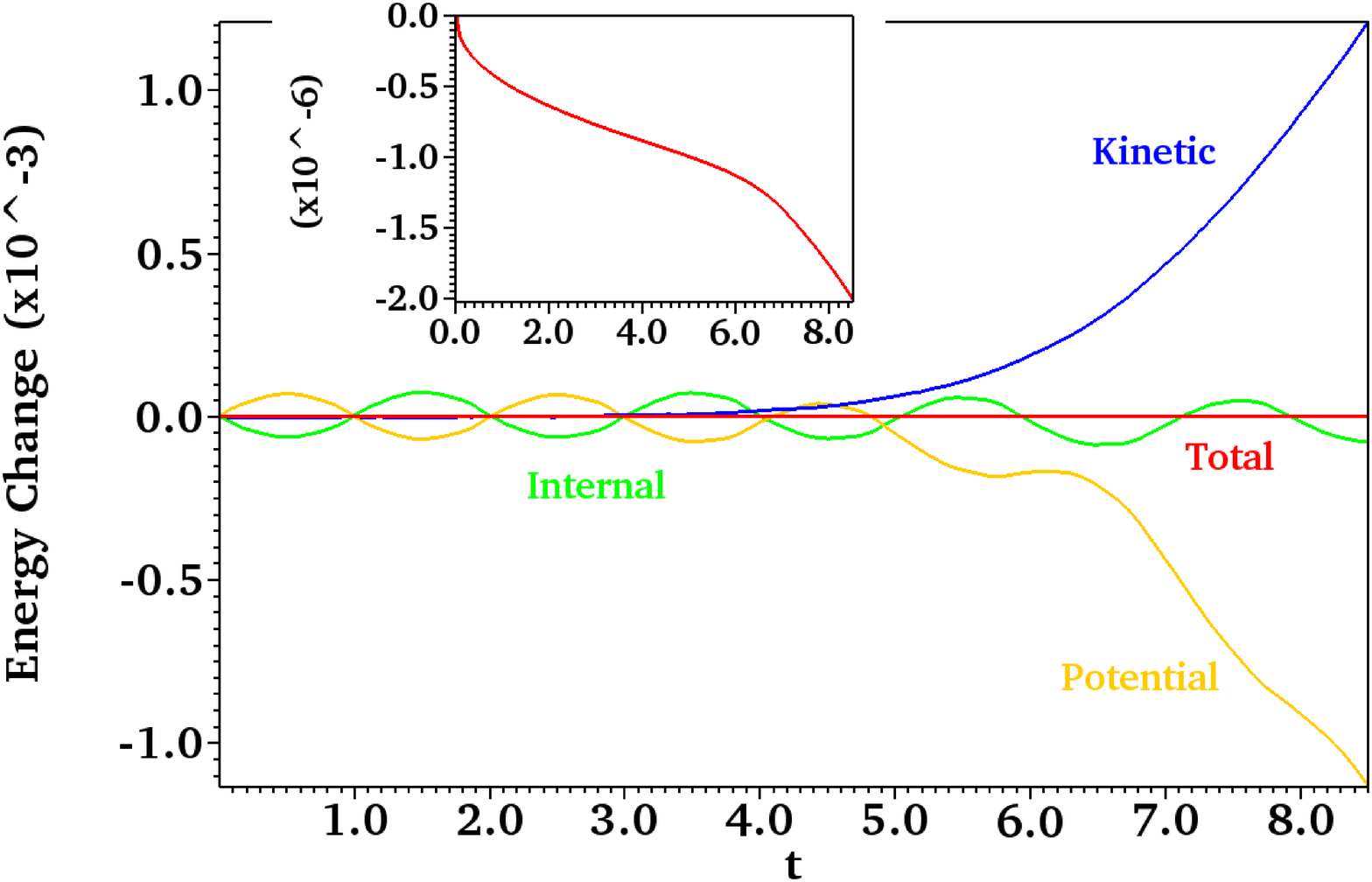}
  \caption{Absolute energies (left) and energy changes (right) in the \texttt{RayleighTaylor} problem for the $512 \times 1536$ computation.
Unlike previous problems, the energy scales---internal (green), potential (orange), and kinetic (blue)---are so disparate that they can only be visualized together on a logarithmic plot (left; the total energy would be visually coincident with the internal energy and is not shown).
The generation of kinetic energy at the expense of potential energy is better visualized in terms of energy \textit{changes} on a linear scale (right).
Energy conservation---see the total change (red)---is evidently under control relative to the scale of kinetic energy generation, though not to machine precision (inset).  
}
  \label{fig:RayleighTaylorTally}
\end{centering}
\end{figure}

\begin{figure}
\begin{centering}
  \includegraphics[width=0.49\textwidth]{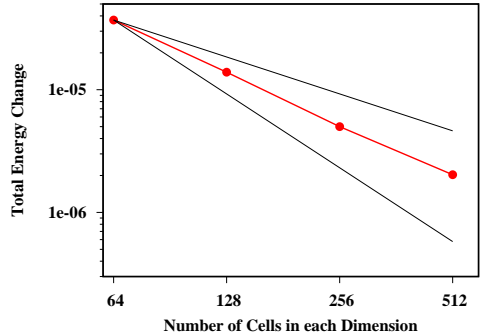}
  \caption{The total energy change in the \texttt{RayleighTaylor} problem should tally to zero; in practice, as resolution increases the numerical solution (red) approaches this ideal between first order (upper black) and second order (lower black).}
  \label{fig:RayleighTaylorConvergence}
\end{centering}
\end{figure}

The class used to compute global diagnostics for the \texttt{RiemannProblem} in Section~\ref{sec:RiemannProblem}---\texttt{Tally\_P\_P\_Form}---does not include potential energy, so the extension \texttt{Tally\_RT\_Form} (see Fig.~\ref{fig:FluidDynamics}) adds a tally of potential energy for this specific problem, as shown in Fig.~\ref{fig:RayleighTaylorTally}.
The custom tally objects, which are members of \texttt{Fluid\_ASC\_Form}, must be allocated between the allocation and initialization of the fluid (cf. lines~25-29 of Listing~\ref{lst:Program_SW_Outline}); we refer the reader to the source code for further details. 
Unlike previous problems, the energy scales in \texttt{RayleighTaylor} are so disparate that they can only be visualized together on a logarithmic plot (left panel of Fig.~\ref{fig:RayleighTaylorTally});
the generation of kinetic energy at the expense of potential energy is better visualized in terms of energy \textit{changes} on a linear scale (right panel).
The initial perturbation induces an oscillatory exchange of internal and potential energy, with the kinetic energy hardly registering until the exponential development of the instability begins to take off.
Energy conservation is evidently under control relative to the scale of kinetic energy generation, though not to machine precision (right panel, inset); 
as a function of resolution it converges between first and second order, as shown in Fig.~\ref{fig:RayleighTaylorConvergence}.

%%%%%%%%%%%%%%%%%%%%%%%%
\subsection{Sedov-Taylor Blast Wave}
\label{sec:SedovTaylor}

A Sedov-Taylor blast wave results from the release of thermal energy $E$ at single point in a medium of uniform density $\rho_0$ and negligible pressure $p_0$ (see e.g. Ref.~\cite{Cardall2014GENASIS:-Genera} and references therein).
We take $E = 1$, $\rho_0 = 1$, and $p_0 = 0.0$ on a domain of radius $R_\mathrm{max} = 0.35$ and evolve to time $t = 0.05$.  
Again the adiabatic index is $1.4$.
We approximate the ideal of energy release at a single point by spreading $E$ uniformly in a region of radius $R_E = 0.03\, R_\mathrm{max}$, employing a subgrid of 20 subcells in each dimension to determine the average energy density in cells that intersect the boundary of the detonation region.

The major new element introduced in this example is the use of curvilinear coordinates.
As in previous problems, the class \texttt{SedovTaylorForm}---an extension of the \texttt{Mathematics} class \texttt{Integrator\_C\_Template}---has an \texttt{Initialize} method with a structure similar to Listing~\ref{lst:Subroutine_Initialize_PW}.
This class sets up the same problem---a spherical blast wave in 3D \textit{geometry}---but with different coordinate systems, depending on the dimensionality of the \textit{solution}: spherical coordinates on a 1D chart, cylindrical coordinates on a 2D chart, and Cartesian coordinates on a 3D chart.\footnote{These choices avoid coordinate singularities---and therefore excessively restrictive Courant time step conditions---at the origin of spherical coordinates, and on the vertical axis of spherical and cylindrical coordinates.}
In 1D, a local character variable 
\begin{lstlisting}
  CoordinateSystem = 'SPHERICAL'
\end{lstlisting}
is set, and the inner and outer boundary conditions on the radial coordinate $r$ chart are set to \texttt{'REFLECTING'} and \texttt{'OUTFLOW'} (i.e. zero gradient) respectively via the mechanism described in the second paragraph of Section~\ref{sec:RiemannProblem}.
In 2D, we have
\begin{lstlisting}
  CoordinateSystem = 'CYLINDRICAL'
\end{lstlisting}
with reflecting and outflow boundary conditions in cylindrical $r$ and outflow boundaries in vertical $z$.
In 3D, we confirm the default
\begin{lstlisting}
  CoordinateSystem = 'CARTESIAN'
\end{lstlisting}
and set outflow boundaries in all dimensions.
The subsequent call
\begin{lstlisting}
  call PS % CreateChart &
    ( CoordinateSystemOption = CoordinateSystem, &
      MinCoordinateOption = MinCoordinate, &
      MaxCoordinateOption = MaxCoordinate, &
      nCellsOption = nCells )
\end{lstlisting}
now contains the optional argument \texttt{CoordinateSystemOption}, with local variables \texttt{MinCoordinate}, \texttt{MaxCoordinate}, and \texttt{nCells} having been also set to appropriate values depending on dimensionality.

Unlike \texttt{RayleighTaylor}, there are no physical sources in this problem; but there are fictitious forces arising from the use of curvilinear coordinates, and they are applied using the same mechanism discussed in Section~\ref{sec:RayleighTaylor}.
In this case, however, the user does not need to supply a subroutine to compute source terms;
instead, after initialization of the \texttt{Step} member (cf. line~35 of Listing~\ref{lst:Subroutine_Initialize_PW}), the line
\begin{lstlisting}
  S % ApplySources => ApplySourcesCurvilinear_Fluid_P
\end{lstlisting}
points this procedure pointer to a routine available in \texttt{Fluid\_P\_Template} (see Fig.~\ref{fig:FluidDynamics}).

\begin{figure}
\begin{centering}
  \includegraphics[width=0.51\textwidth]{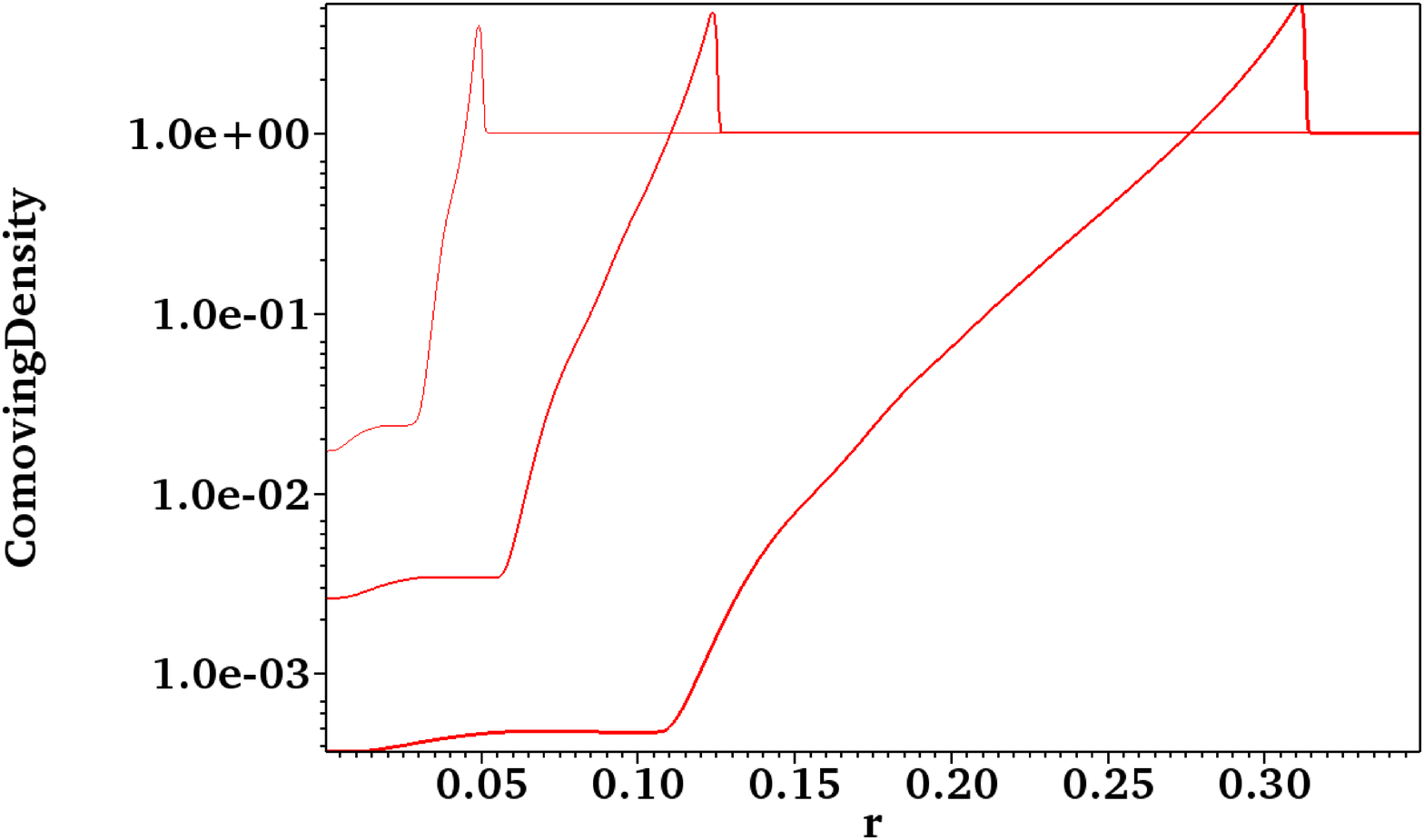}
  \includegraphics[width=0.51\textwidth]{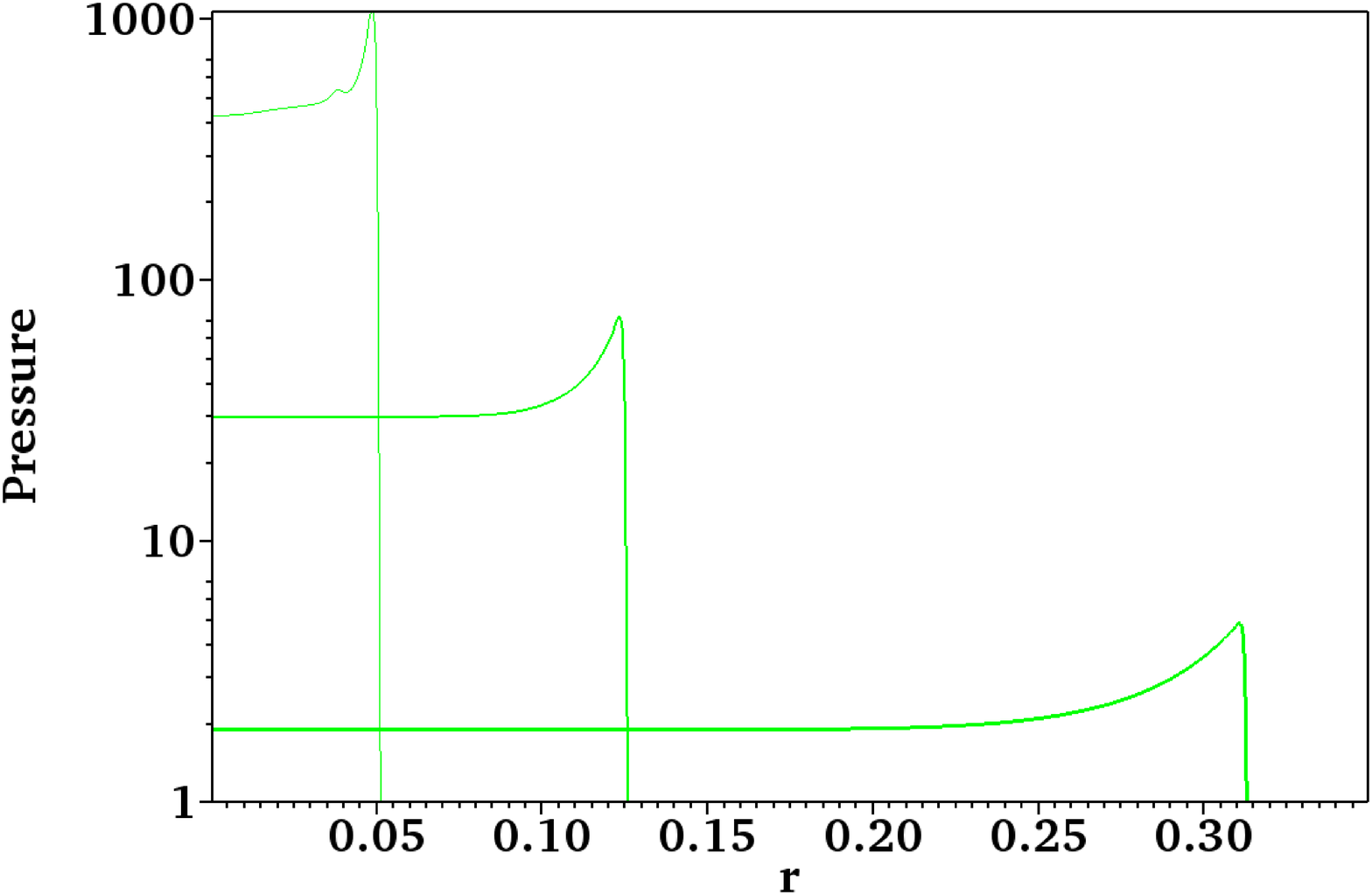}
  \includegraphics[width=0.51\textwidth]{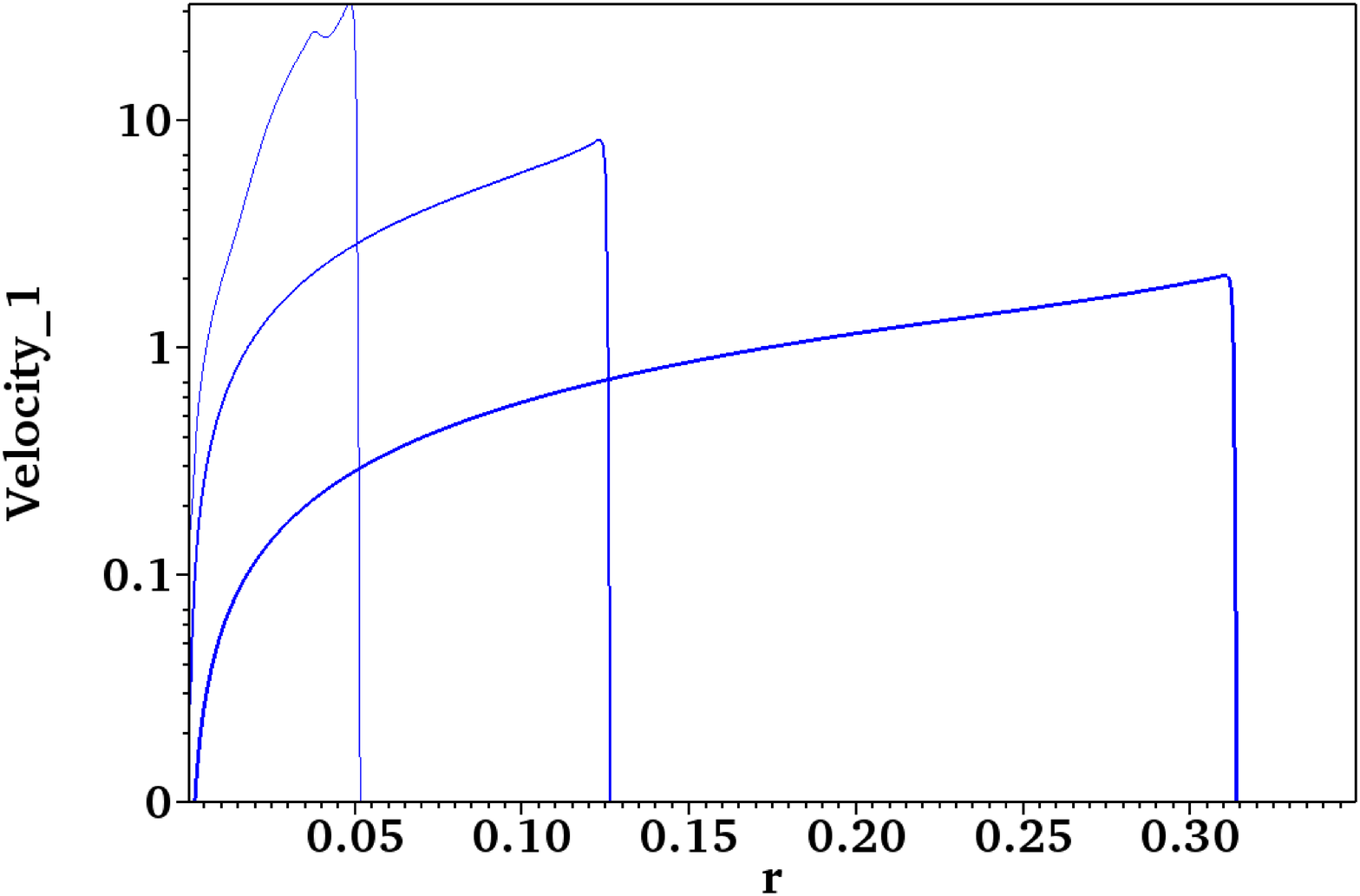}
  \caption{The \texttt{SedovTaylor} blast wave in 1D (spherical coordinates): radial profiles of density (upper), pressure (center), and radial velocity (lower), at times $t = 0.0005$, $0.005$, and $0.05$ (left to right and increasing thickness). 
Computed with 512 cells in spherical coordinate $r$.}
  \label{fig:SedovTaylor_1D}
\end{centering}
\end{figure}

\begin{figure}
\begin{centering}
  \includegraphics[width=0.32\textwidth]{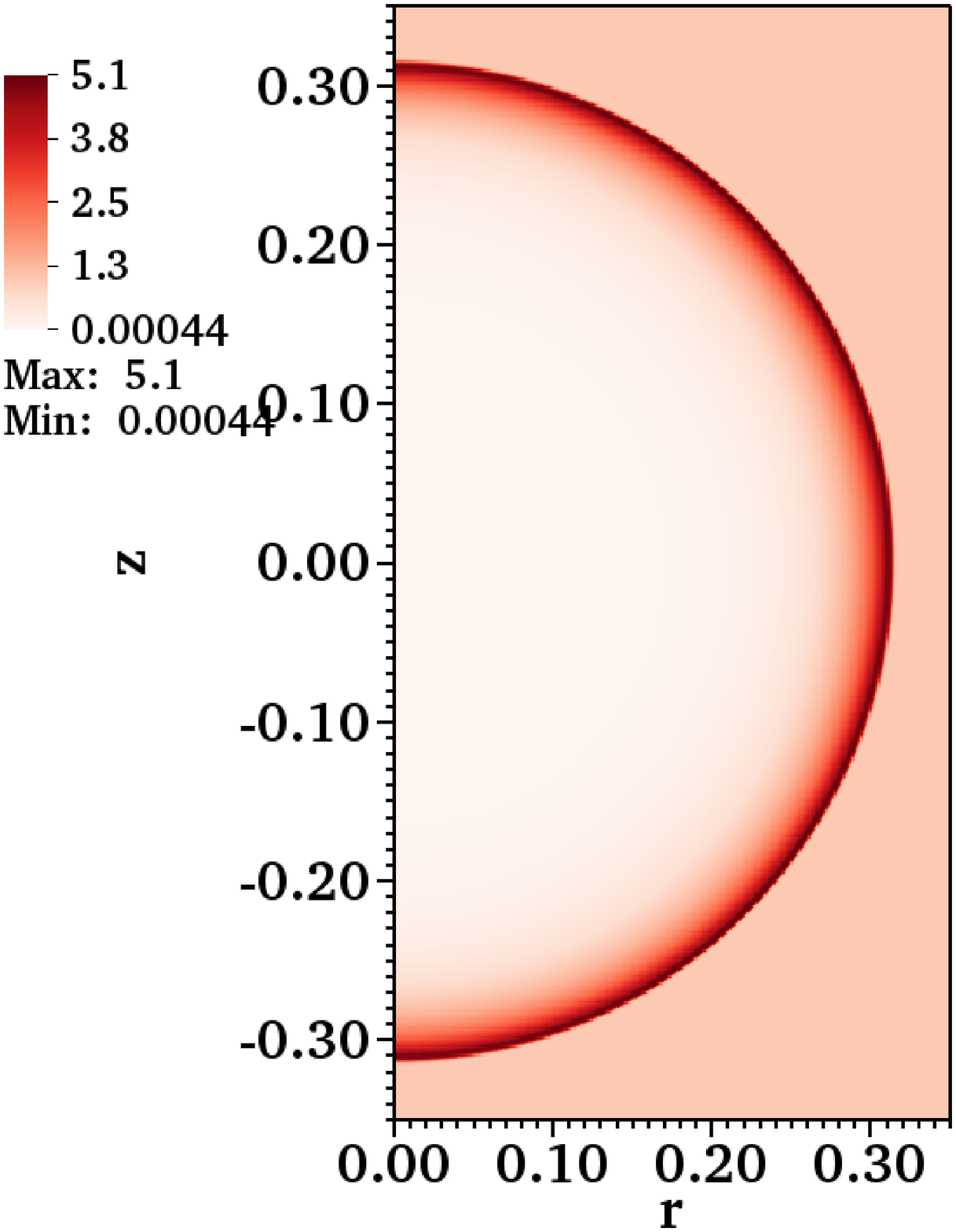}
  \includegraphics[width=0.32\textwidth]{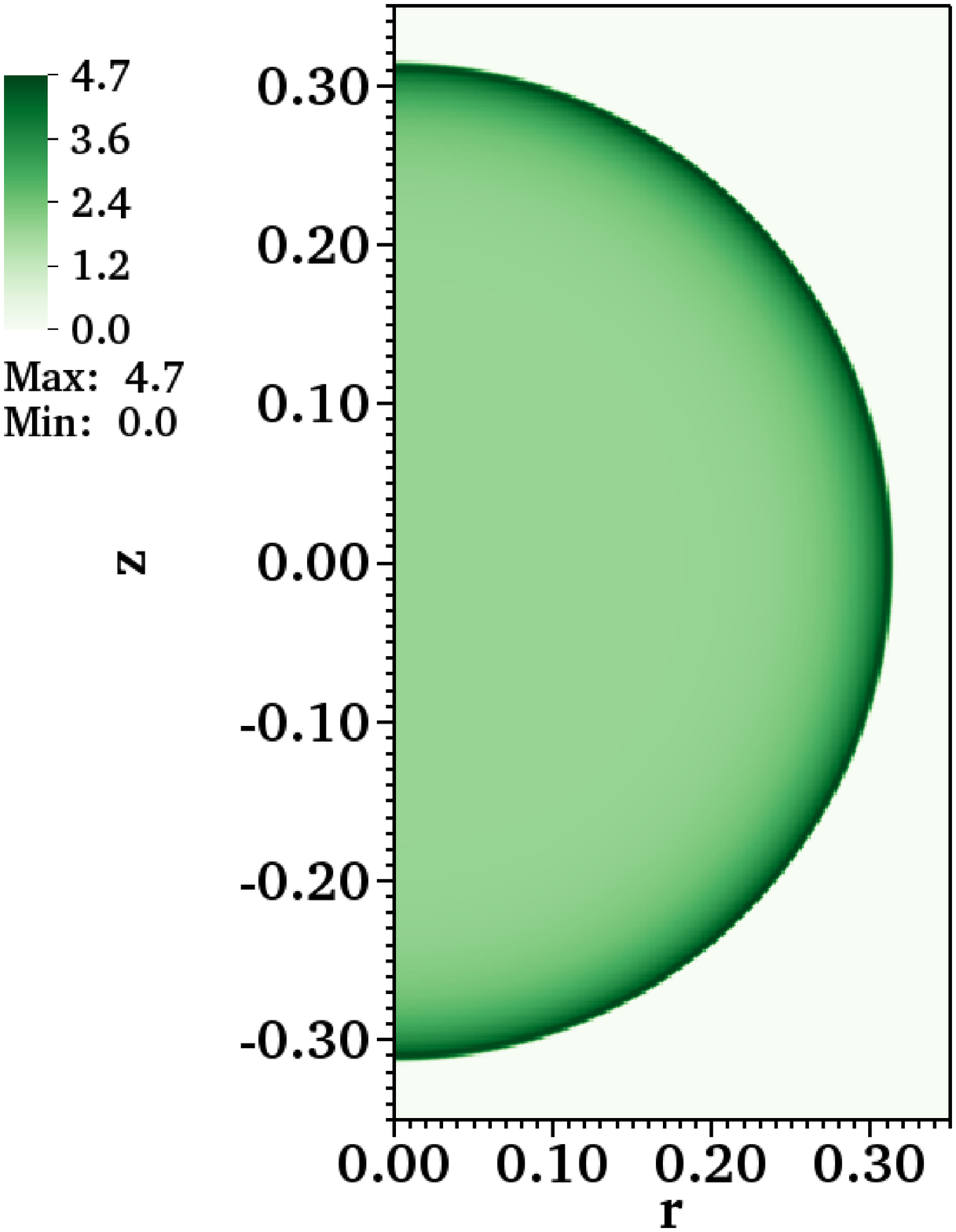}
  \includegraphics[width=0.32\textwidth]{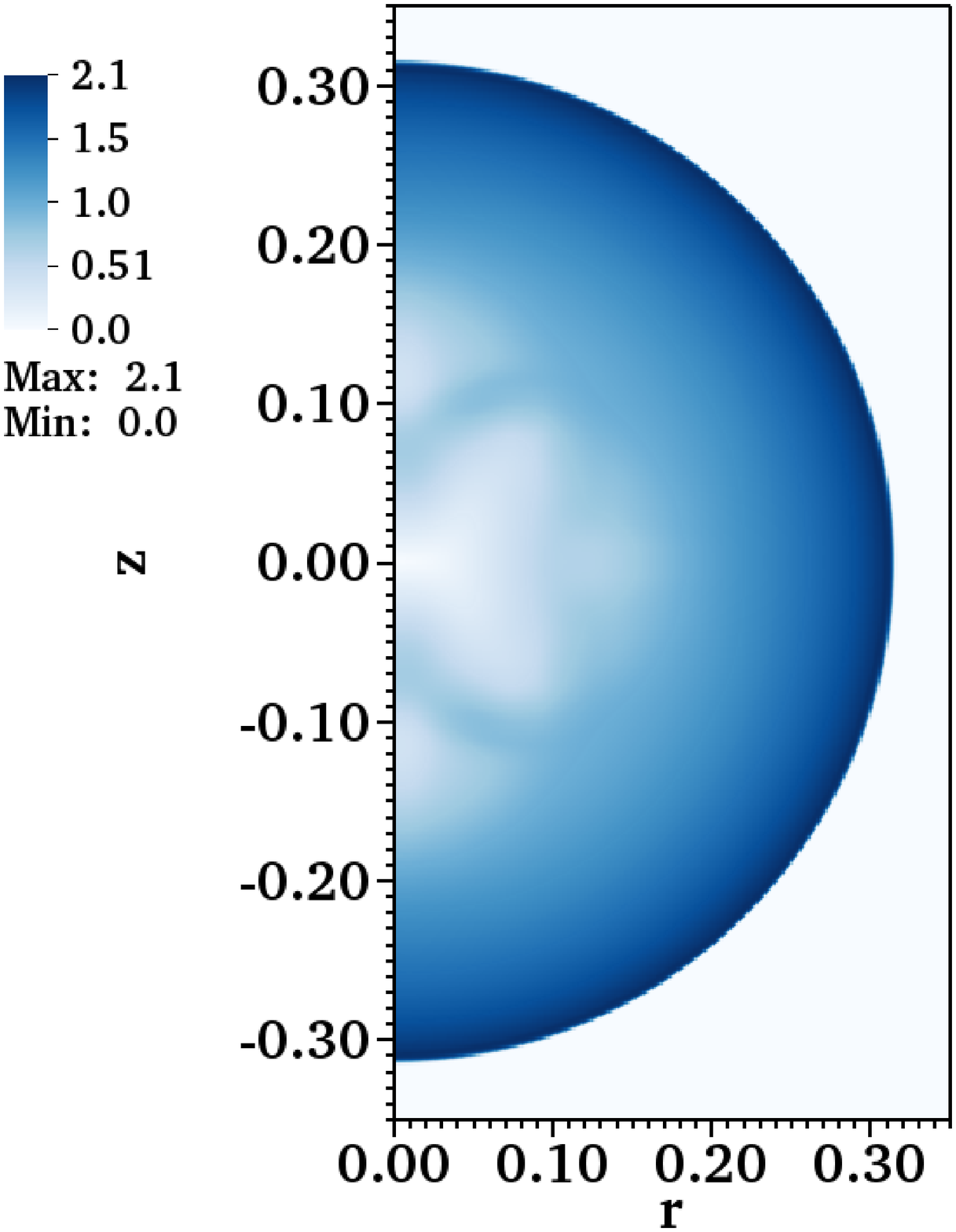}
  \includegraphics[width=0.32\textwidth]{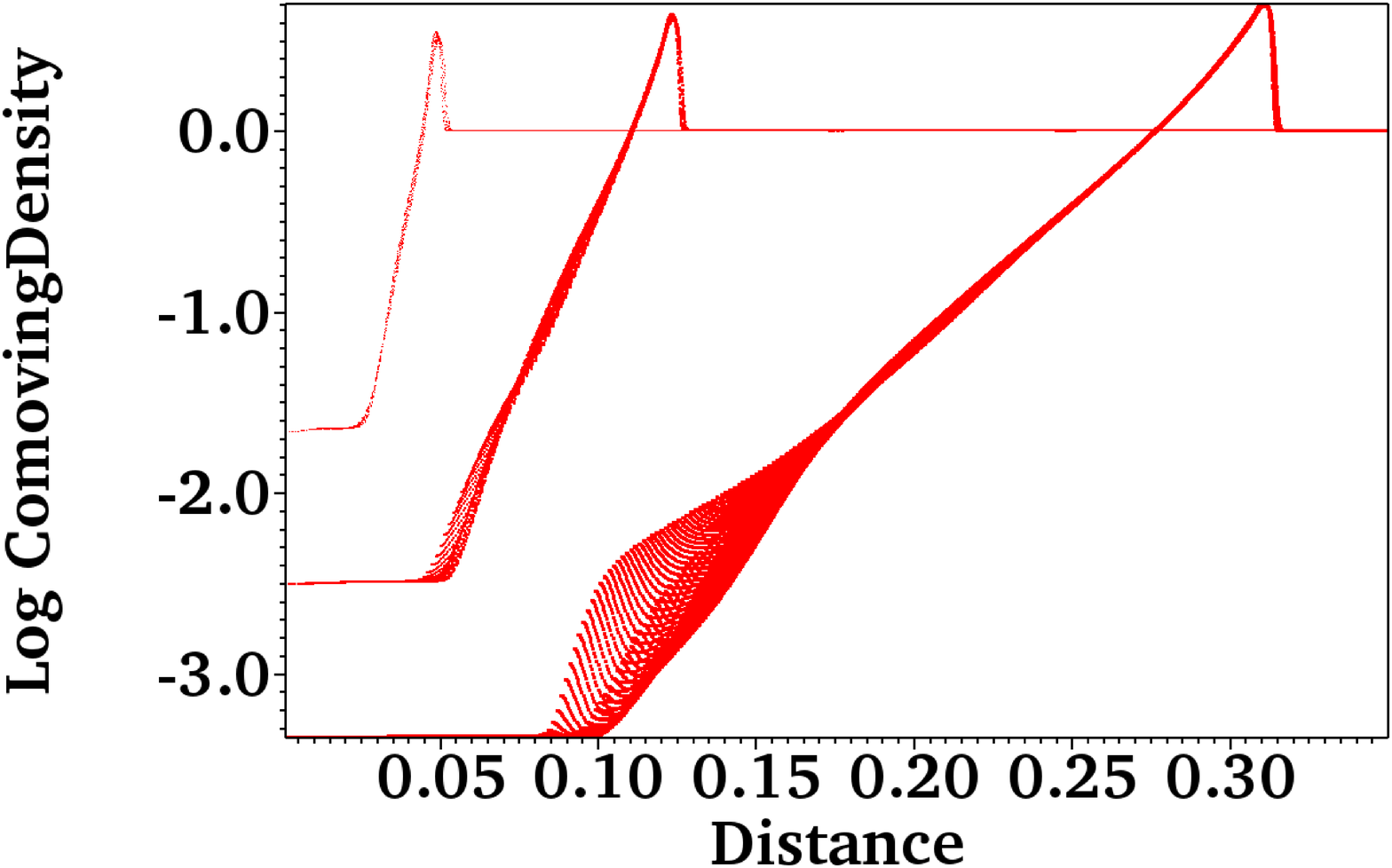}
  \includegraphics[width=0.32\textwidth]{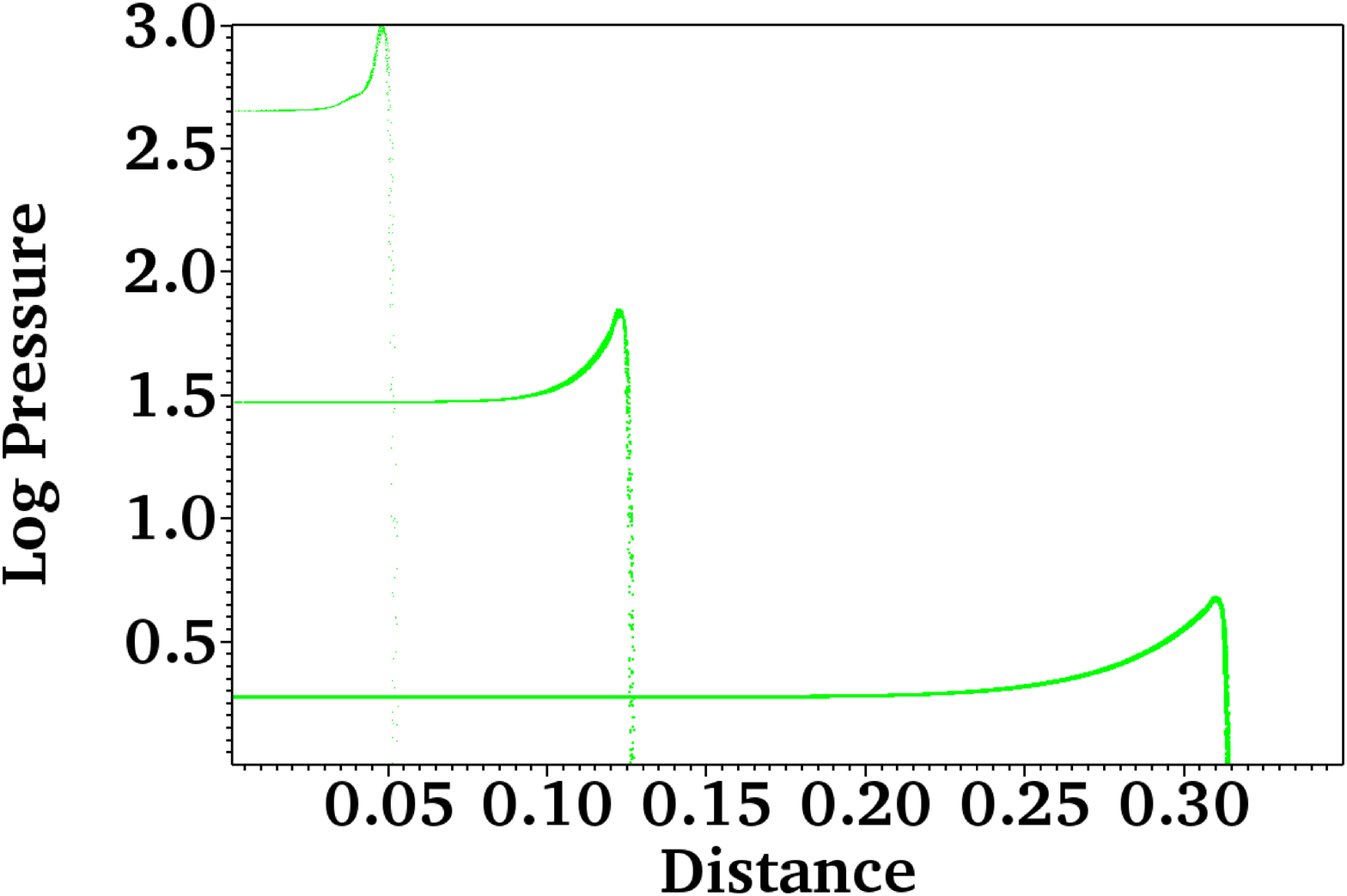}
  \includegraphics[width=0.32\textwidth]{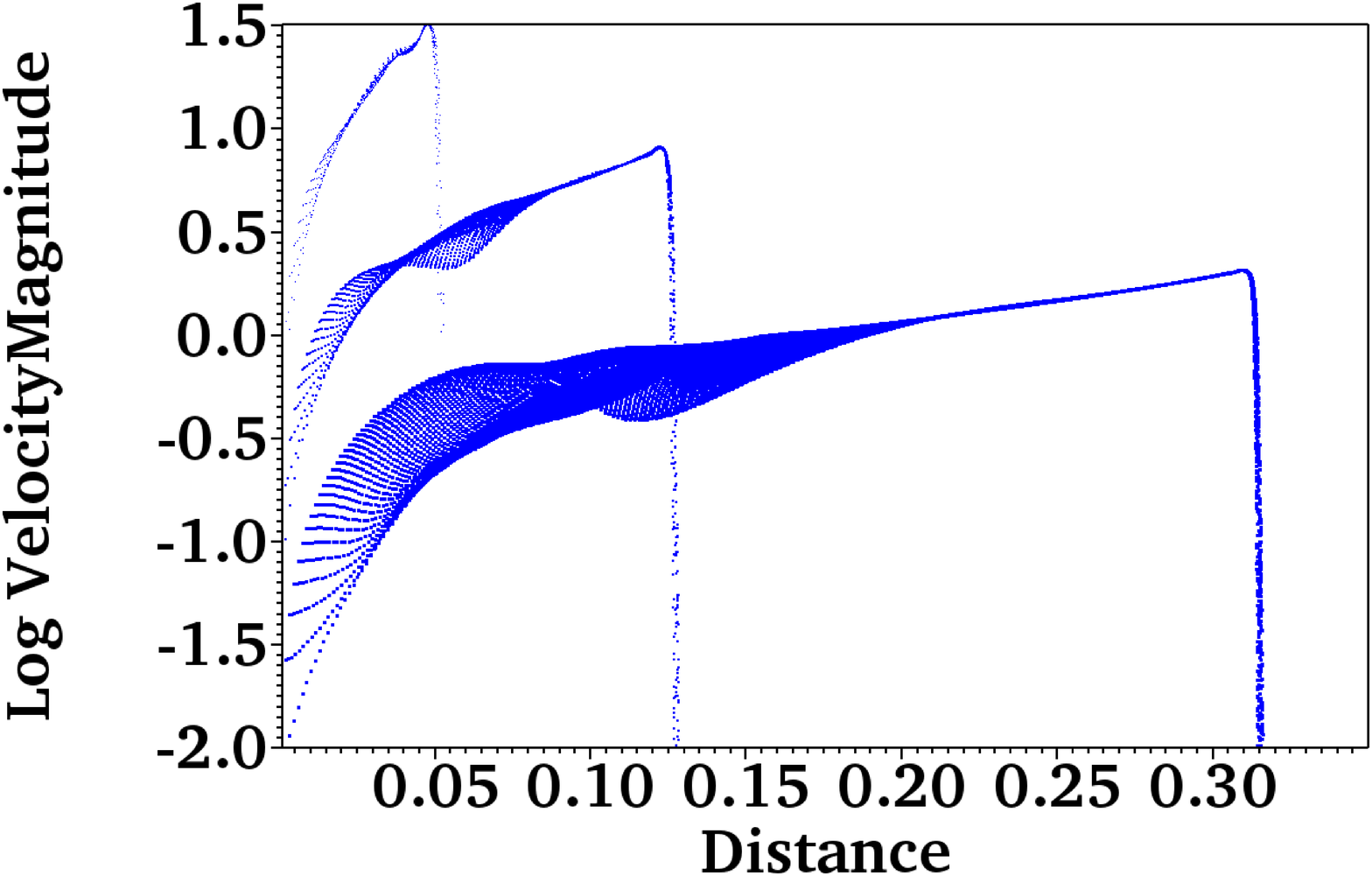}
  \caption{The \texttt{SedovTaylor} blast wave in 2D (cylindrical coordinates): density (left), pressure (center), and velocity magnitude (right), visualized in 2D at $t = 0.05$ (upper) and as scatter plots as a function of radial distance at $t = 0.0005$, $0.005$, and $0.05$ (lower, compare Fig.~\ref{fig:SedovTaylor_1D}).
Computed with $[256,512]$ cells in cylindrical coordinates $[r, z]$.}
  \label{fig:SedovTaylor_2D}
\end{centering}
\end{figure}

\begin{figure}
\begin{centering}
  \includegraphics[width=0.32\textwidth]{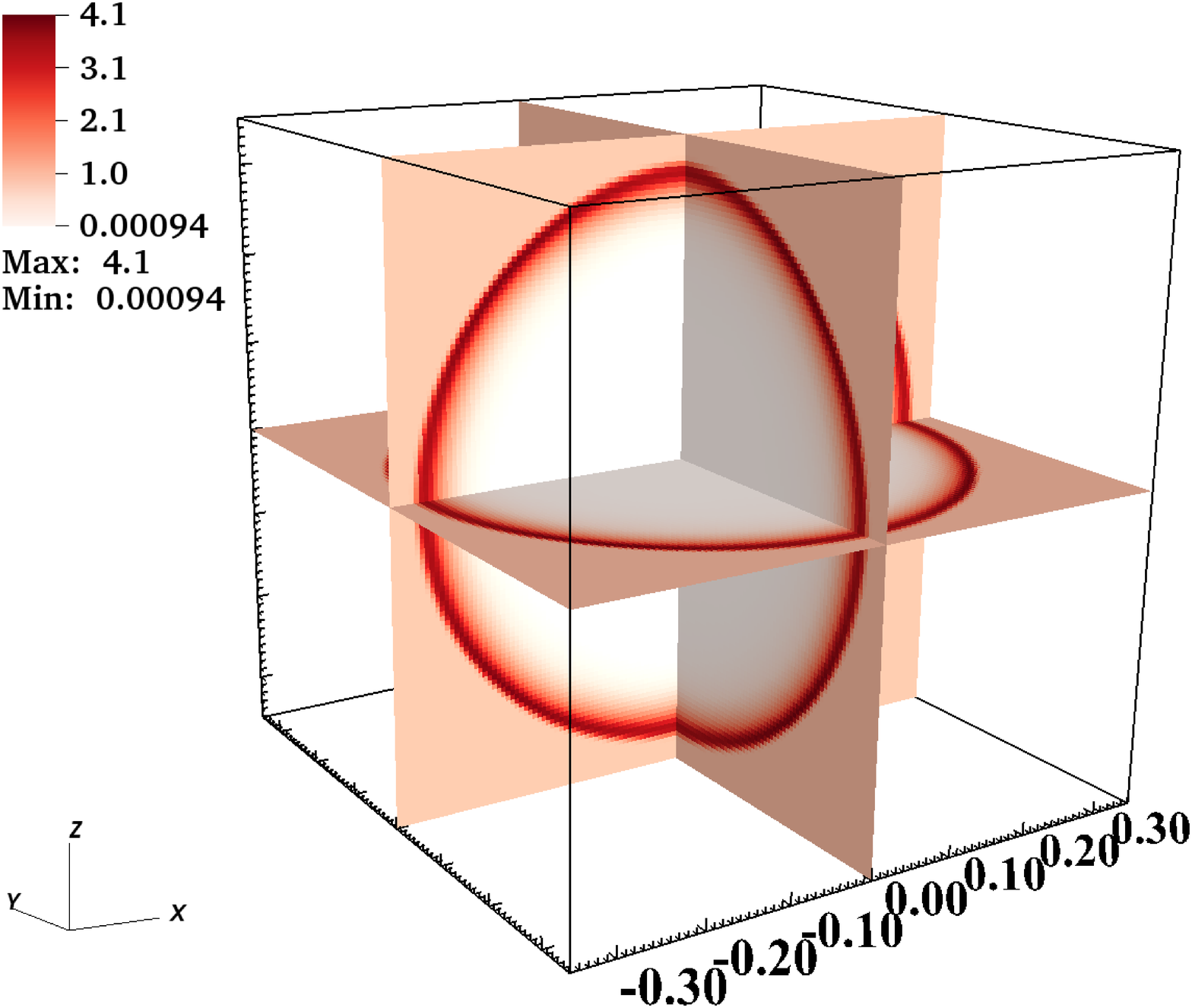}
  \includegraphics[width=0.32\textwidth]{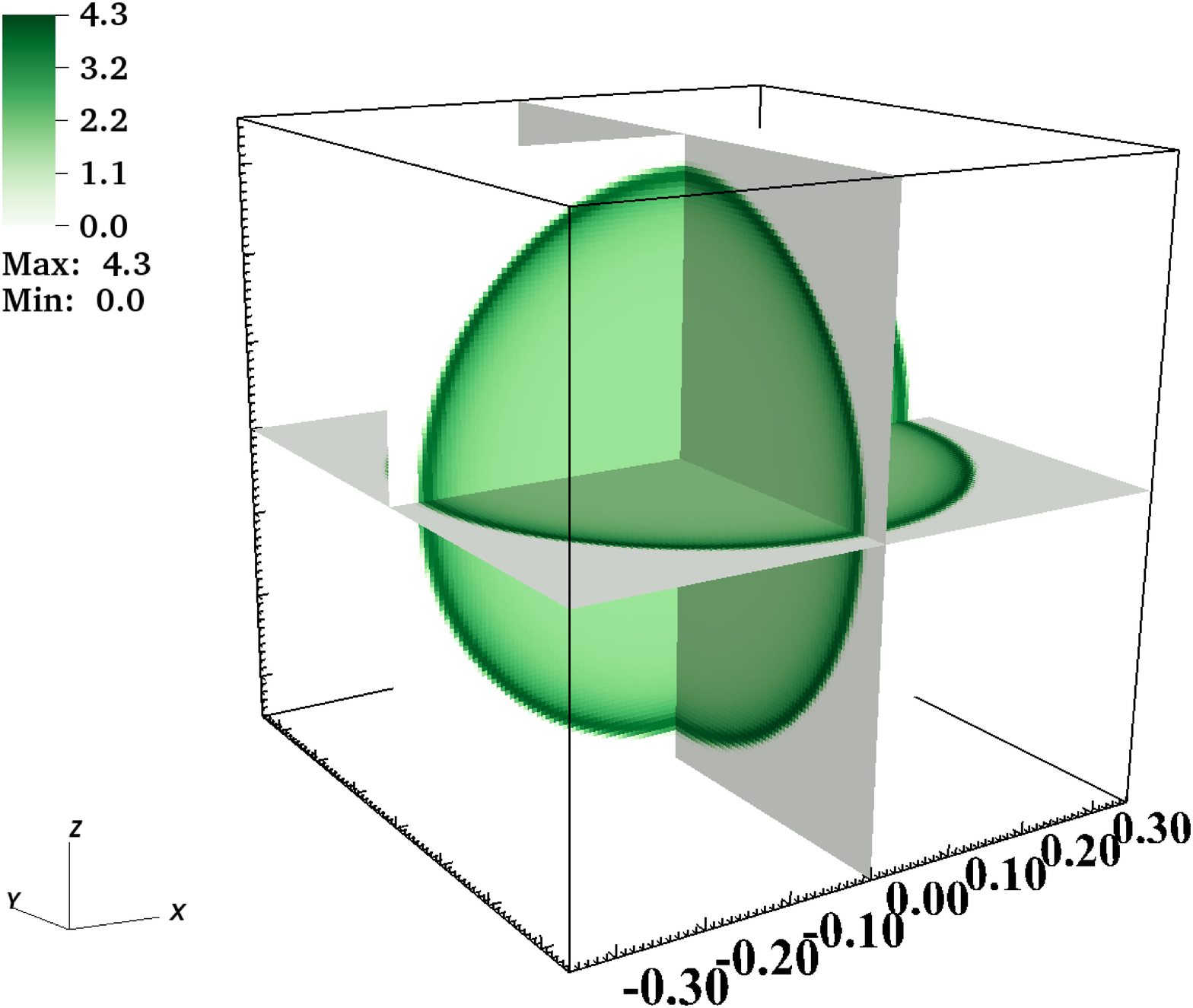}
  \includegraphics[width=0.32\textwidth]{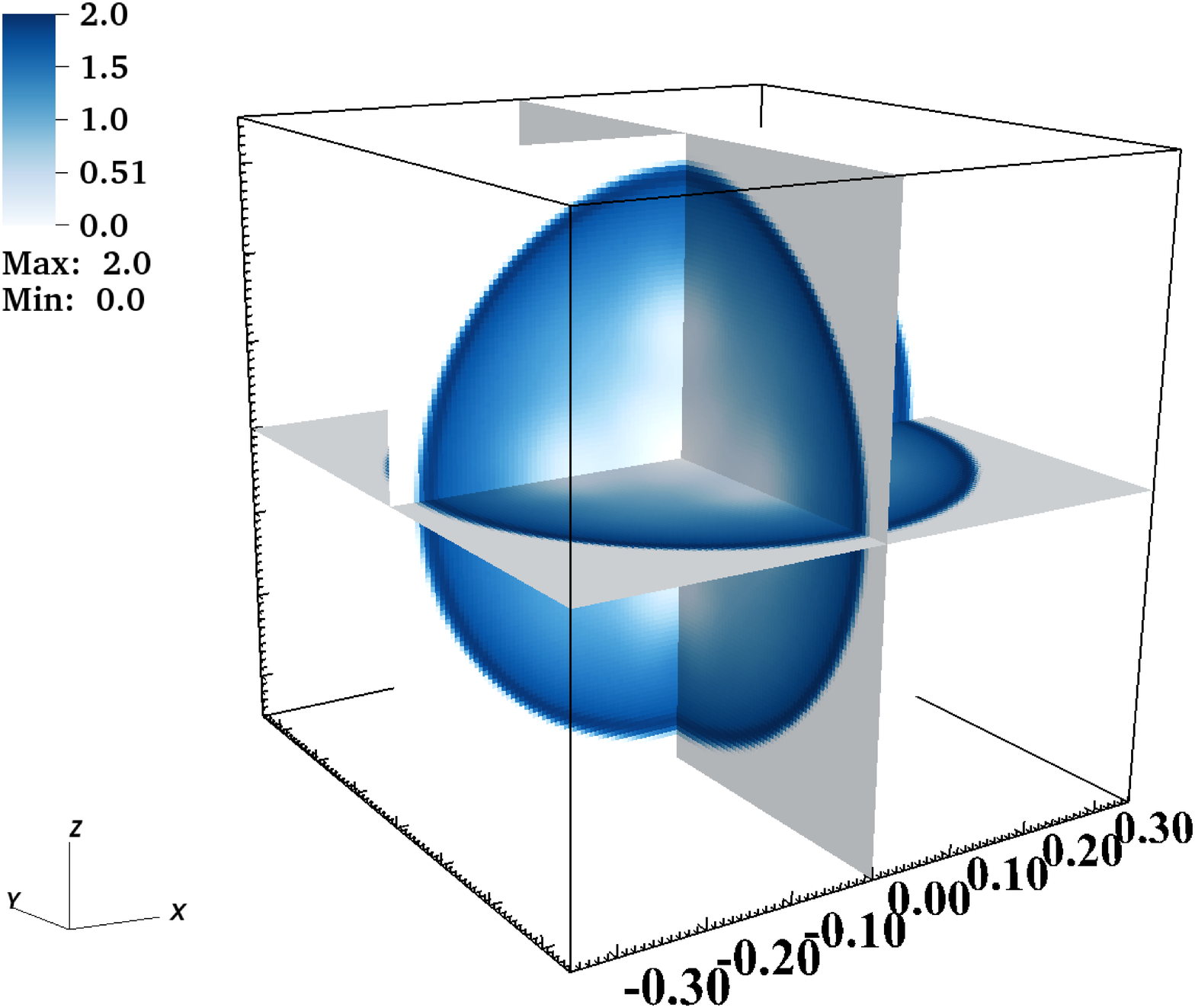}
  \includegraphics[width=0.32\textwidth]{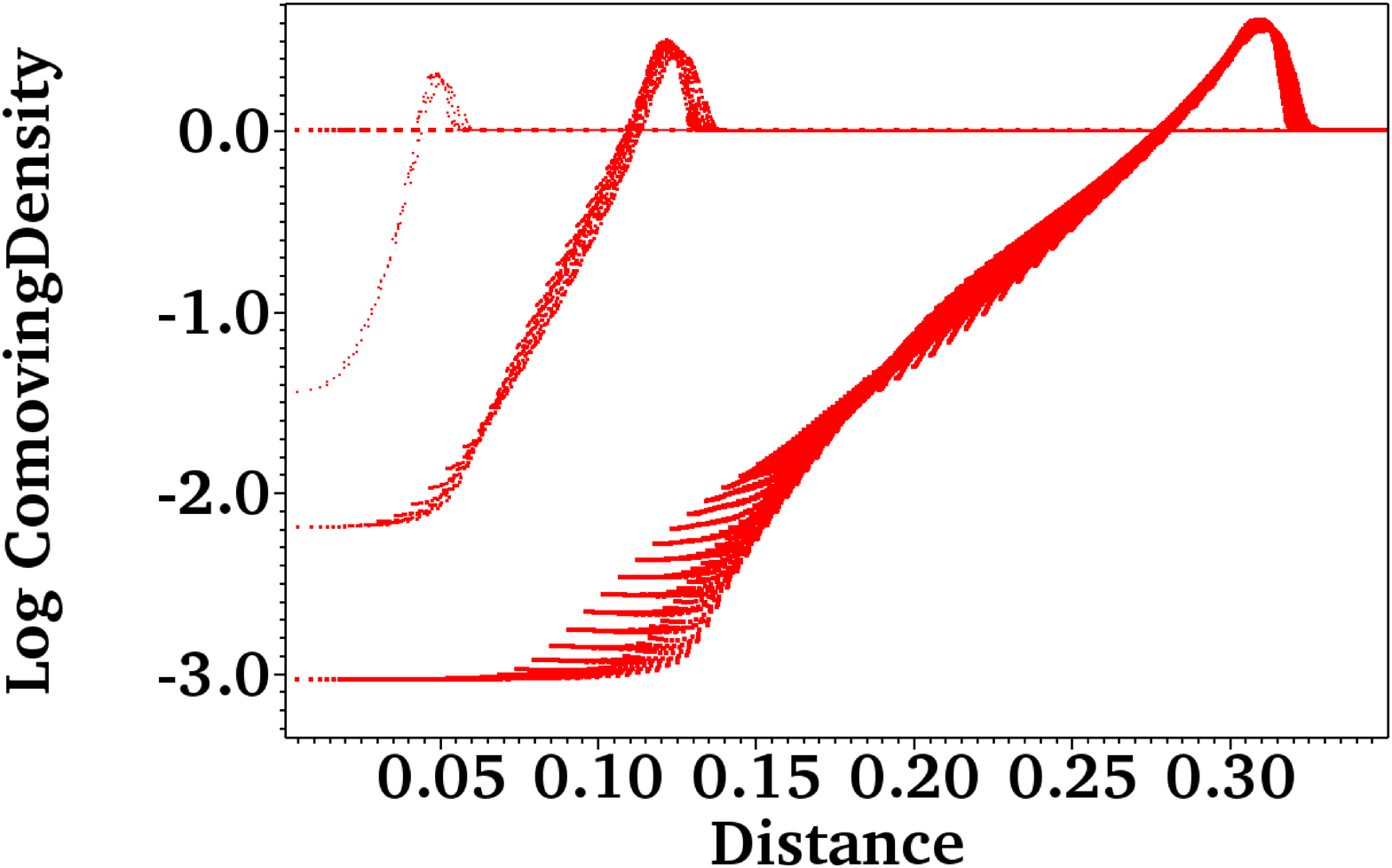}
  \includegraphics[width=0.32\textwidth]{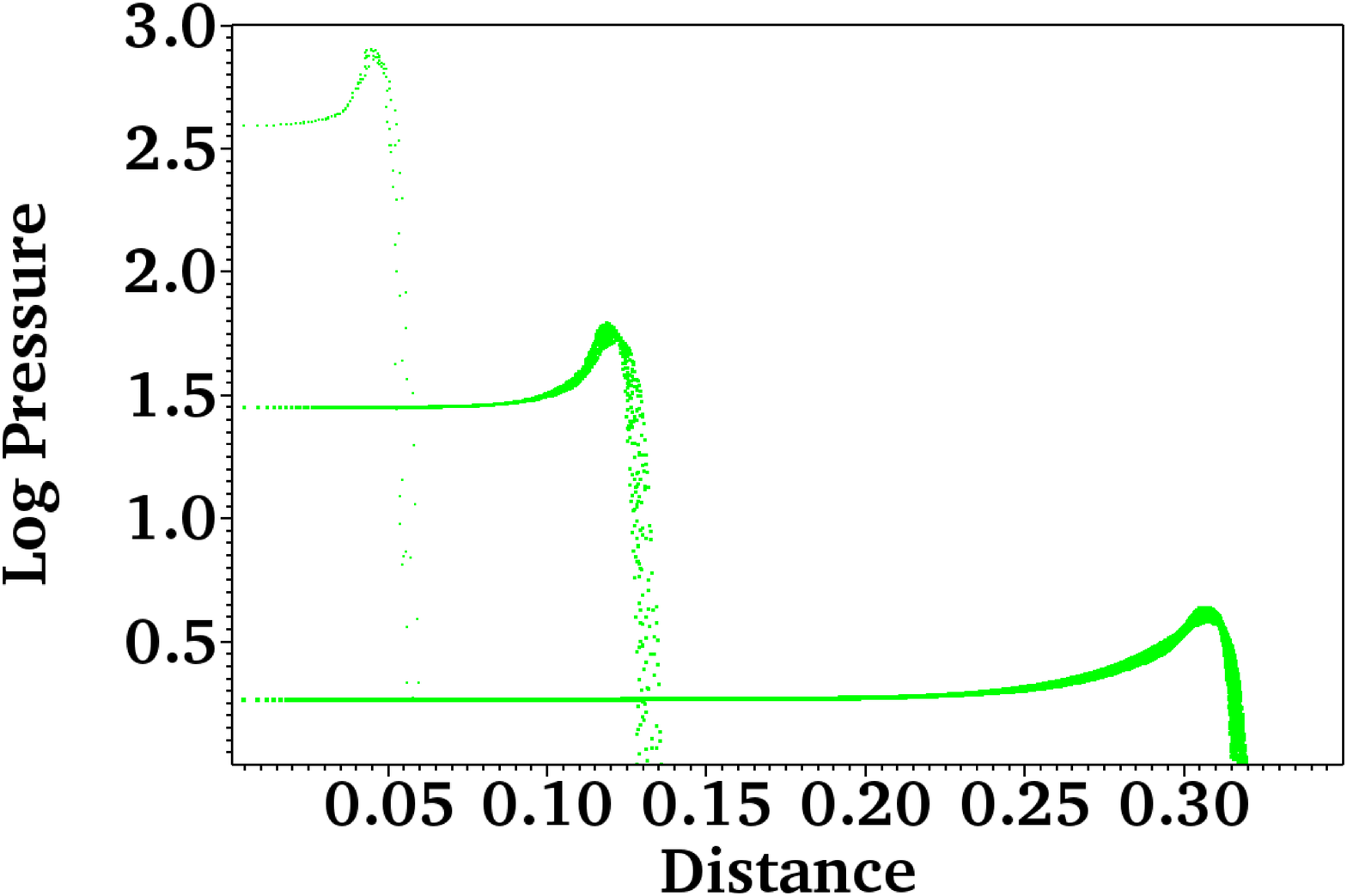}
  \includegraphics[width=0.32\textwidth]{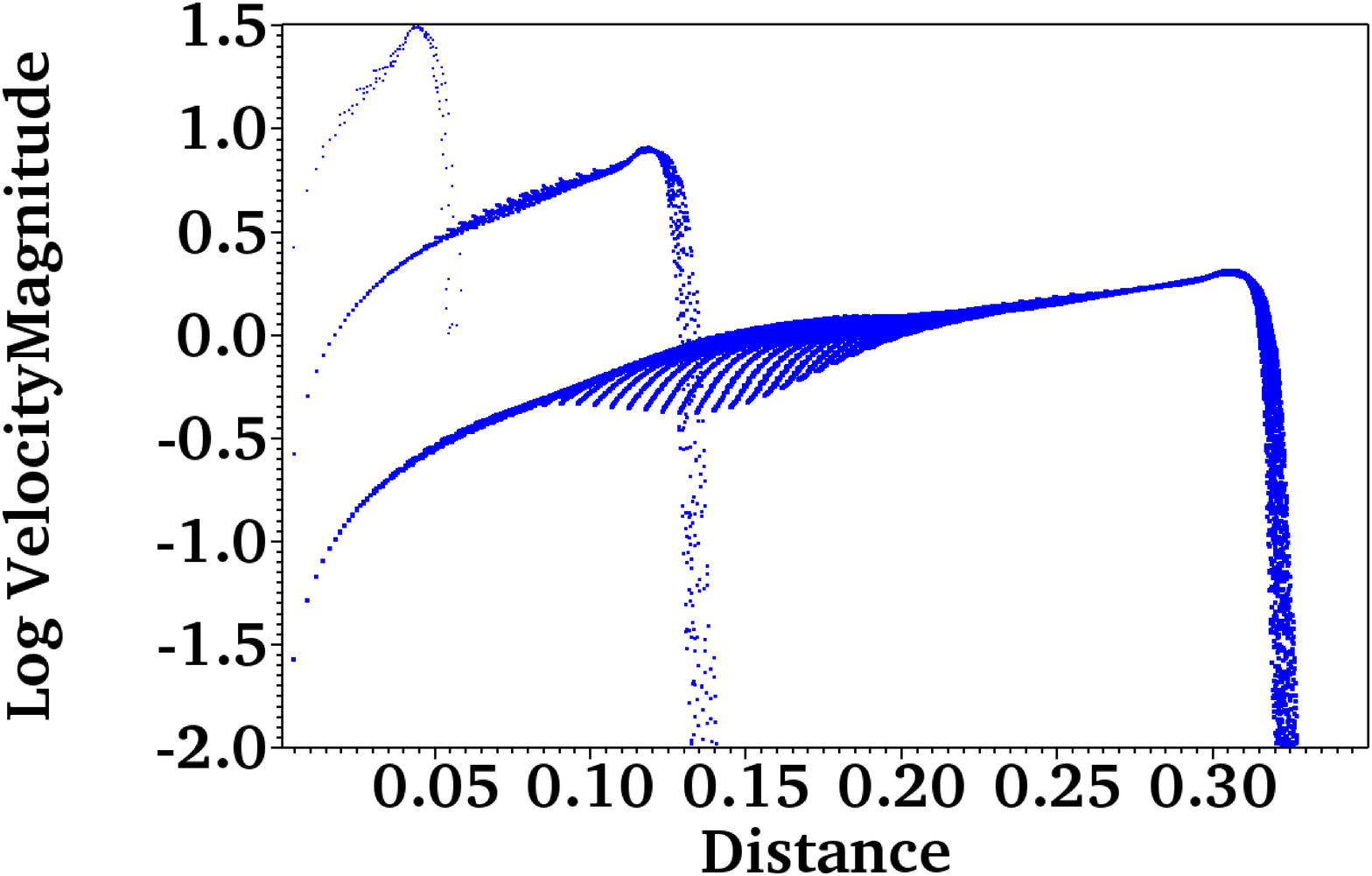}
  \caption{The \texttt{SedovTaylor} blast wave in 3D (Cartesian coordinates): density (left), pressure (center), and velocity magnitude (right), visualized as slices in 3D at $t = 0.05$ (upper) and as scatter plots as a function of radial distance at $t = 0.0005$, $0.005$, and $0.05$ (lower, compare Figs.~\ref{fig:SedovTaylor_1D} and \ref{fig:SedovTaylor_2D}).
Computed with $[128,128,128]$ cells in Cartesian coordinates $[x, y, z]$.}
  \label{fig:SedovTaylor_3D}
\end{centering}
\end{figure}

Output from the \texttt{SedovTaylor} problem is displayed in Figs.~\ref{fig:SedovTaylor_1D}-\ref{fig:SedovTaylor_3D}.
One notable feature in the 1D and 2D curvilinear cases in Figs.~\ref{fig:SedovTaylor_1D} and \ref{fig:SedovTaylor_2D} is the flatness of the pressure curves near the origin: this relies on a cancellation achieved by making the discrete representation of certain curvilinear source terms consistent with the representation of the divergence operator in the finite volume approach.

%%%%%%%%%%%%%%%%%%%%%%%%
\subsection{Fishbone-Moncrief Equilibrium Torus}
\label{sec:FishboneMoncrief}

Our final example problem is a Newtonian version of the Fishbone-Moncrief equilibrium torus \cite{Fishbone1976Relativistic-fl}.
This isentropic axisymmetric configuration has constant specific angular momentum $l$ about the polar axis, and therefore angular velocity 
\begin{equation}
\omega = \frac{l}{r^2 \sin^2 \theta}
\end{equation}
that falls off as the square of the distance from the axis ($r$ and $\theta$ are the radial distance and polar angle in spherical coordinates).
Along with pressure gradients, the angular momentum supports the torus against the gravity of a central point mass $M$; self-gravity is neglected.
The set of allowed configurations is spanned by the range $1 < \kappa < 2$ of the angular momentum parameter 
\begin{equation}
\kappa = \frac{l^2}{G M R_\mathrm{in}},
\end{equation}
where $G$ is the gravitational constant and $R_\mathrm{in}$ is the radius of the inner equatorial edge of the torus.
The fluid distribution is given in terms of the dimensionless enthalpy
\begin{equation}
w = \frac{e + p}{\rho c^2},
\end{equation}
where is $e$ is the internal energy density, $p$ is the pressure, $\rho$ is the mass density, and $c$ is the speed of light.
Remarkably, the expression
\begin{equation}
w(r,\theta) = \frac{G M}{c^2 R_\mathrm{in}} \left[ \frac{R_\mathrm{in}}{r} -1 - \frac{\kappa}{2} \left( 1 - \frac{R_\mathrm{in}}{r^2 \sin^2 \theta} \right) \right]
\end{equation}
characterizes the fluid configuration, regardless of the equation of state, with $w = 0$ defining the boundary of the torus.
The maximum enthalpy is
\begin{equation}
w_\mathrm{max} = \frac{G M}{c^2 R_\mathrm{in}} \left[ \frac{1}{2}\left( \kappa + \frac{1}{\kappa}\right) - 1 \right],
\end{equation}
and it turns out that 
\begin{equation}
R_\mathrm{out} = R_\mathrm{in}\left( \frac{\kappa}{2 - \kappa}\right) 
\end{equation}
is the outer equatorial radius.

Specific choices must of course be made for purposes of numerical evolution.
We utilize a polytropic equation of state with adiabatic index 1.4, and choose $\kappa = 1.85$, $M = 3\, M_\odot$, $R_\mathrm{in} = 6\, G M \,c^{-2}$, and $\rho_\mathrm{max} = 10^{12}\,\mathrm{g \, cm^{-3}}$.
We solve the problem in 2D spherical coordinates $[r, \theta]$, specifying the coordinate system by the mechanism described in the second paragraph of Section~\ref{sec:SedovTaylor}.
To avoid numerical problems associated with the (not fully resolved) steep density gradient at the surface of an isolated gravitationally bound object, we fill the region outside the torus with a tenuous and cold (pressureless) gas in free fall, specifically,
\begin{eqnarray}  
\rho &=& 10^{-6}\, \rho_\mathrm{max} \left( \frac{R_\mathrm{in}}{r}\right)^{3/2}, \\
v^r &=& -\left(\frac{2 G M}{r} \right)^{1/2}
\end{eqnarray}
for the density and radial velocity respectively.
In the cells exterior to the outer radial boundary of the computational domain, these conditions are held fixed throughout the simulation by specification of the \texttt{'INFLOW'} boundary condition 
via the mechanism described in the second paragraph of Section~\ref{sec:RiemannProblem}.
The outflow boundary condition is used on the inner $r$ boundary, and reflecting boundaries are specified at the inner and outer $\theta$ boundaries on the polar axis.

Unlike the previous examples in this paper, we use physical constants and units in this problem.
Constants and units have been discussed in the description of \textsc{GenASiS} \texttt{Basics} \cite{Cardall2015GenASiS-Basics:}, but for completeness we briefly describe their use in this problem.
One use is in parameter specification within the code: for instance, the line
\begin{lstlisting}
  M  =  3.0_KDR  *  UNIT % SOLAR_MASS
\end{lstlisting}
sets the value of the central gravitational mass specified in the previous paragraph.
Units can also be used in parameter specification in parameter file or command line arguments.
For instance, the simulation shown in the figures below was launched with the command line argument 
\begin{lstlisting}
FinishTime=0.1~SECOND
\end{lstlisting}
to run for a longer time than that specified by default.
Another use is in reporting values to standard output.
For instance, the inner equatorial radius and specific angular momentum are specified ($R_\mathrm{in}$) or calculated ($l$) according to expressions in previous paragraphs:
\begin{lstlisting}
  R_In  =  6.0_KDR * G * M  /  c ** 2
  L     =  sqrt ( Kappa * G * M * R_In )
\end{lstlisting}
in which \texttt{G} and \texttt{c} have been aliased by the \texttt{associate} construct to \texttt{CONSTANT \% GRAVITATIONAL} and \texttt{CONSTANT \% SPEED\_OF\_LIGHT} respectively; 
but when reported to standard output using the \texttt{Show} command from \textsc{GenASiS} \texttt{Basics}, more concrete units are used:
\begin{lstlisting}
  call Show ( R_In,  UNIT % KILOMETER, 'R_In' )
  call Show ( L,     UNIT % KILOMETER * UNIT % SPEED_OF_LIGHT, 'L' )
\end{lstlisting}
which results in the display of
\begin{lstlisting}
 R_In  =  2.657931E+001 km
    L  =  1.475890E+001 km c
\end{lstlisting}
to standard output.
Units are also specified as optional arguments in the initialization of the mesh and fluid, so that these are included in output to disk and automatically utilized by the visualization software used to produce the figures below; for details we refer the reader to \texttt{FishboneMoncrief\_Form} in the accompanying code.

\begin{figure}
\begin{centering}
  \includegraphics[width=0.24\textwidth]{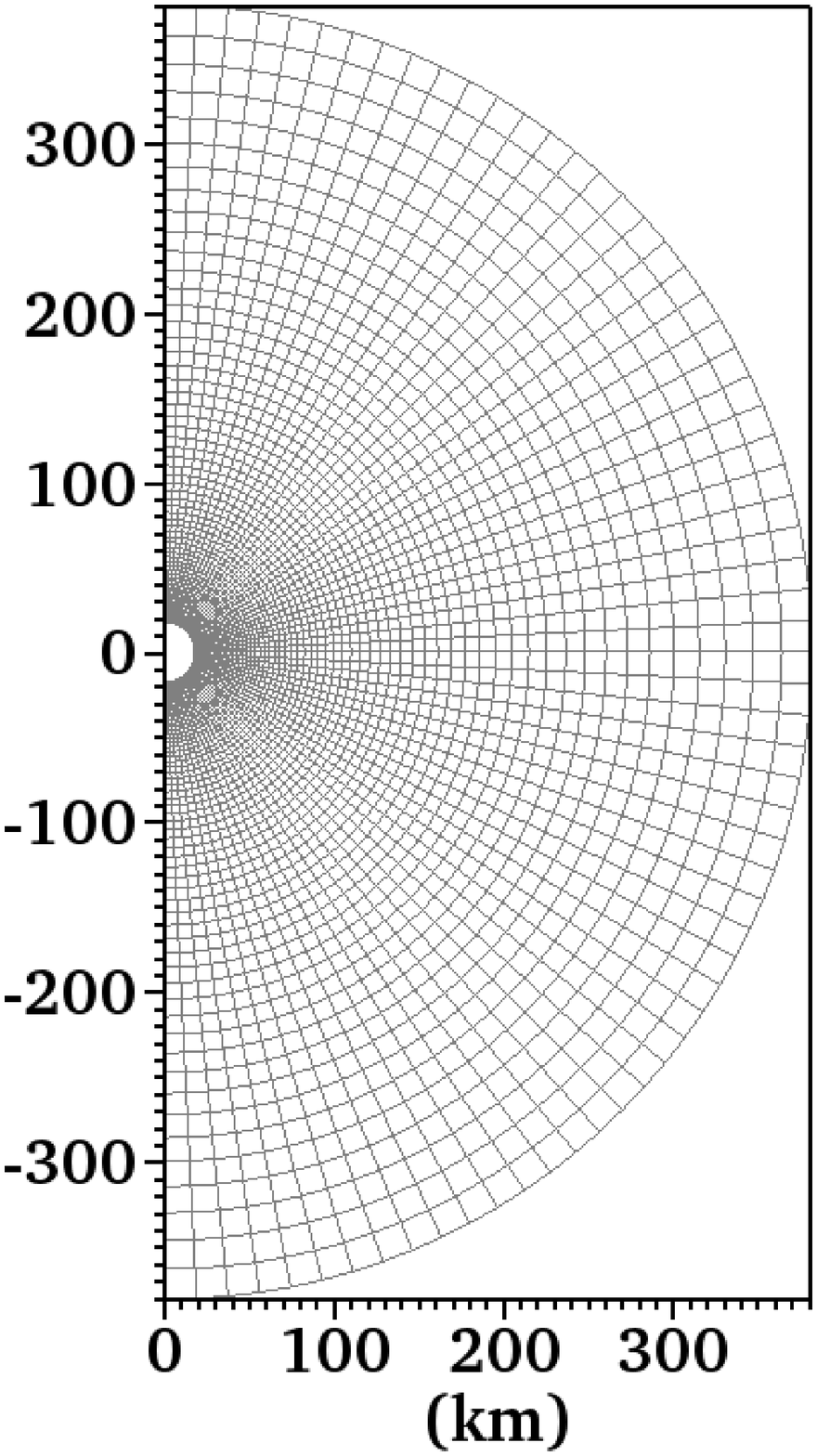}
  \includegraphics[width=0.24\textwidth]{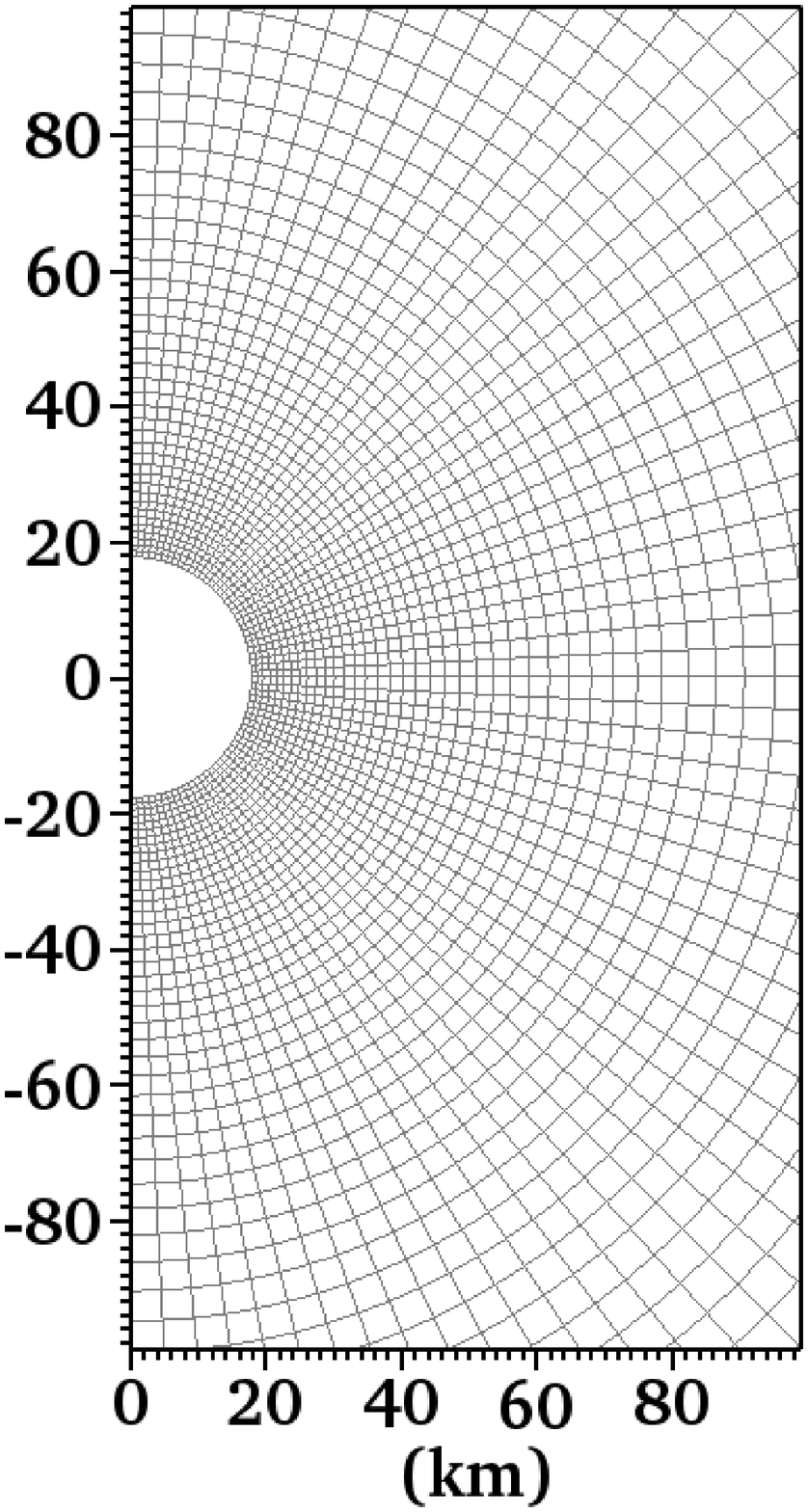}
  \caption{A coarse example of the type of mesh used in the axisymmetric \texttt{FishboneMoncrief} problem: 2D spherical coordinates, with an inner boundary at finite radius, and radial cell width proportional to radius so as to maintain a uniform aspect ratio. 
Shown here are the full computational domain (left) and the inner region (right), with $[64,64]$ cells in $[r, \theta]$.}
  \label{fig:FishboneMoncrief_2D_Mesh}
\end{centering}
\end{figure}

Another difference from our other examples is the use of non-uniform mesh spacing.
A coarse version of the mesh used in this problem is shown in Fig.~\ref{fig:FishboneMoncrief_2D_Mesh}.
A chosen finite inner radial boundary coordinate provides a length scale which, together with a chosen number of equally spaced cells in polar angle, allows construction of a mesh with increasing resolution towards the center while maintaining a cell aspect ratio $\approx 1$ throughout.
In particular, beginning from the inner boundary outward, the expression
\begin{equation}
dr_\leftrightarrow = r_\leftarrow \, d\theta
\end{equation}
relates a cell's radial width $dr_\leftrightarrow$ to its inner radial boundary $r_\leftarrow$ and the uniform angular spacing $d\theta$.

We call this `proportional spacing,' and note that other possibilities for concentrating cells at smaller radius---such as geometric and logarithmic spacing---do not maintain a uniform cell aspect ratio.
To prepare to use this option, the lines
\begin{lstlisting}
Spacing = [ 'PROPORTIONAL', 'EQUAL       ', 'EQUAL       ' ]
Ratio   = [ dTheta, 0.0_KDR, 0.0_KDR ]
\end{lstlisting}
set a couple of local variables.
These are used in the call
\begin{lstlisting}
call PS % CreateChart &
       ( SpacingOption = Spacing, &
         CoordinateSystemOption = CoordinateSystem, &
         CoordinateUnitOption = CoordinateUnit, &
         MinCoordinateOption = MinCoordinate, &
         MaxCoordinateOption = MaxCoordinate, &
         RatioOption = Ratio, &
         nCellsOption = nCells )
\end{lstlisting}
which differs from other calls we have seen to the \texttt{CreateChart} method by the additional inclusion of the \texttt{SpacingOption}, \texttt{RatioOption}, and \texttt{CoordinateUnitOption} arguments (the latter sets the units associated with the coordinates). 
We also note that with proportional spacing the maximum radial coordinate of the mesh is not fixed independently, but follows from the choice of inner boundary and the number of radial cells, the latter of which can be tweaked to give an approximately desired outer boundary radius;
thus the radial component of \texttt{MaxCoordinate} is overwritten by the mesh generation infrastructure to whatever it turns out to be.

\begin{figure}
\begin{centering}
  \includegraphics[width=0.32\textwidth]{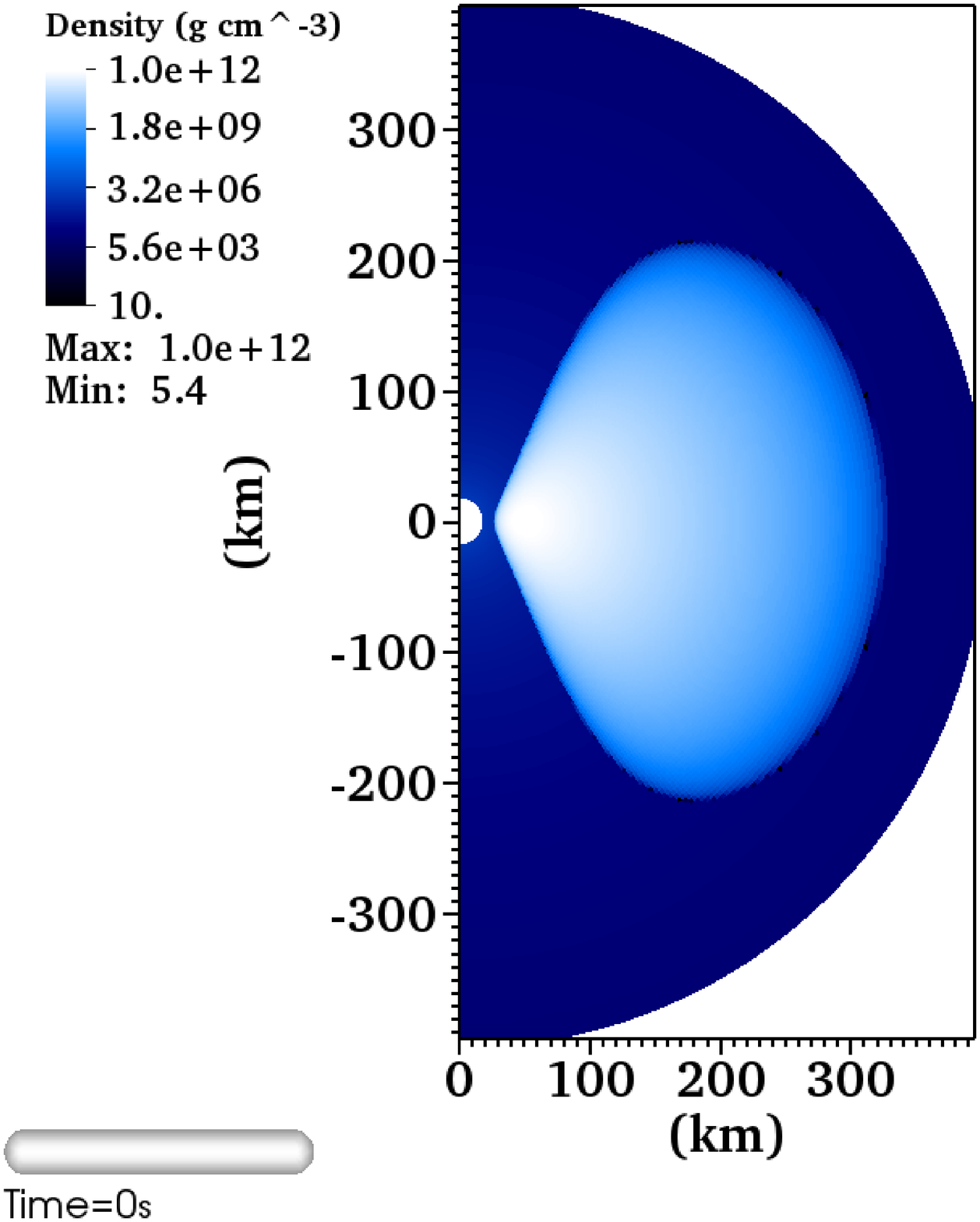}
  \includegraphics[width=0.32\textwidth]{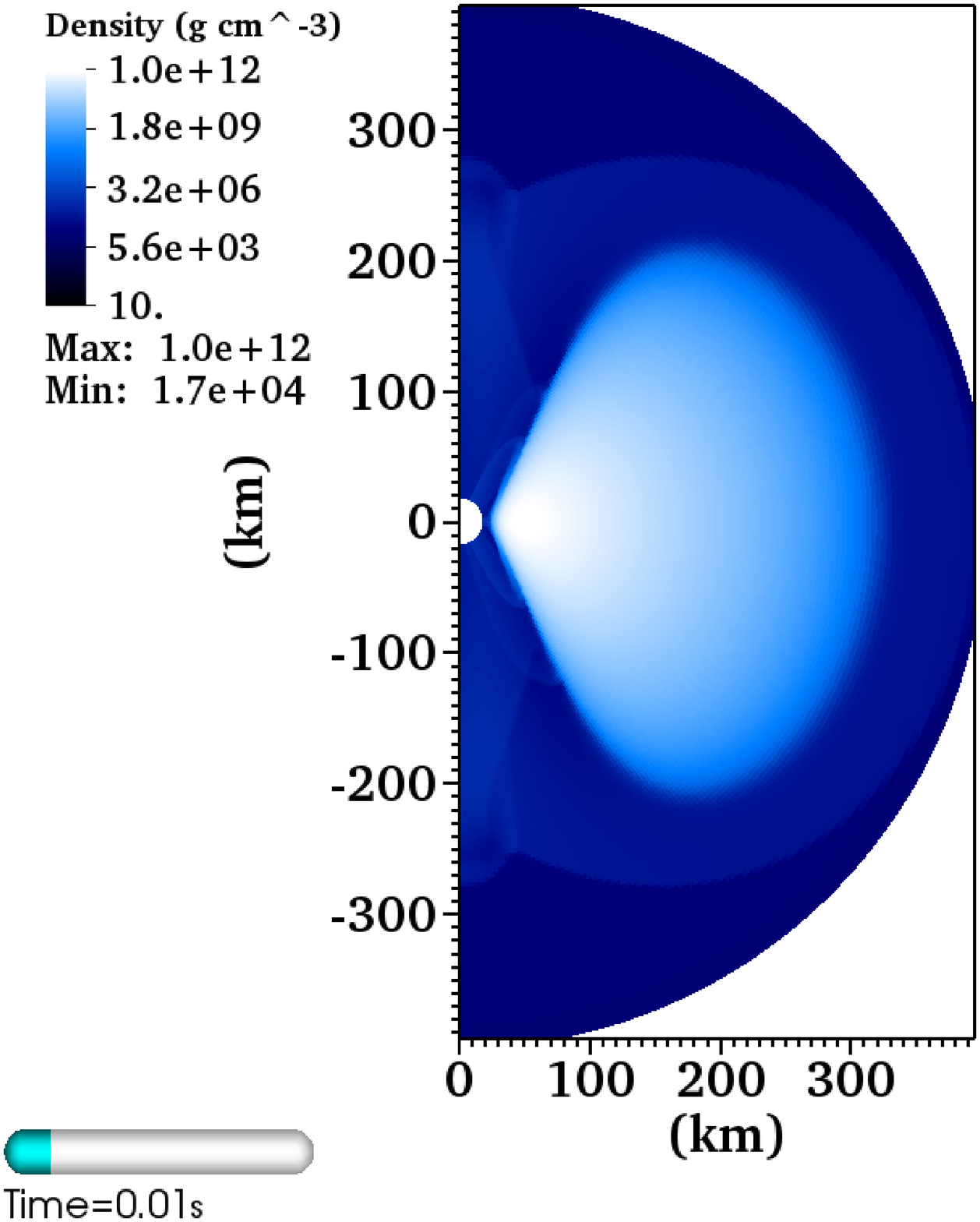}
  \includegraphics[width=0.32\textwidth]{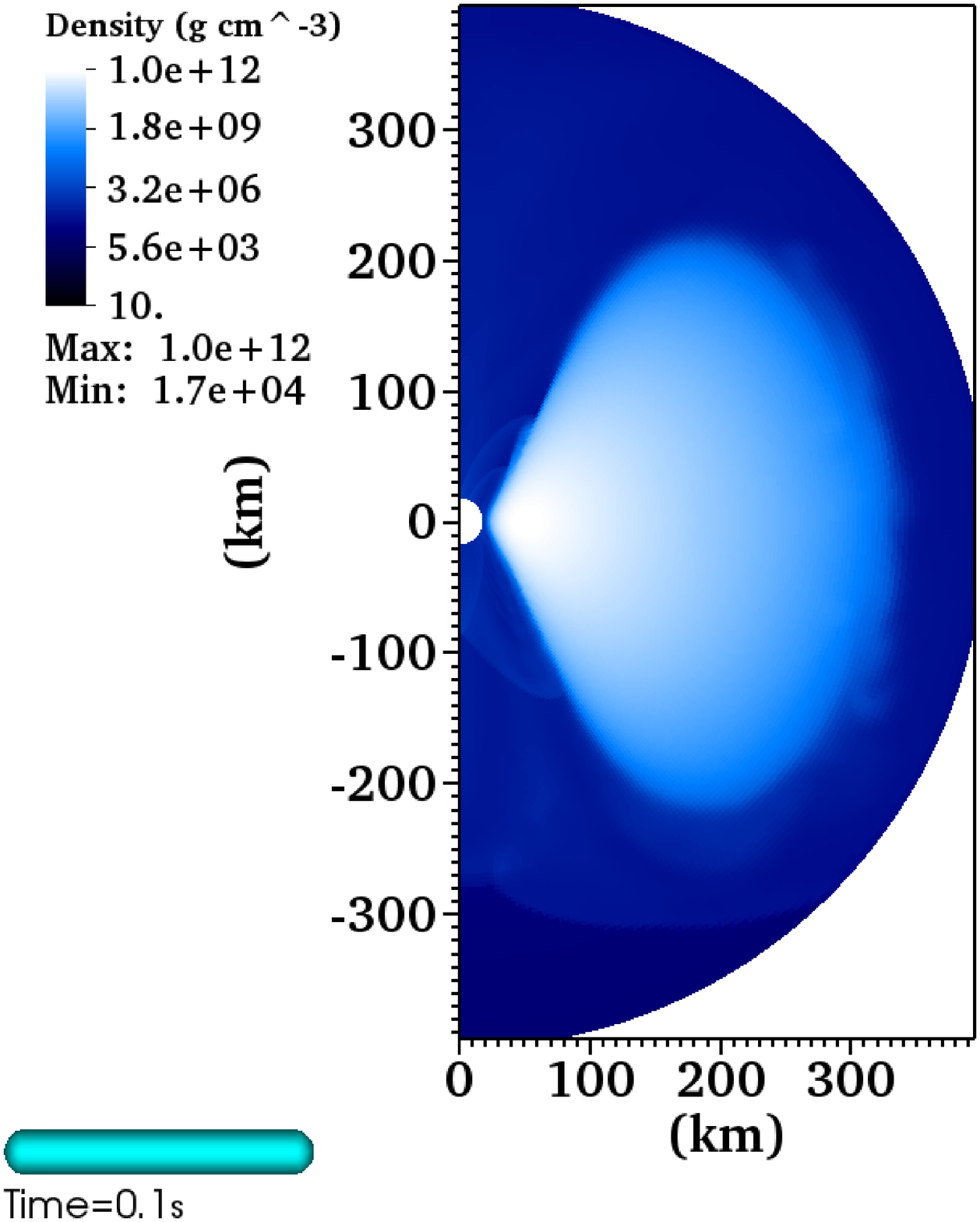}
  \includegraphics[width=0.32\textwidth]{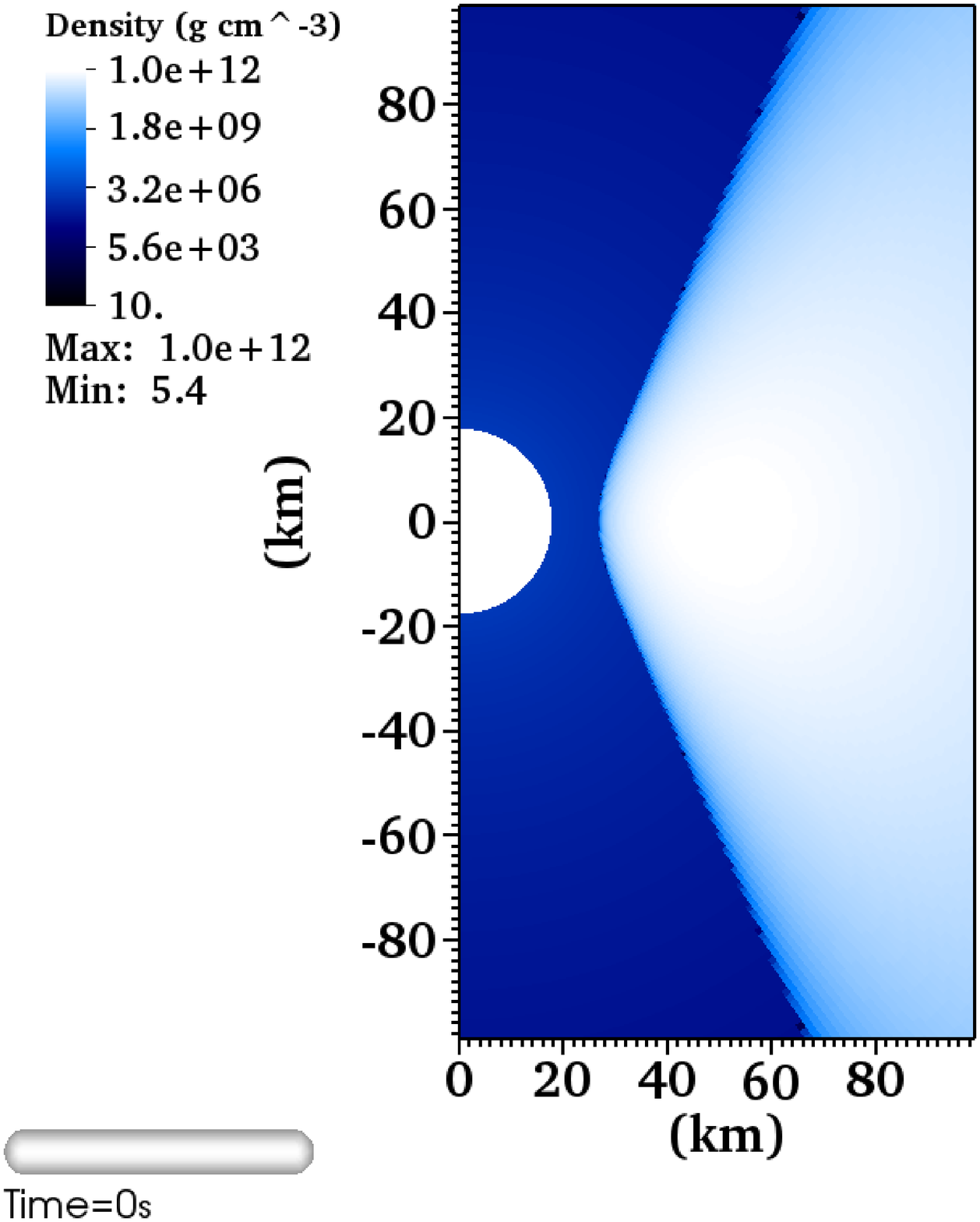}
  \includegraphics[width=0.32\textwidth]{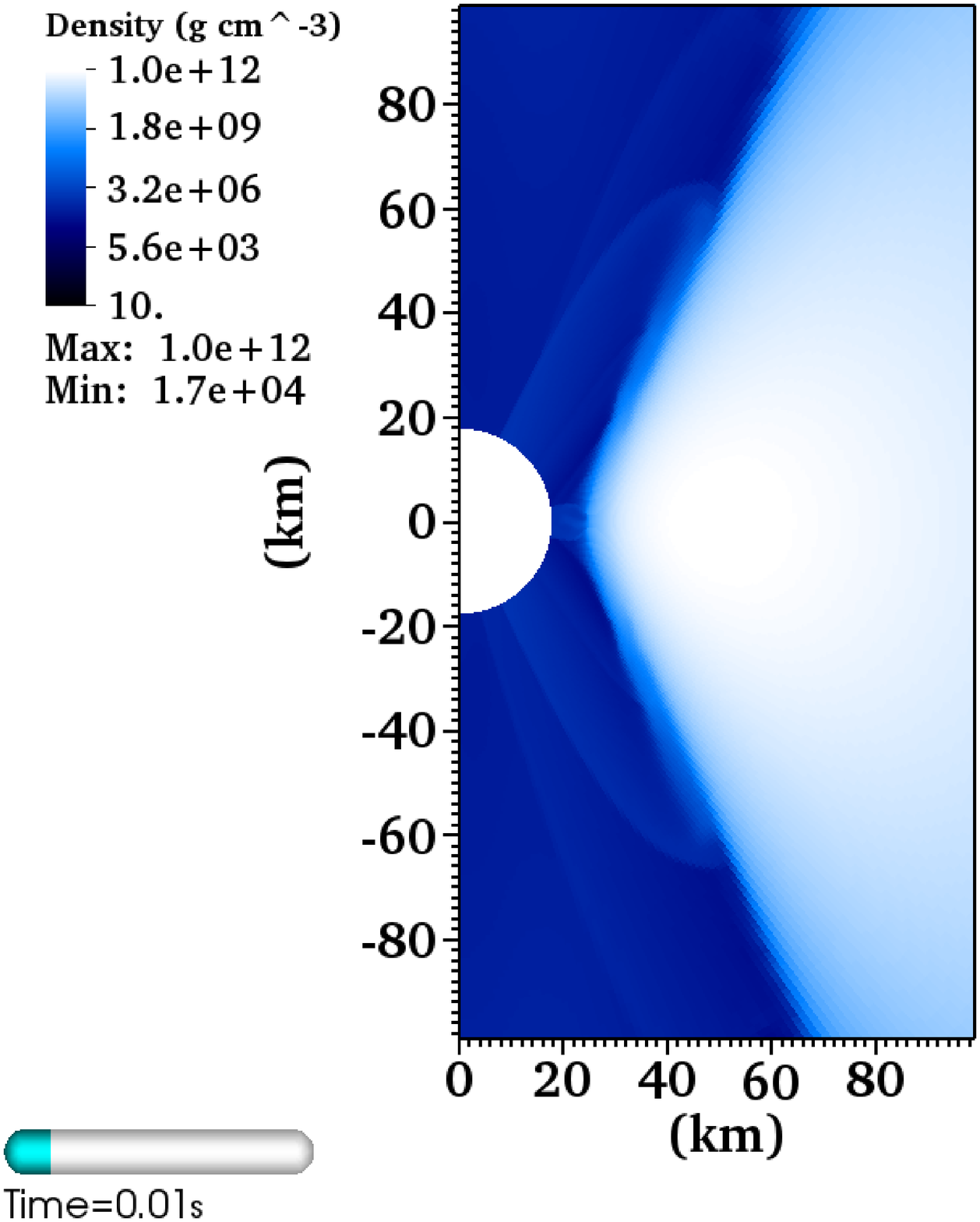}
  \includegraphics[width=0.32\textwidth]{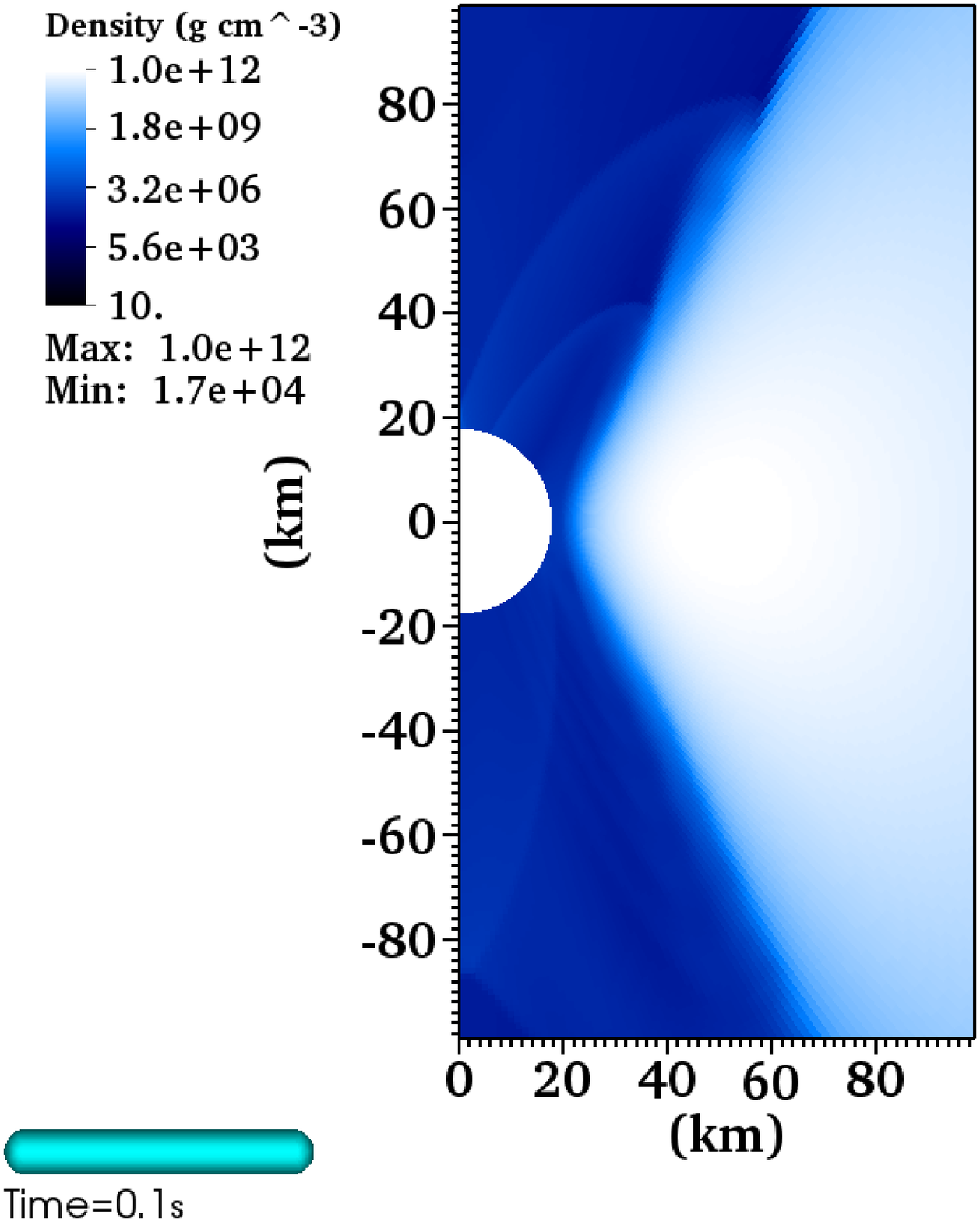}
  \caption{Log density in the \texttt{FishboneMoncrief} problem, at times $t = 0.0\,\mathrm{s}$ (left), $t = 0.01\,\mathrm{s}$ (center), and $t = 0.1\,\mathrm{s}$ (right), showing the full computational domain (upper) and magnifying the inner region (lower).
Computed with $[256,256]$ cells in spherical coordinates $[r,\theta]$.}
  \label{fig:FishboneMoncrief_2D_Density}
\end{centering}
\end{figure}

\begin{figure}
\begin{centering}
  \includegraphics[width=0.32\textwidth]{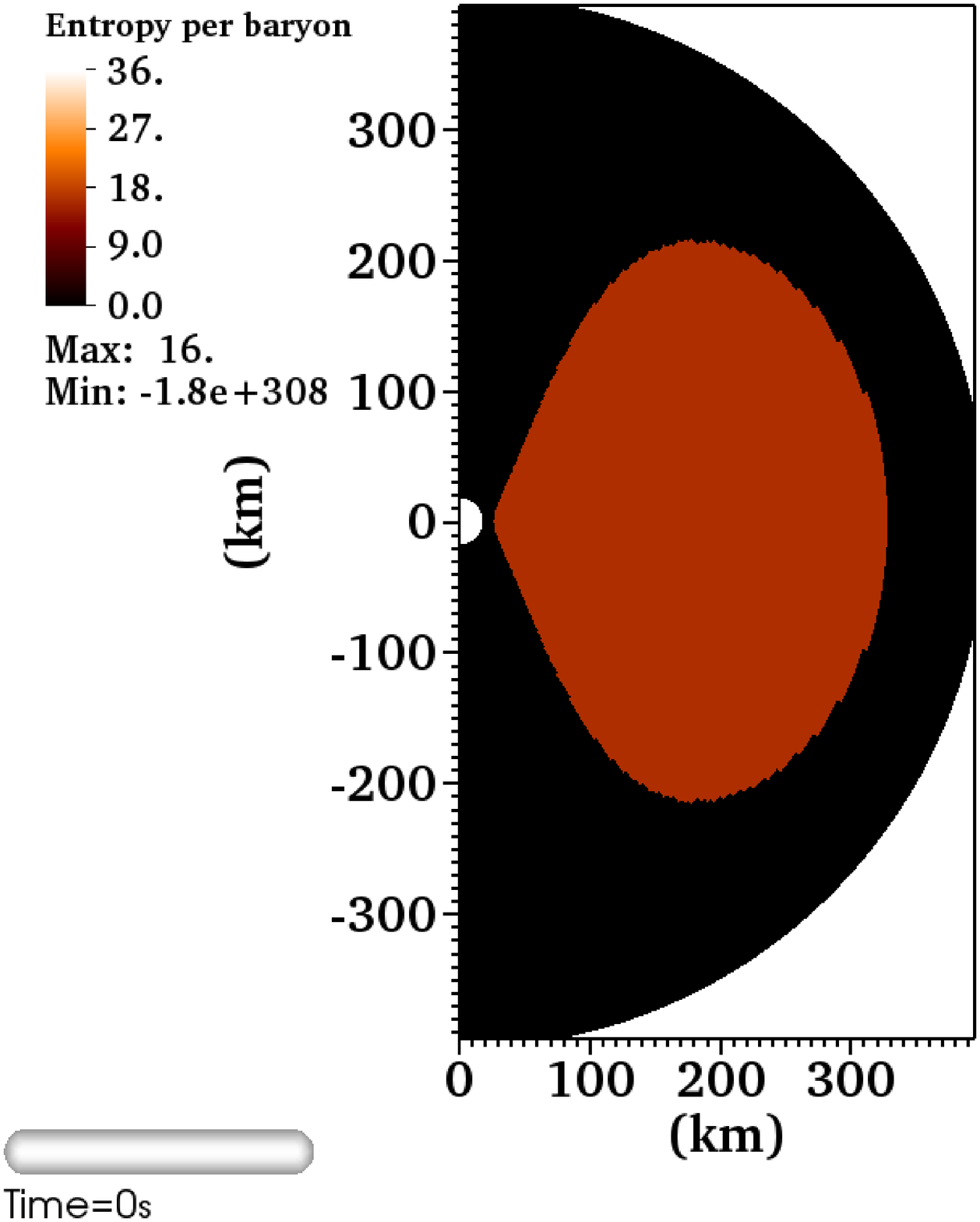}
  \includegraphics[width=0.32\textwidth]{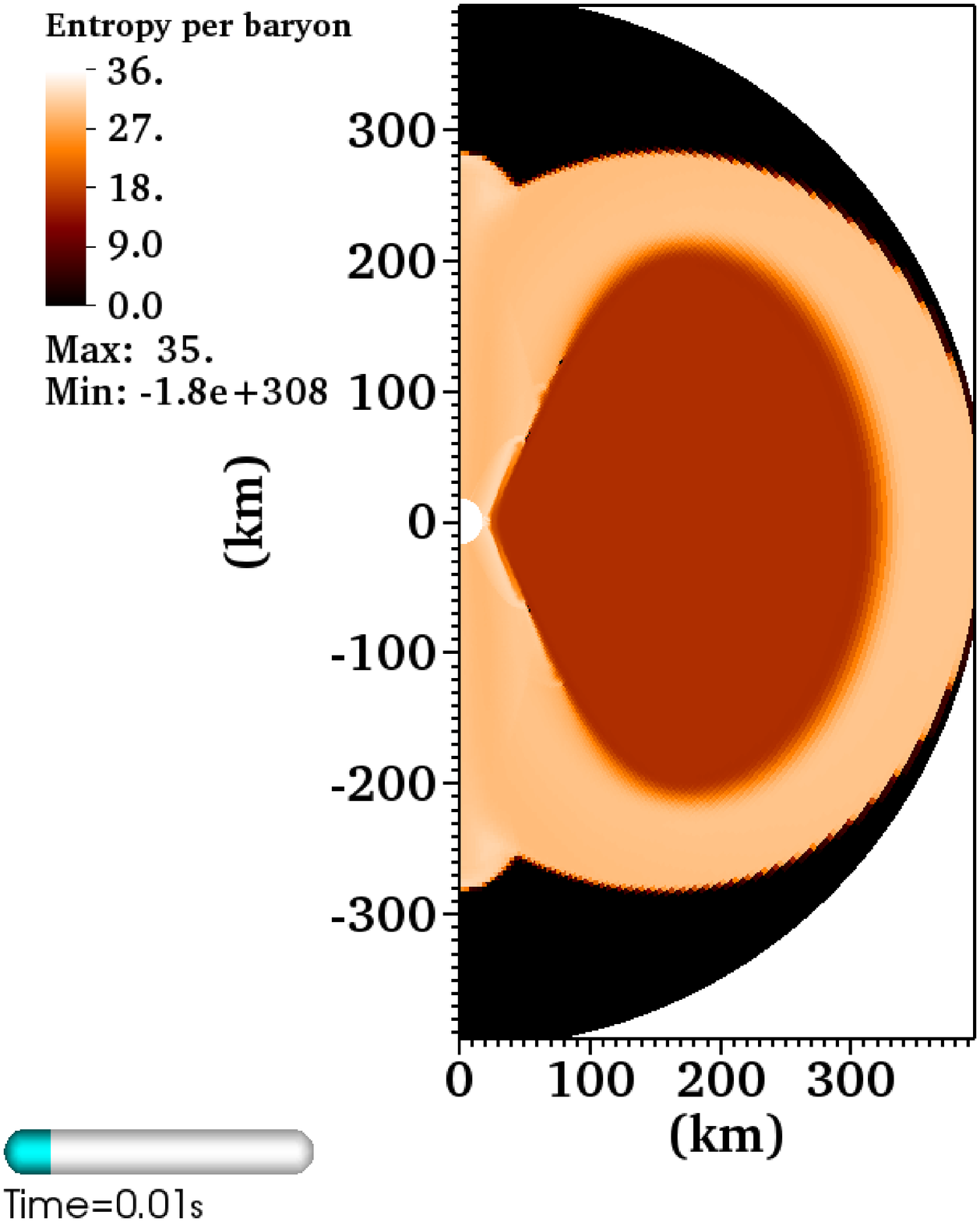}
  \includegraphics[width=0.32\textwidth]{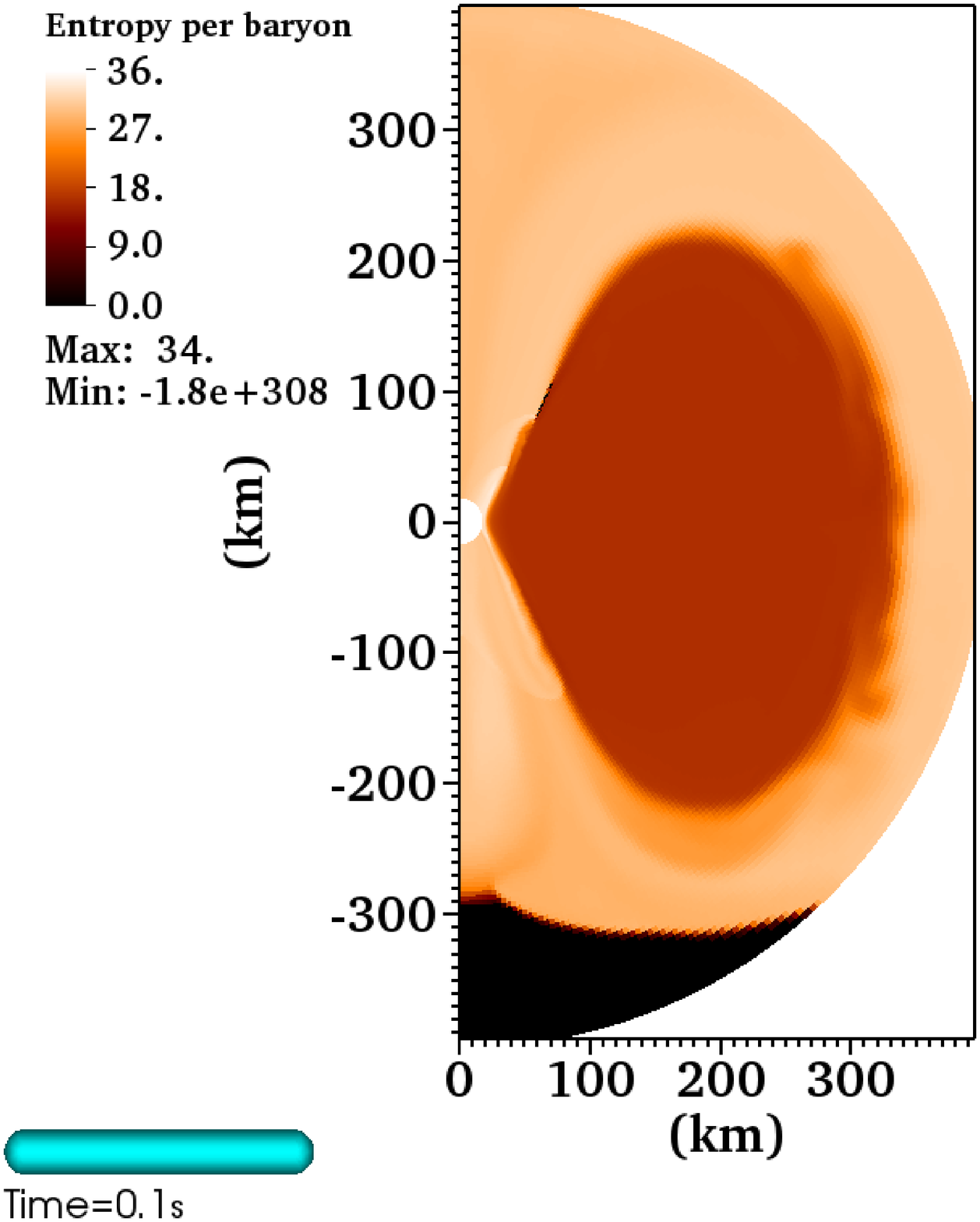}
  \includegraphics[width=0.32\textwidth]{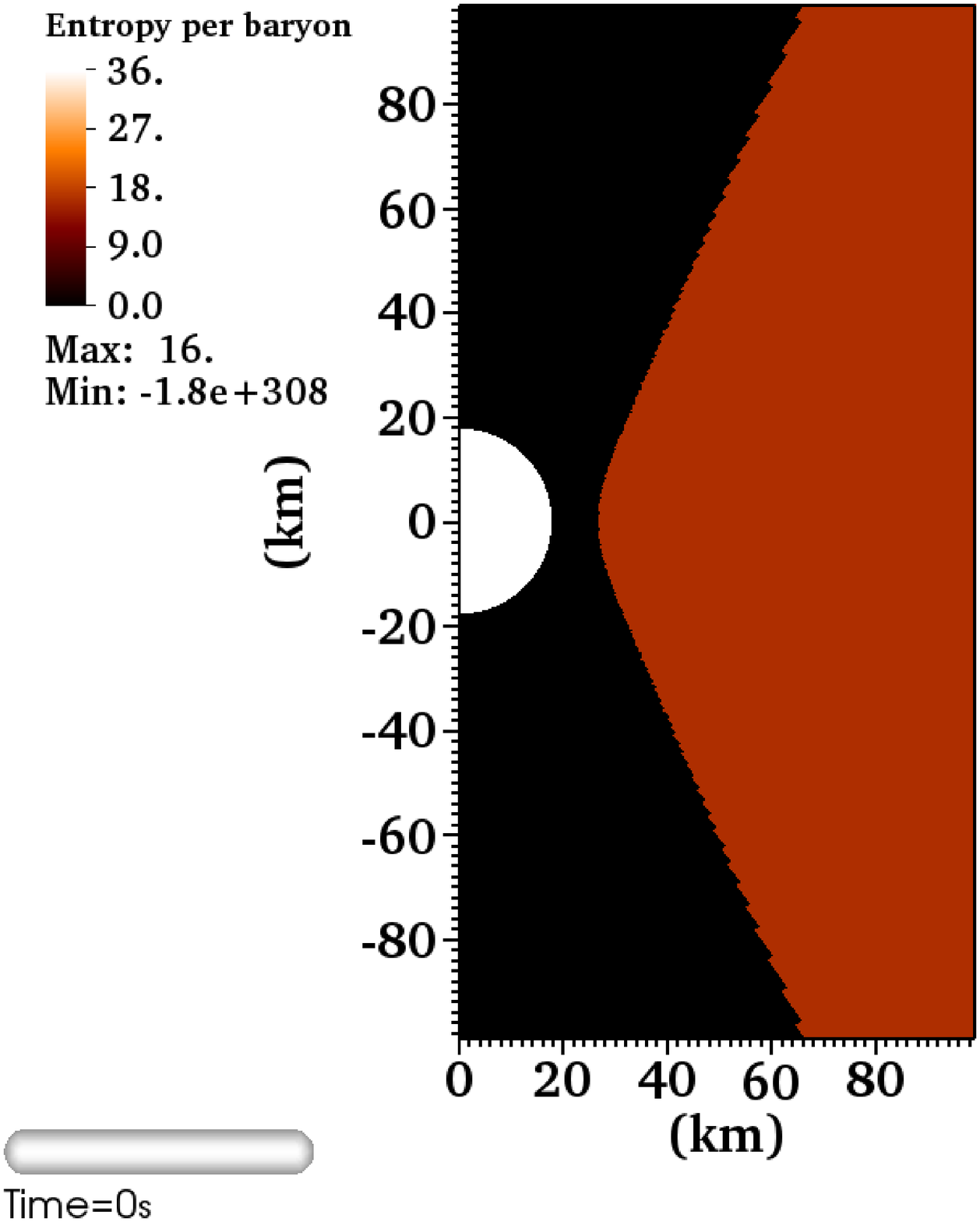}
  \includegraphics[width=0.32\textwidth]{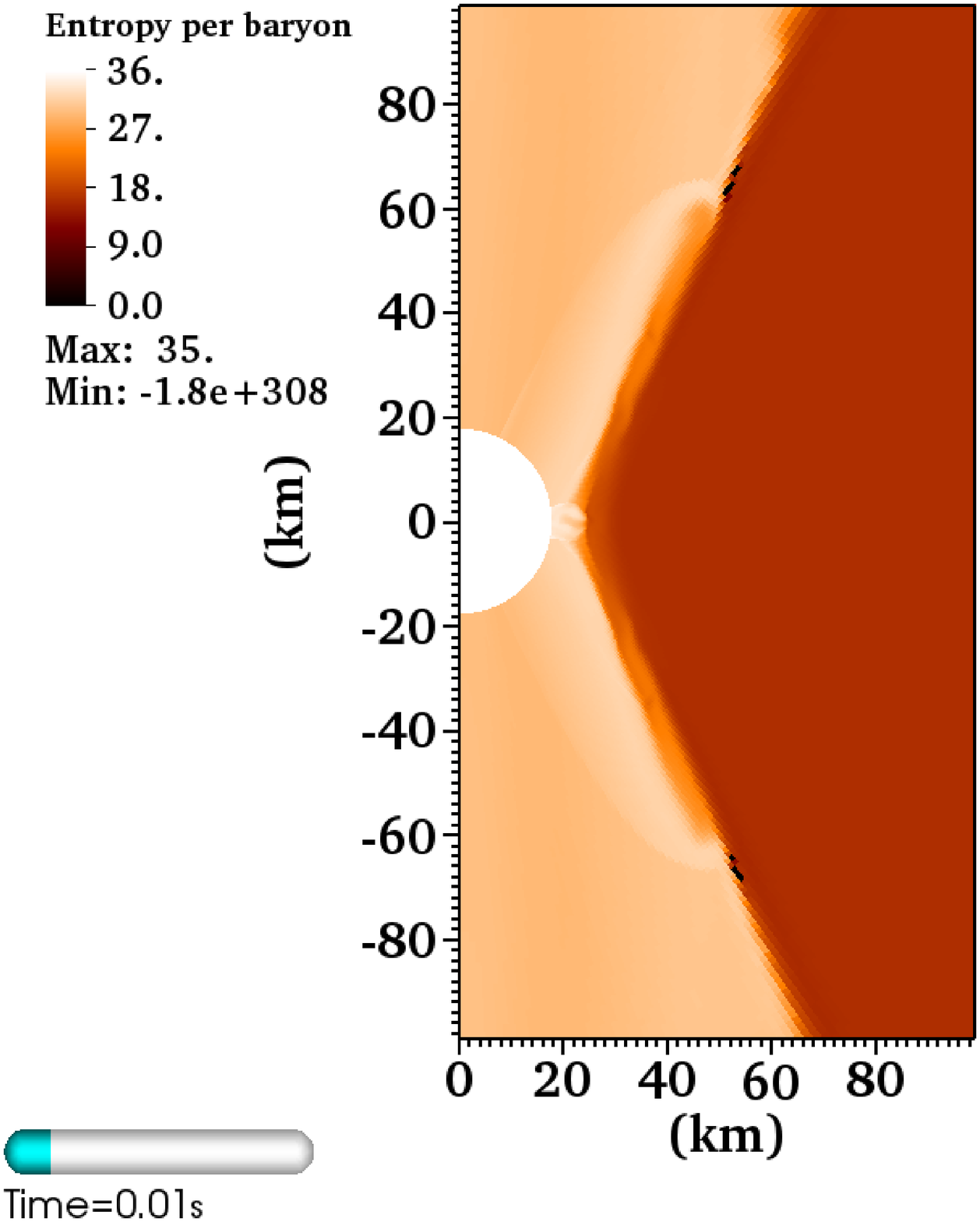}
  \includegraphics[width=0.32\textwidth]{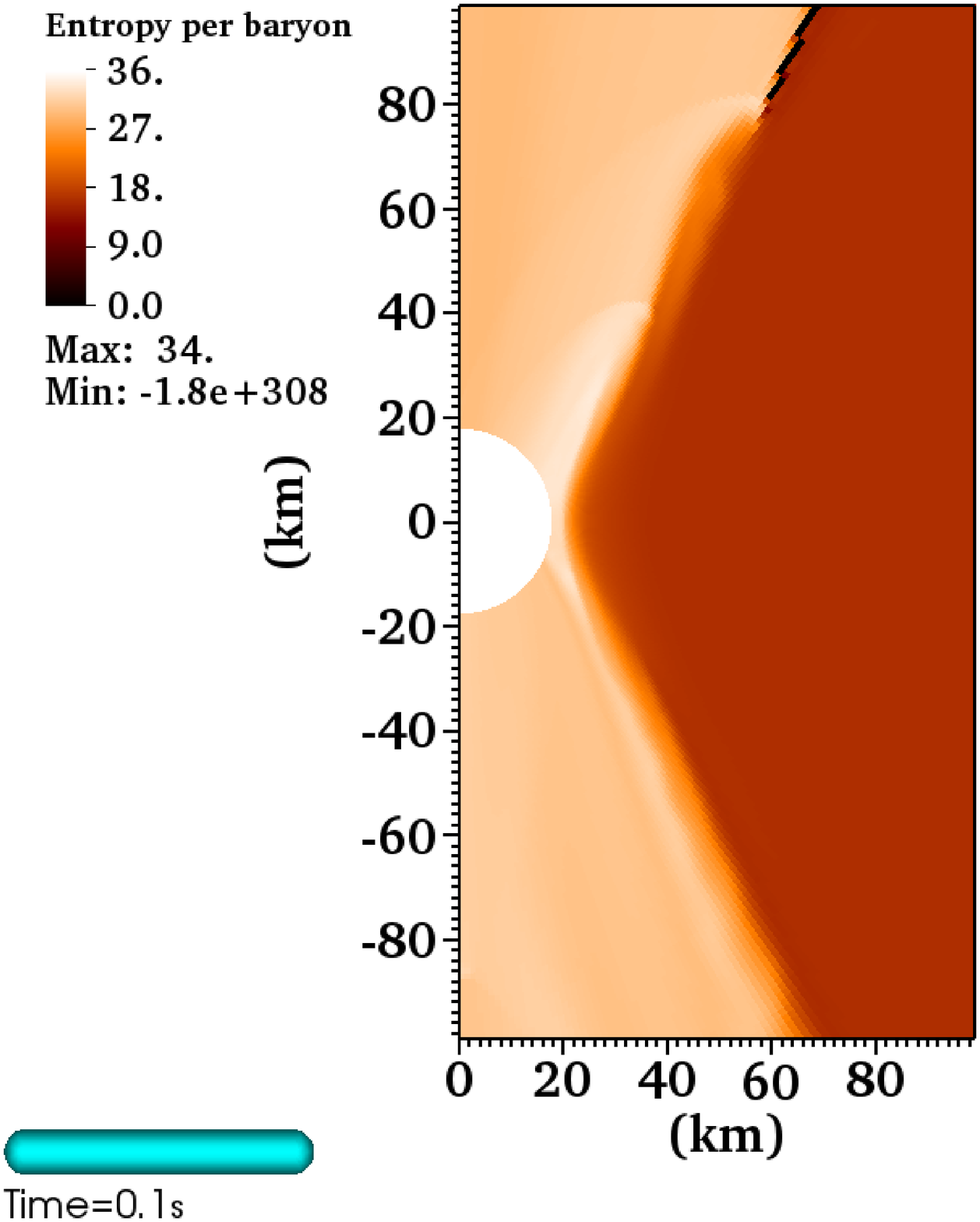}
  \caption{Entropy in the \texttt{FishboneMoncrief} problem, at times $t = 0.0\,\mathrm{s}$ (left), $t = 0.01\,\mathrm{s}$ (center), and $t = 0.1\,\mathrm{s}$ (right), showing the full computational domain (upper) and magnifying the inner region (lower).
  Computed with $[256,256]$ cells in spherical coordinates $[r,\theta]$.}
  \label{fig:FishboneMoncrief_2D_Entropy}
\end{centering}
\end{figure}

\begin{figure}
\begin{centering}
  \includegraphics[width=0.32\textwidth]{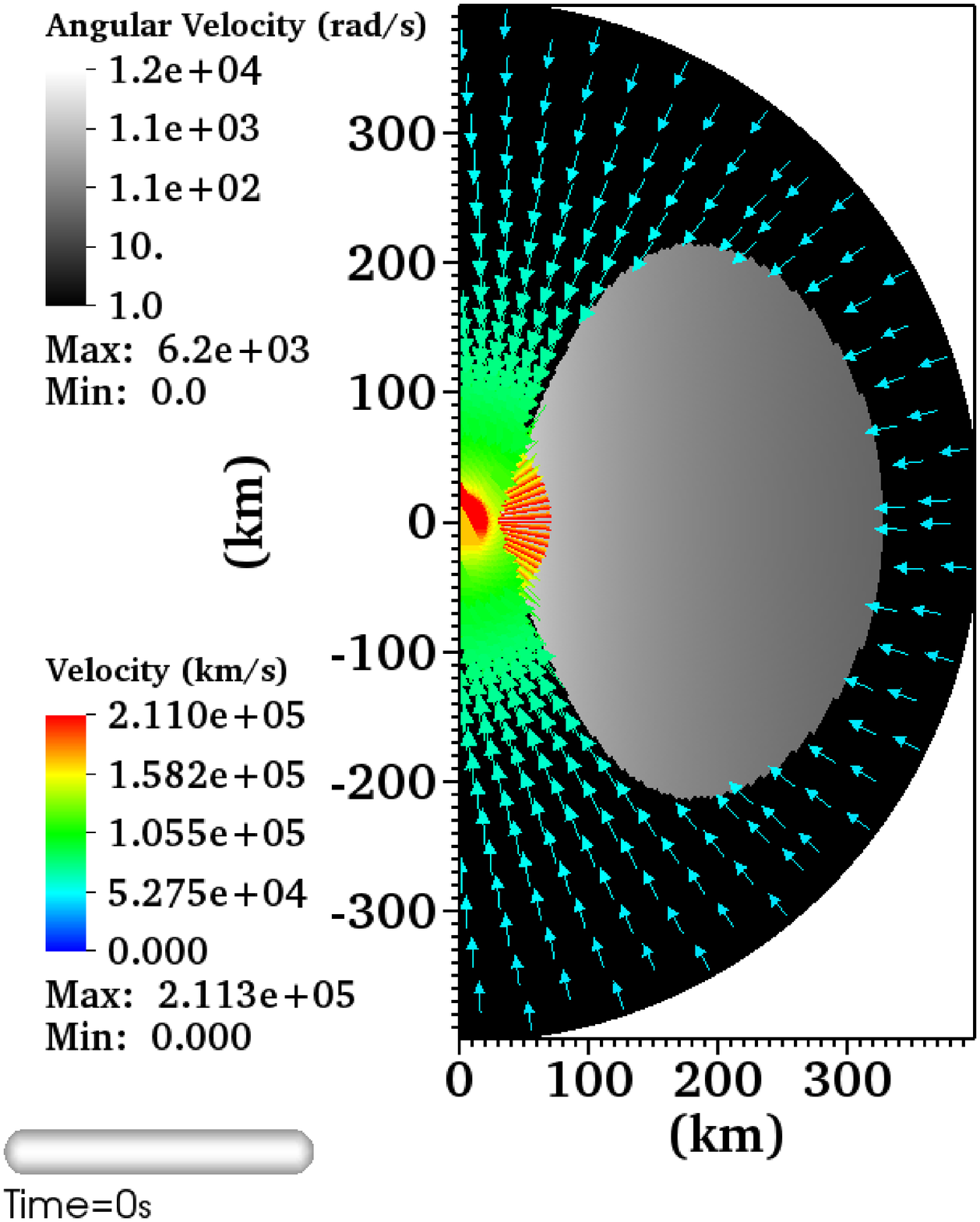}
  \includegraphics[width=0.32\textwidth]{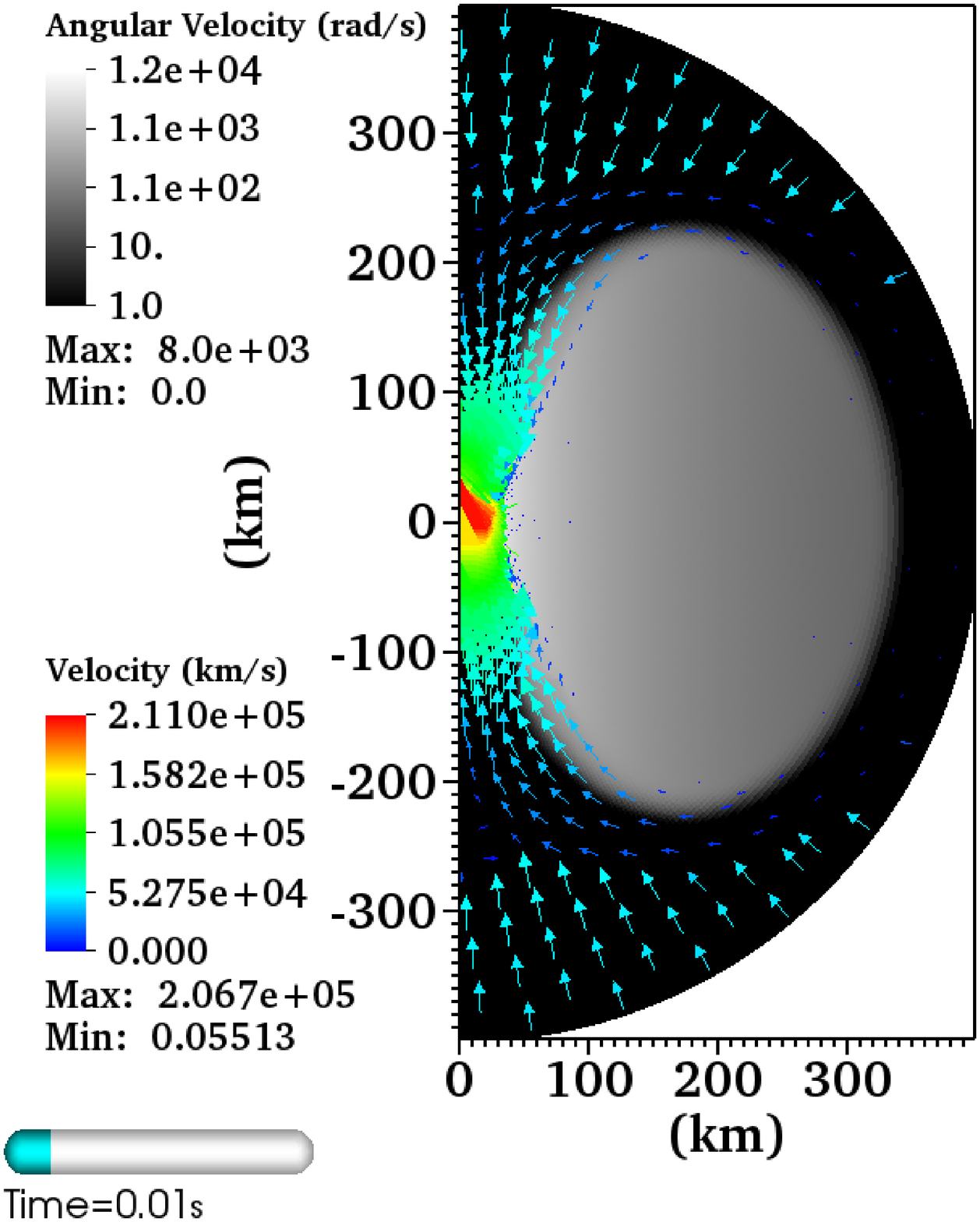}
  \includegraphics[width=0.32\textwidth]{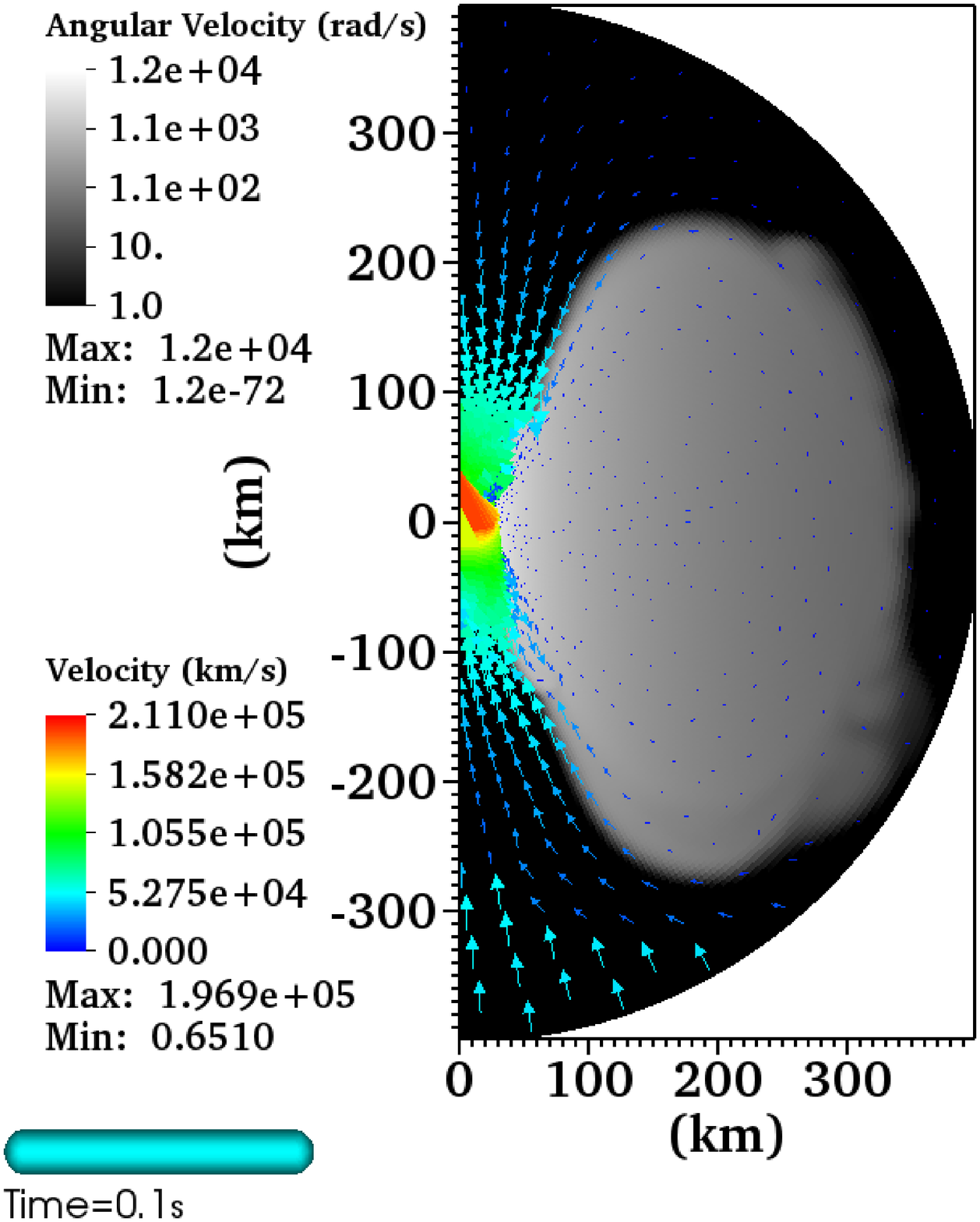}
  \includegraphics[width=0.32\textwidth]{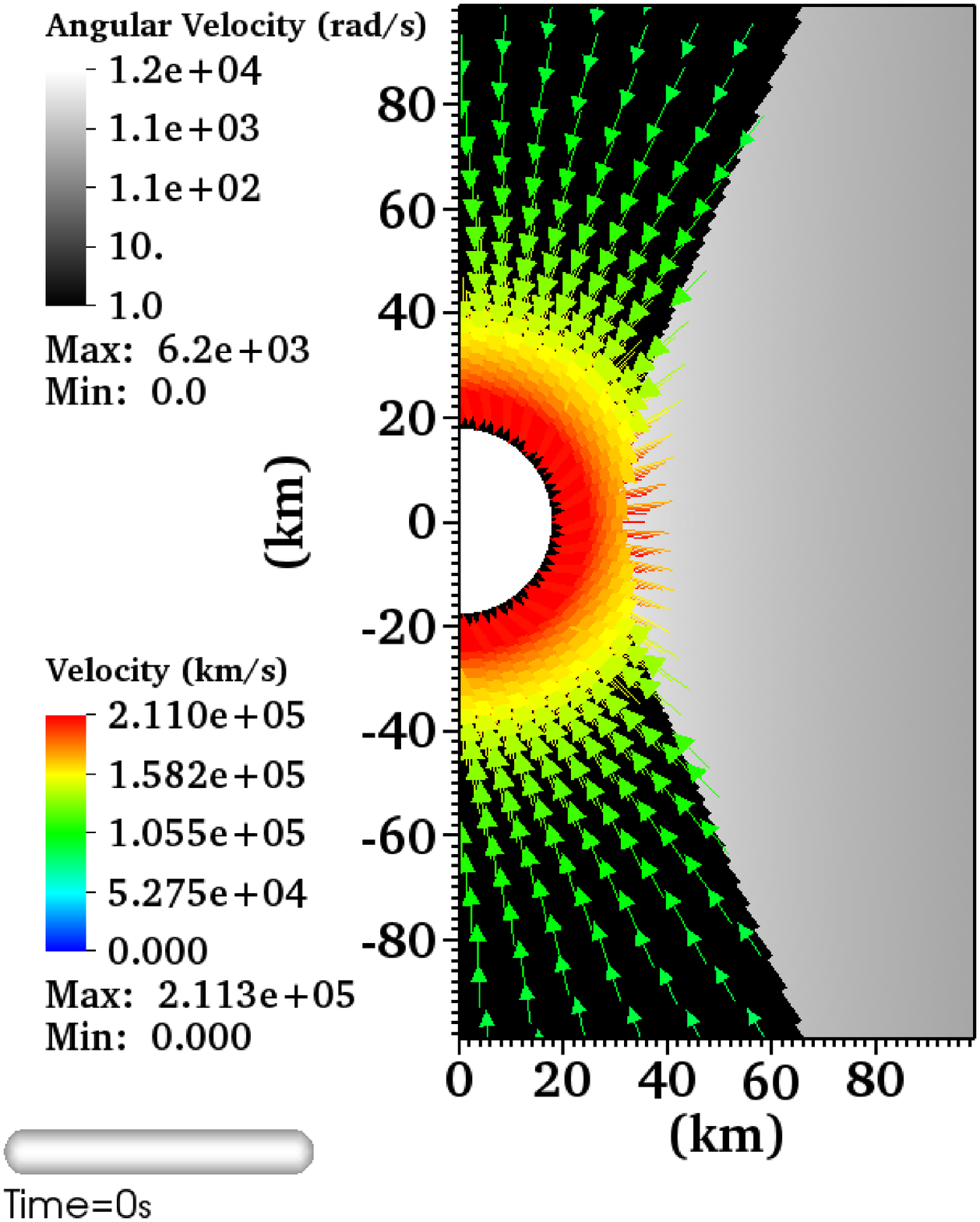}
  \includegraphics[width=0.32\textwidth]{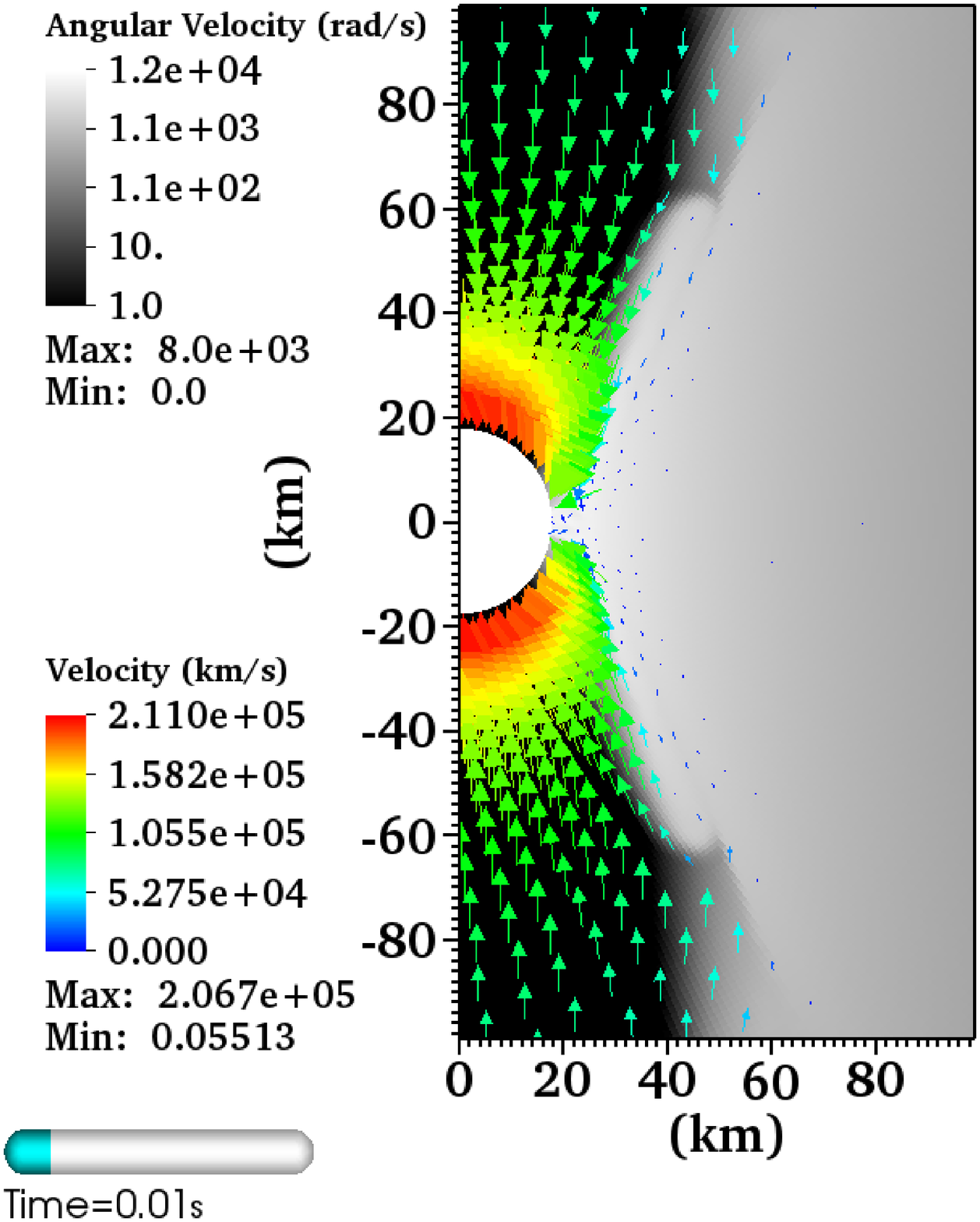}
  \includegraphics[width=0.32\textwidth]{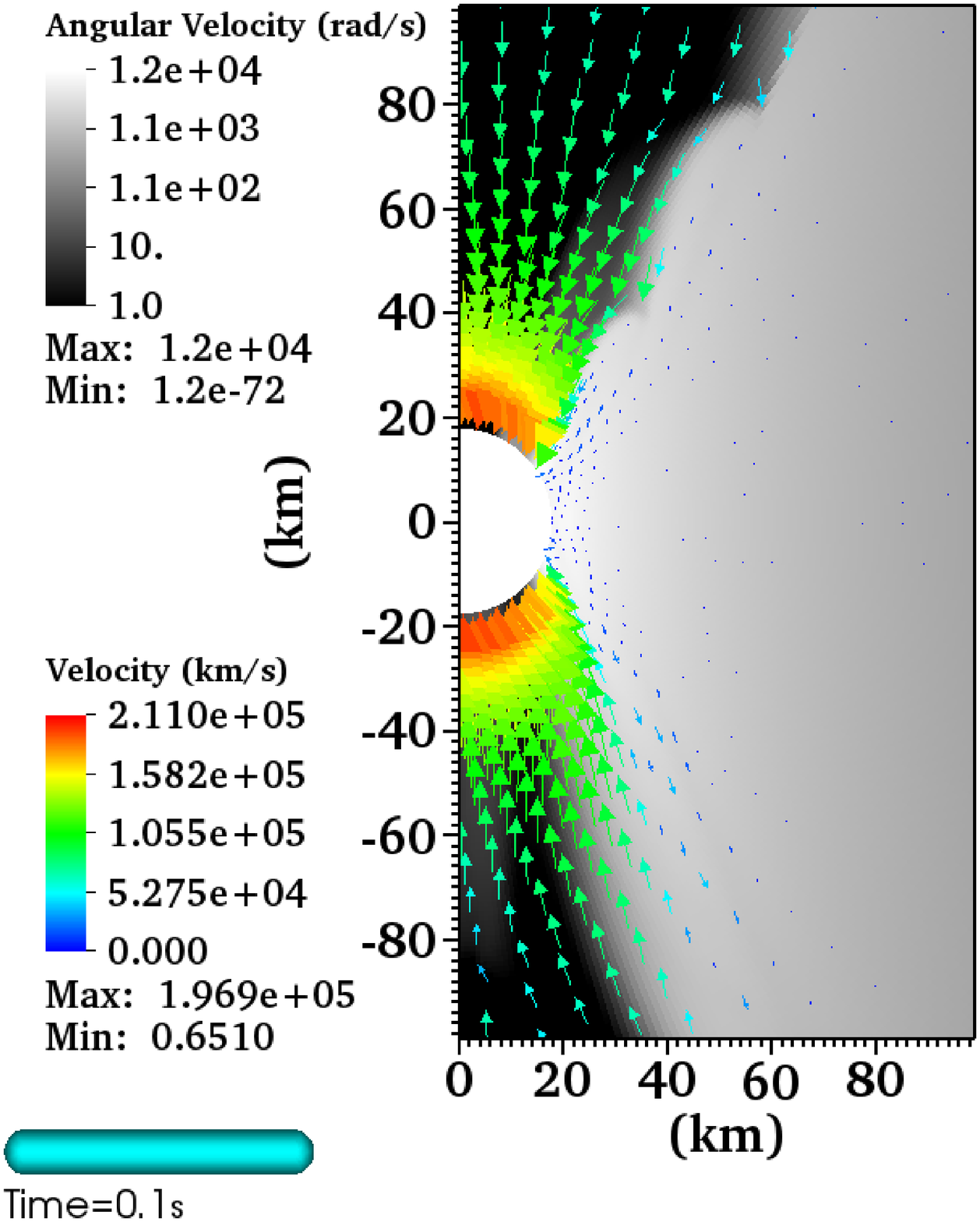}
  \caption{Angular velocity about the symmetry axis (greyscale) and poloidal velocity (vector field)  in the \texttt{FishboneMoncrief} problem, at times $t = 0.0\,\mathrm{s}$ (left), $t = 0.01\,\mathrm{s}$ (center), and $t = 0.1\,\mathrm{s}$ (right), showing the full computational domain (upper) and magnifying the inner region (lower).
  Computed with $[256,256]$ cells in spherical coordinates $[r,\theta]$.}
  \label{fig:FishboneMoncrief_2D_Velocity}
\end{centering}
\end{figure}

The torus maintains itself quite satisfactorily during evolution from $t = 0.0\,\mathrm{s}$ to $t = 0.1\,\mathrm{s}$, an interval comprising some $26$ dynamical times $(G \rho_\mathrm{max})^{-1/2}$ and corresponding to almost 100 orbits at the inner edge; see Figs.~\ref{fig:FishboneMoncrief_2D_Density} (density), \ref{fig:FishboneMoncrief_2D_Entropy} (entropy), and \ref{fig:FishboneMoncrief_2D_Velocity} (velocity).
The surface of the torus is initially sharp in the left panels of these figures at $t = 0.0\,\mathrm{s}$, with the free fall of the tenuous cold gas apparent in Fig.~\ref{fig:FishboneMoncrief_2D_Velocity}.
The imperfectly resolved steep density gradient and strong angular velocity shearing at the torus surface, together with the diffusivity generated at discontinuities by the finite volume scheme, result in the spread of tiny amounts of surface material. 
This is barely visible---even on a logarithmic scale---as wisps of density in Fig.~\ref{fig:FishboneMoncrief_2D_Density}, which emphasizes the relative durability of the torus itself, the object of interest in this test.
But for completeness and out of curiosity we also note how the code handles the material outside the torus. The solver-discontinuity-heated material leaving the torus surface is more apparent as higher entropy matter in Fig.~\ref{fig:FishboneMoncrief_2D_Entropy}, which as a `hot bubble' resists the infall of the surrounding cold gas in Fig.~\ref{fig:FishboneMoncrief_2D_Velocity}.  
In the middle panels at $t = 0.01\,\mathrm{s}$ this stray torus material has spread some distance from the surface; it goes on to fill out the computational volume, but is itself infalling toward the origin, such that infall of cold material from the constant outer boundary condition at times comes back into play, as in the bottom of the upper right panels of Figs.~\ref{fig:FishboneMoncrief_2D_Entropy} and \ref{fig:FishboneMoncrief_2D_Velocity}.
By the end of simulation at $t = 0.1\,\mathrm{s}$ the torus has suffered some disruption at its outer edge.

\begin{figure}
\begin{centering}
  \includegraphics[width=0.51\textwidth]{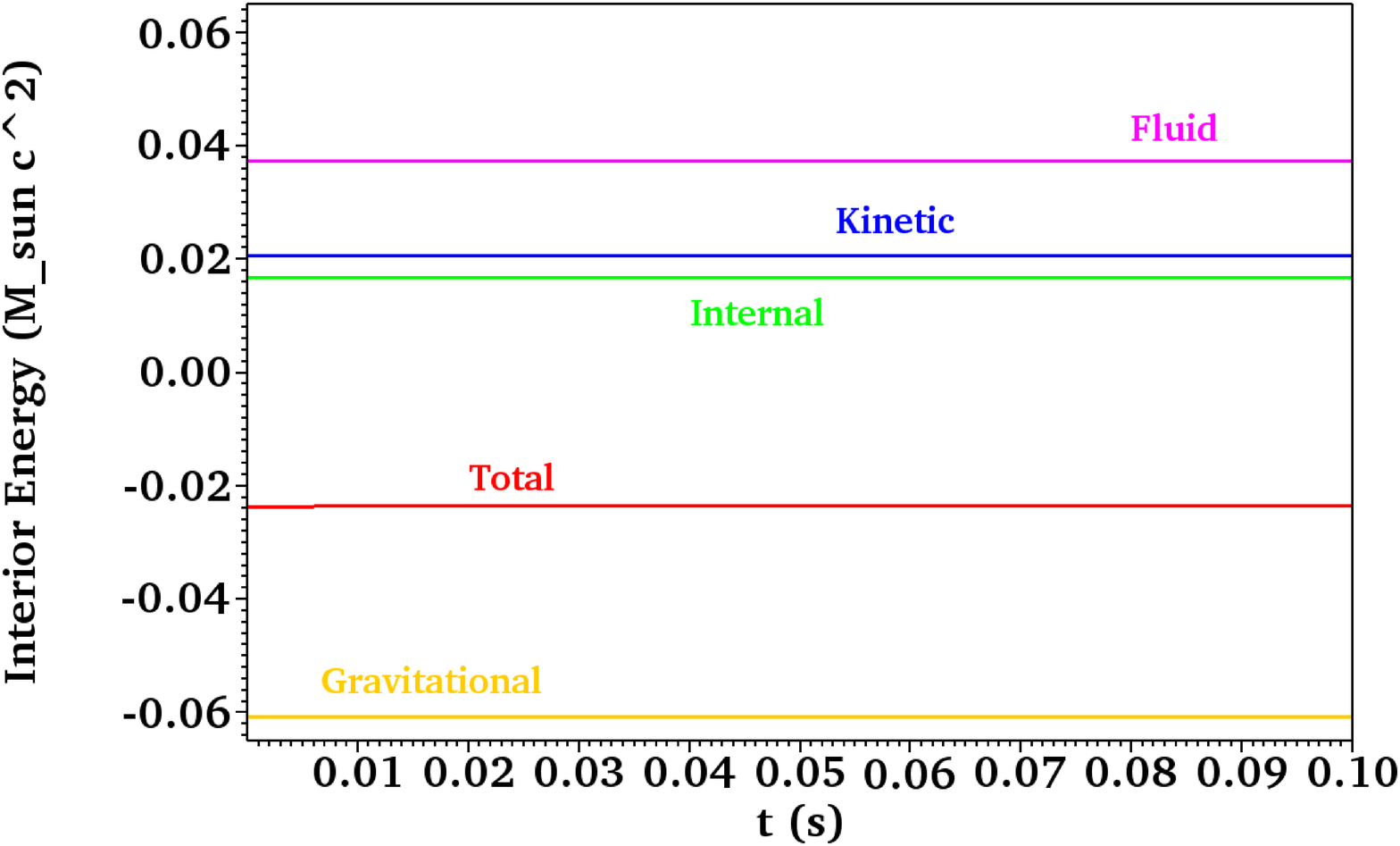}
  \includegraphics[width=0.51\textwidth]{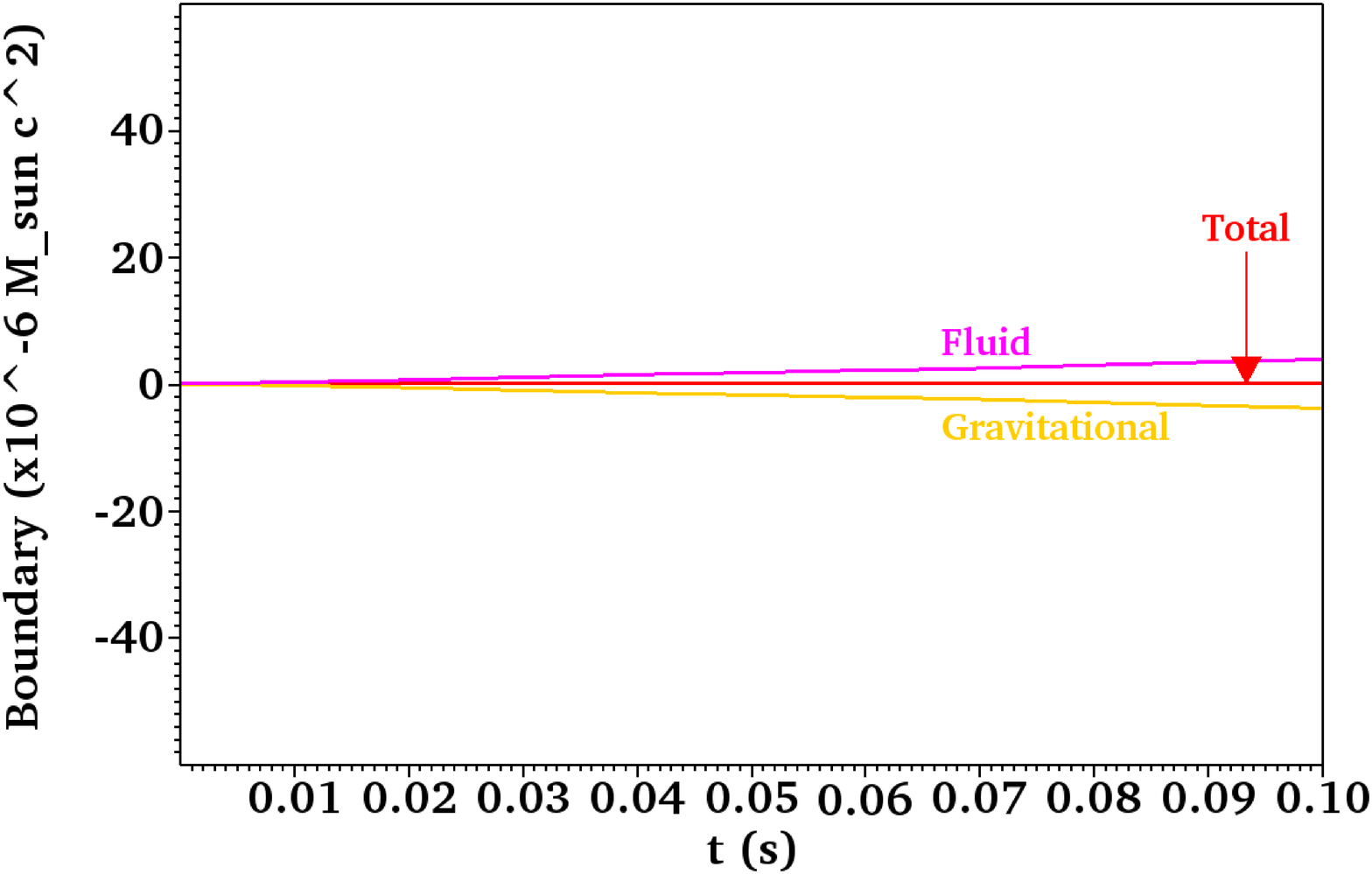}
  \includegraphics[width=0.51\textwidth]{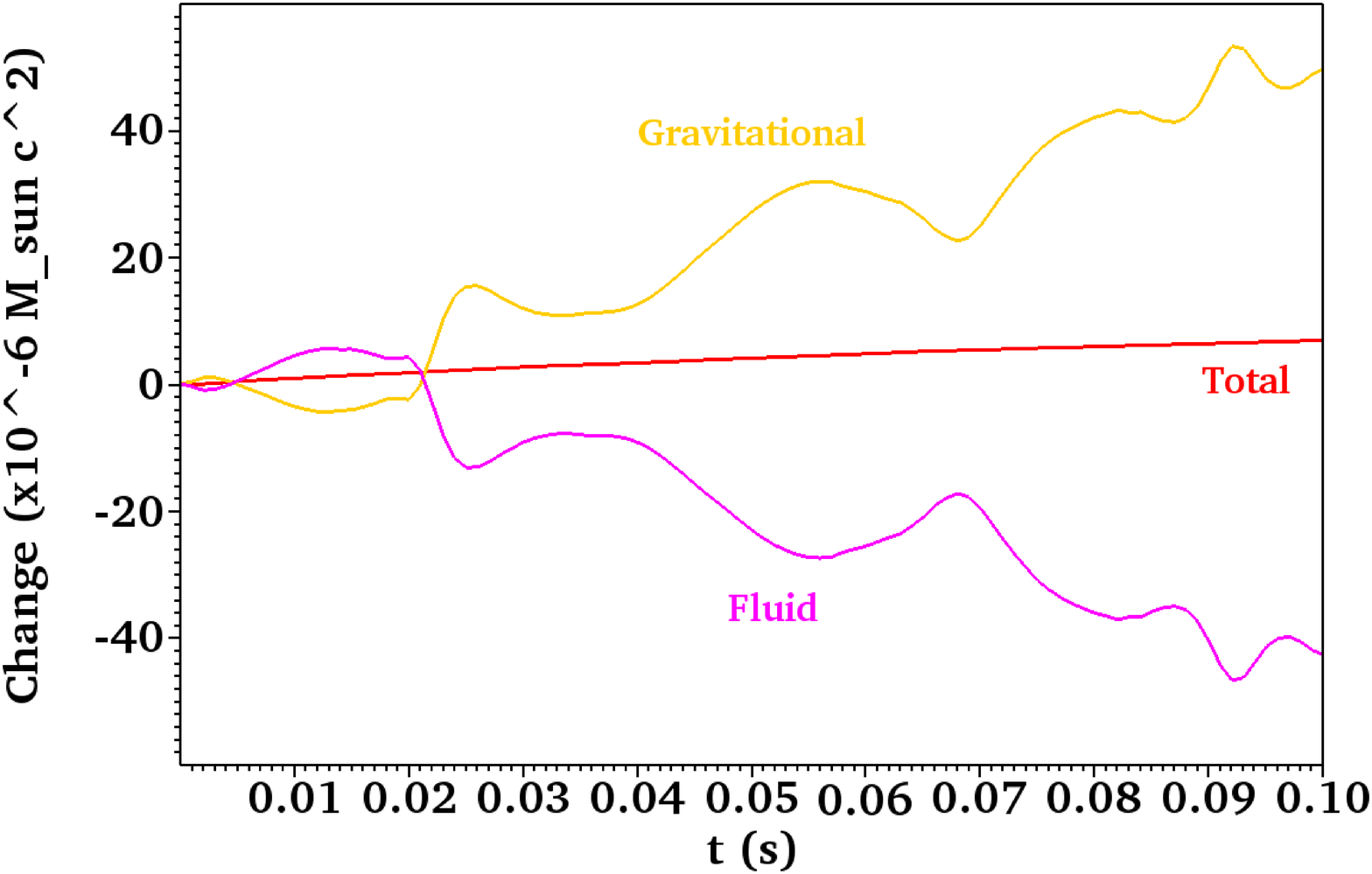}
  \caption{Energy tallies as a function of time in the \texttt{FishboneMoncrief} problem: integrals over the computational domain (upper); integrals of fluxes through the boundary surfaces (center); and changes in the sums of these (lower).
Note that the scale of the upper panel is orders of magnitude larger than that of the lower panels.  
The fluid energy (magenta) is the sum of internal and kinetic energies (green and blue respectively in the upper panel); the total energy (red) is the sum of the fluid and gravitational (orange) energies.}
  \label{fig:FishboneMoncrief_2D_Tally}
\end{centering}
\end{figure}

The energy tallies displayed in Fig.~\ref{fig:FishboneMoncrief_2D_Tally} provide another measure of the stability of the torus under the numerical scheme.
As expected, the fluid (internal plus kinetic), gravitational, and total energies on the computational domain remain constant throughout the simulation---at least to the naked eye, as visualized in the upper panel of Fig.~\ref{fig:FishboneMoncrief_2D_Tally}.
Evidently the matter entering and leaving the domain via inflow and outflow boundary conditions has a minimal impact on the scales of interest.
But a net amount of matter does leave the domain, as the small (note the different scale from the upper panel) positive fluid energy that has passed out of the computational domain attests in the center panel. 
Because matter flowing in through the outer boundary would contribute negatively to the boundary tally, we attribute the small positive value to matter that has diffused from the torus surface and subsequently fallen through the inner boundary.
Visible on the same scale, but of larger impact, are the gradual changes in overall fluid and gravitational energy during the simulation shown in the lower panel.
Apparently the torus undergoes some oscillations, but these are superposed on trends of decreasing fluid energy and increasing (less negative) gravitational energy.
We attribute this to a gradual diffusion of torus material, slowly causing expansion of the torus as a whole, and gradually increasing its potential energy at the expense of internal energy and kinetic energy of rotation.
This shifting balance is visible in the complementary shapes of these curves; that it is imperfectly accomplished is manifest in the gradual drift of the sum of these curves, the total energy change.  
This reflects the fact that the exchange of gravitational and fluid energy occurs via a source term, and is not captured to machine precision, but only (at the resolution presented) to about one part in $10^3$. (This is in contrast to the baryon number and angular momentum about the symmetry axis, which are conserved to machine precision.\footnote{The angular momentum about the symmetry axis is conserved to machine precision because, in the coordinate basis we use in our solver, this component of angular momentum density is the momentum variable solved for, without any source terms.})

%%%%%%%%%%%%%%%%%%%%%%%%%%%%%%%%%%%%%%%%%%%%%%%%%%
\section{Overview of \texttt{Mathematics} Functionality}
\label{sec:MathematicsFunctionality}

With the example problems in Sec.~\ref{sec:ExampleProblems} in mind, including the more detailed top-down discussion of the \texttt{PlaneWave} example, we briefly describe some of the functionality available in \texttt{Mathematics}.
In particular we discuss in bottom-up fashion (i.e. in order of compilation) the code divisions appearing in Fig.~\ref{fig:Mathematics_Structure}.

%%%%%%%%%%%%%%%%%%%%%%%%
\subsection{Manifolds}
\label{sec:Manifolds}

As a drama plays out upon a stage, so a system governed by partial differential equations evolves in a space known mathematically as a `manifold.'
Because it is the stage underlying the drama, it is not surprising that \texttt{Manifolds} is the first code division appearing in the diagram on the left side of Fig.~\ref{fig:Mathematics_Structure}. 
It is one of the \texttt{Mathematics} divisions that is not a leaf division. 
Its subdivisions \texttt{Atlases} and \texttt{Bundles} are shown in the rightmost diagram of Fig.~\ref{fig:Mathematics_Structure}. 
We discuss each in turn.

%%%%%%%%%%%
\subsubsection{Atlases}
\label{sec:Atlases}

\begin{figure}
\centering
\includegraphics[width=2.2in]{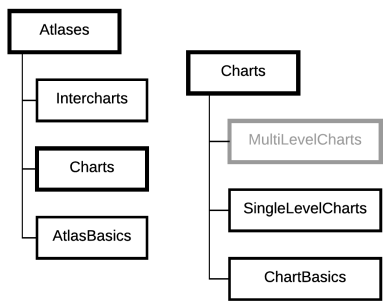}
\caption{\textit{Left:} Structure of \texttt{Atlases}. 
\textit{Right:} Substructure of \texttt{Charts}.
For the place of \texttt{Atlases} in the overall structure of \texttt{Mathematics}, see the rightmost panel of Fig.~\ref{fig:Mathematics_Structure}.
}
\label{fig:Atlases_Structure}
\end{figure}

The first division of \texttt{Manifolds} is \texttt{Atlases}.
Borrowing the idea of a book of maps as a guide to finding one's way through some region,
in mathematical terms an `atlas' is a collection of `charts' or coordinate patches allowing specification of points in the manifold. 
As shown in the left diagram of Fig.~\ref{fig:Atlases_Structure}, \texttt{Atlases} has subdivisions \texttt{AtlasBasics}, \texttt{Charts}, and \texttt{Intercharts}, which we now describe.

%%%%%%%%%%%
\textit{\textbf{AtlasBasics.}}
This division contains classes handling some basic information about an atlas.
The members of the \texttt{ATLAS} singleton specify the maximum numbers of dimensions, charts, sets of fields, and I/O streams an atlas class can have.
\texttt{ConnectivityForm} is a class that indexes connections (faces and edges) between adjacent segments (1D),
quadrilaterals (2D), or hexahedra (3D); this is used to, for example, select the desired manifold boundary in an array of boundary conditions.
Some atlas metadata---dimensionality, numbers of charts and fields, and members and methods for declaring boundary conditions---are handled by \texttt{AtlasHeaderForm}.
Finally, \texttt{FieldAtlasTemplate} is a generic template for a set of related fields on a manifold.

%%%%%%%%%%%
\textit{\textbf{Charts.}}
The classes embodying a single coordinate patch are found in this division.
As shown in the right diagram of Fig.~\ref{fig:Atlases_Structure}, \texttt{Charts} currently has subdivisions \texttt{ChartBasics} and \texttt{SingleLevelCharts}.
In future releases we plan to include \texttt{MultiLevelCharts} for adaptive mesh refinement.

%%%%%
\textit{ChartBasics.}
This division includes classes handling some basic information about a chart.
Some of this---such as dimensionality; number of fields; and members and methods for setting up the coordinate grid, including decomposition into bricks for parallelization---is handled by \texttt{ChartTemplate}.
Similar to \texttt{FieldAtlasTemplate} mentioned above, \texttt{FieldChartTemplate} is a generic template for a set of related fields on a chart.
This also has a specialization \texttt{Field\_CSL\_Template}, where \texttt{CSL} stands for `chart single level.' 
Its members include scalar polymorphic instances of \texttt{VariableGroupForm} \cite{Cardall2015GenASiS-Basics:} called \texttt{Field} and \texttt{FieldOutput}; these are intended to respectively contain a full set of related fields, and a smaller subset thereof for I/O purposes.
The class \texttt{ChartHeader\_SL\_Form} is an extension of \texttt{ChartTemplate} that adds some metadata members relevant to a single-level (\texttt{SL}) chart, as well as an array of polymorphic pointers to \texttt{Field\_CSL\_Template} objects representing fields on a single-level chart.

Classes implementing the basic geometry of flat space, including with curvilinear coordinates, are also included in \texttt{ChartBasics}.
The first of these is \texttt{GeometryFlatForm}---an extension of \texttt{VariableGroupForm}---which stores the coordinate centers and widths of the cells in a grid, along with the limited number of metric functions needed for flat space curvilinear coordinates.
\texttt{GeometryFlat\_CSL\_Form} represents the geometry on a single-level chart; it is the first example of a concrete extension of \texttt{Field\_CSL\_Form}, and is responsible for allocating the polymorphic \texttt{Field} member of the latter to type \texttt{GeometryFlatForm} and calling its initialization.

%%%%%
\textit{SingleLevelCharts.} The main functionality of single-level charts is implemented in this division.
The class \texttt{ChartStream\_SL\_Form} handles I/O for a series of output files (e.g. time slices or cycles).
The abstract class \texttt{Chart\_SL\_Template} contains functionality common to local and distributed single-level charts (i.e. those contained with a single parallel process, and those distributed by domain decomposition among several parallel processes in a distributed-memory parallel computing environment).
This class is an extension of \texttt{ChartHeader\_SL\_Form}, and adds members and methods for dealing with the chart geometry, I/O, and data at the chart boundaries. 
The extensions \texttt{Chart\_SLL\_Form} (`single level local') and \texttt{Chart\_SLD\_Form} (`single level distributed') of \texttt{Chart\_SL\_Template} provide some specifics relevant to those cases, including the exchange of data at the boundaries of adjacent `bricks' of cells belonging to neighboring processes in the distributed case (`ghost exchange'). 

%%%%%%%%%%%
\textit{\textbf{Intercharts.}}
The classes in this division draw on \texttt{AtlasBasics} and \texttt{Charts} to represent atlases.
In the present release only the simplest case of an atlas with a single chart is implemented, but we expect multi-chart atlases in the future, for instance to avoid coordinate singularities in curvilinear coordinate systems. 
A class \texttt{Field\_ASC\_Template}, where \texttt{ASC} stands for `atlas single chart,' extends \texttt{FieldAtlasTemplate}; it represents a field on a manifold with a single chart and as such has a scalar polymorphic member \texttt{Chart} of class \texttt{FieldChartTemplate} for this purpose.
The abstract class \texttt{Atlas\_SC\_Template}---an extension of \texttt{AtlasHeaderForm}---also has a scalar polymorphic member \texttt{Chart}, but of class \texttt{ChartTemplate}, together with an array of polymorphic pointers to \texttt{Field\_ASC\_Template} objects representing fields on a single-chart atlas.
\texttt{Atlas\_SC\_Template} also has methods for creating the chart, adding fields, and applying boundary conditions.
For the geometry of a single-chart atlas, the class \texttt{Geometry\_ASC\_Form} extends \texttt{Field\_ASC\_Template} and, for a single-level chart, allocates its polymorphic \texttt{Chart} member to \texttt{GeometryFlat\_CSL\_Form}.
Members and methods for dealing with geometry and I/O are the final pieces of functionality added in extending \texttt{Atlas\_SC\_Template} to the fully functional \texttt{Atlas\_SC\_Form}.

In the example problems in Section~\ref{sec:ExampleProblems}, it is instances of the \texttt{Intercharts} classes \texttt{Atlas\_SC\_Form} and \texttt{Geometry\_ASC\_Form} that represent position space and its geometry.

%%%%%%%%%%%
\subsubsection{Bundles}
\label{sec:Bundles}

As indicated by its light grey coloring in the rightmost diagram of Fig.~\ref{fig:Mathematics_Structure}, \texttt{Bundles} is not included in the present code release, but we nevertheless point it out by way of information on the planned overall structure of \textsc{GenASiS} \texttt{Mathematics}.
A `fiber bundle' is a manifold which at least locally behaves like a product space $B \times F$, where $B$ is a `base manifold' and $F$ is another manifold called the `fiber.' 
The bundle can be thought of as an infinite number of copies of $F$, one for each point in $B$.
In a `tangent bundle' the fibers are the tangent spaces of the (in general curved) base manifold---the vector spaces, one at each point of the base manifold, containing all the tangent vectors to the manifold at each point.
The tangent bundle\footnote{Or technically, the cotangent bundle, if momenta are regarded as covectors in a canonical description.} is a suitable description of the momentum space needed for particle distribution functions, consisting of a bundle of momentum space fibers over the (possibly curved) position space base manifold.

%%%%%%%%%%%%%%%%%%%%%%%%
\subsection{Operations}
\label{sec:Operations}

After \texttt{Manifolds} comes \texttt{Operations} (see the leftmost diagram of Fig.~\ref{fig:Mathematics_Structure}), which implements some mathematical operators.
The middle right diagram of Fig.~\ref{fig:Mathematics_Structure} shows divisions devoted to \texttt{Algebra} and \texttt{Calculus}.

%%%%%%%%%%%
\subsubsection{Algebra}
\label{sec:Algebra}

This leaf division provides overloaded interfaces to matrix operations such as \texttt{Add} and \texttt{MultiplyAdd}. 
This exposes elemental variables to the compiler and provides a lower-level context to implement threading.

%%%%%%%%%%%
\subsubsection{Calculus}
\label{sec:Calculus}

\begin{figure}
\centering
\includegraphics[width=2.2in]{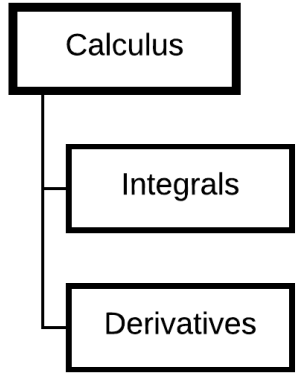}
\caption{Structure of \texttt{Calculus}. 
For the place of \texttt{Calculus} in the overall structure of \texttt{Mathematics}, see the middle right panel of Fig.~\ref{fig:Mathematics_Structure}.
}
\label{fig:Calculus_Structure}
\end{figure}

As shown in Fig.~\ref{fig:Calculus_Structure}, \texttt{Calculus} contains classes implementing \texttt{Derivatives} and \texttt{Integrals} on \texttt{Manifolds}.
In the present release, as we have only single-chart atlases with single-level charts, these operations are performed on instances of \texttt{Chart\_SL\_Template}, either local (\texttt{Chart\_SLL\_Form}) or distributed (\texttt{Chart\_SLD\_Form}).

%%%%%%%%%%%
\textit{\textbf{Derivatives.}}
The two classes in this division so far are \texttt{DifferenceForm} and \texttt{GradientForm}. 
Both of these take an instance of \texttt{VariableGroupForm} as input, treating its data as cell-centered values.
On output, \texttt{DifferenceForm} yields `left' differences (a center value minus its previous neighbor value).
\texttt{GradientForm}, on the other hand, yields centered derivatives, with an option to use the slope limiter described in Ref.~\cite{Cardall2014GENASIS:-Genera}. 

%%%%%%%%%%%
\textit{\textbf{Integrals.}}
Given appropriate \texttt{real} arrays as integrands, \texttt{VolumeIntegralForm} and \texttt{SurfaceIntegralForm} use the geometry of a chart to compute integrals over the interior and the bounding surface of a chart respectively.

%%%%%%%%%%%%%%%%%%%%%%%%
\subsection{Solvers}
\label{sec:Solvers}

The last division of \texttt{Mathematics} is \texttt{Solvers} (see the diagram on the left side of Fig.~\ref{fig:Mathematics_Structure}).
Its subdivisions \texttt{Fields}, \texttt{Constraints}, and \texttt{Evolutions} are shown in the middle left diagram of Fig.~\ref{fig:Mathematics_Structure}; only \texttt{Fields} and \texttt{Evolutions} are included in this release and discussed here.

The classes included in this release are aimed at time-explicit evolution of hyperbolic conservation laws (continuity equations).
In spacetime, these involve the covariant 4-divergences of 4-currents, for instance, a 4-vector $N^\mu$, or the four 4-currents in a symmetric rank-2 tensor $T^{\mu\nu}$:
\begin{eqnarray}
\nabla_\mu \, N^\mu &=& A, \label{eq:VectorDivergence} \\
\nabla_\mu \, T^{\mu\nu} &=& B^\nu, \label{eq:TensorDivergence}
\end{eqnarray}
where $A$ and $B^\nu$ are scalar and 4-vector sources respectively (spacetime indices are Greek letters).
Assuming flat spacetime, and labeling the time coordinate as $t$ and the spatial coordinates as $x^i$ (spatial indices are latin letters), a collection of such equations can be cast in the form
\begin{equation}
\frac{\partial \, \mathcal{U}}{\partial t}  
+  \frac{1}{\sqrt{\gamma}} \frac{\partial }{\partial x^i} 
    \left[ \sqrt{\gamma} \, \mathcal{F}^i (\mathcal{U}) \right]
=  \mathcal{S}(\mathcal{U}). \label{eq:ConservationLaw}
\end{equation}
Here $\gamma$ is the determinant of the 3-metric $\gamma_{ij}$ of a spacelike slice; in flat space the form $\left(\gamma_{ij}\right) = \mathrm{diag}[1,\gamma_{22},\gamma_{33}]$ is sufficient to accommodate common coordinate systems.
The evolution of the set of conserved variables denoted $\mathcal{U}$ is governed by the position space divergence of the fluxes $\mathcal{F}^i (\mathcal{U})$ and the sources $\mathcal{S}(\mathcal{U})$.
The latter include not only terms arising from external sources like $A$ and $B^\nu$ in Eqs.~\ref{eq:VectorDivergence} and \ref{eq:TensorDivergence}, but also geometric terms (`fictitious forces,' in the case of flat space curvilinear coordinates) arising from connection coefficient terms in the divergence of tensors of rank $> 1$ as in Eq.~\ref{eq:TensorDivergence}. 

The finite volume approach spatially discretizes the problem by taking a volume average of Eq.~\ref{eq:ConservationLaw} over each cuboid cell in the mesh:
\begin{equation}
\frac{d\langle \mathcal{U} \rangle}{d t} 
= - \frac{1}{\langle \sqrt{\gamma}\rangle} \sum_q \frac{1}{(\Delta x)^q}
    \left[ \langle \sqrt{\gamma} \, \mathcal{F}^q \rangle_{q \rightarrow} 
            - \langle \sqrt{\gamma} \, \mathcal{F}^q \rangle_{\leftarrow q} \right]
    + \langle \mathcal{S} \rangle. \label{eq:FiniteVolume}
\end{equation}
The system is now a set of time-dependent ordinary differential equations which can be integrated with standard explicit techniques.
Angle brackets without a subscript denote a cell volume average, while those with subscripts denote an area average over the outer ($\leftarrow q$) or inner ($q \rightarrow$) cell face in dimension $q$ (we switch the spatial index from $i$ to $q$ as we switch to an explicit sum from the previous implied sum of repeated indices). 
The coordinate width of the cell in dimension $q$ is $(\Delta x)^q$.
We take the flux components $\mathcal{F}^q$ to be given in the coordinate basis.

%%%%%%%%%%%
\subsubsection{Fields}
\label{sec:Fields}

This division contains classes that describe field types acted on by the solver classes.
Of central importance in the present case is the \texttt{abstract} class \texttt{CurrentTemplate} mentioned in Section~\ref{sec:PlaneWave}, an extension of \texttt{VariableGroupForm} \cite{Cardall2015GenASiS-Basics:} that contains members and methods related to solving conservation laws involving the divergence of a 4-current.
Among the members of \texttt{CurrentTemplate} is an array of indices of the \texttt{VariableGroupForm} data member \texttt{Value} containing the conserved fields $\mathcal{U}$ in Eq.~\ref{eq:ConservationLaw}.
In addition to the conserved variables $\mathcal{U}$, the description of the system typically involves 
two other sets of variables: `primitive' variables $\mathcal{W}$, often equal in number to the number of balanced variables; and additional `auxiliary' variables $\mathcal{A}$, some determined by one or more closure relations involving the primitive variables (e.g. an `equation of state' in the case of hydrodynamics).
\texttt{CurrentTemplate} contains an array of indices of the primitive variables, and also \texttt{deferred} methods for computing $\mathcal{U}$ and $\mathcal{A}$ from $\mathcal{W}$, for computing $\mathcal{W}$ and $\mathcal{A}$ from $\mathcal{U}$, and for implementing the functional form of the fluxes $\mathcal{F}^i (\mathcal{U})$.
 
\texttt{Fields} also contains a class \texttt{Tally\_C\_Form} (where \texttt{C} stands for `current') that, by default, contains the members and methods needed to perform volume integrals of the conserved variables over the computational domain and surface integrals of their fluxes through the manifold boundaries.
The class \texttt{Current\_ASC\_Template}, an extension of \texttt{Field\_ASC\_Template} (recall that \texttt{ASC} stands for `atlas single chart'), includes
several instances of \texttt{Tally\_C\_Form} as members, as well as the methods needed to compute and combine them in order to provide the machinery needed to assess that the numerical outcome of conservation law evolution is as expected.

%%%%%%%%%%%
\subsubsection{Evolutions}
\label{sec:Evolutions}

\begin{figure}
\centering
\includegraphics[width=2.2in]{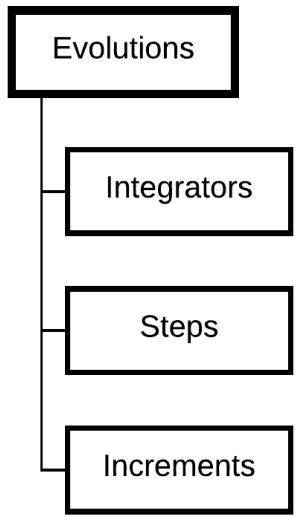}
\caption{Structure of \texttt{Evolutions}. 
For the place of \texttt{Evolutions} in the overall structure of \texttt{Mathematics}, see the middle left panel of Fig.~\ref{fig:Mathematics_Structure}.
}
\label{fig:Evolutions_Structure}
\end{figure}

\texttt{Evolutions} contains solvers that integrate systems of fields forward in time.
As shown in Fig.~\ref{fig:Evolutions_Structure}, it comprises three subdivisions: \texttt{Increments}, \texttt{Steps}, and \texttt{Integrators}.

%%%%%%%%%%%
\textit{\textbf{Increments.}}
By an `increment' we refer to a first-order accurate (in time) change in the dependent variables of a set of ordinary differential equations in a small but finite time step $\Delta t$. 
For the finite volume (\texttt{FV}) case of Eq.~\ref{eq:FiniteVolume}, the \texttt{Increments} division has a class \texttt{IncrementDivergence\_FV\_Form} that computes the increment
\begin{equation}
(\Delta \,\mathcal{U})_\leftrightarrow 
= - \frac{\Delta t}{(\sqrt{\gamma})_\leftrightarrow} \sum_q \frac{1}{(\Delta x)^q}
   \left[ \langle \sqrt{\gamma} \, \mathcal{F}^q \rangle_{q \rightarrow} 
            - \langle \sqrt{\gamma} \, \mathcal{F}^q \rangle_{\leftarrow q} \right]. \label{eq:Increment}
\end{equation}
In this initial implementation we work at second order in space. 
We take cell volume averages to be indistinguishable from cell center values denoted by a double-headed arrow ($\leftrightarrow$).
As described in Ref.~\cite{Cardall2014GENASIS:-Genera},
slope-limited piecewise-linear reconstruction of primitive variables $\mathcal{W}$ is used to obtain face-centered fluxes, which are adopted as the face averages on either side of a cell interface.
These so-called `left' and `right' states are resolved by a `Riemann solver' into the fluxes used in Eq.~\ref{eq:Increment}; in this release, we use only the simple HLL solver \cite{Cardall2014GENASIS:-Genera}.
This increment is time-explicit in that it is evaluated using the current known values of $\mathcal{U}$ (and $\mathcal{W}$ and $\mathcal{A}$ as needed).

%%%%%%%%%%%
\textit{\textbf{Steps.}}
In the Runge-Kutta method of solving ordinary differential equations of the form
\begin{equation}
\frac{dy}{dt} = f ( y ), \label{eq:ODE}
\end{equation}
multiple first-order (in time) increments are combined to obtain a step accurate to higher order in the time step $\Delta t$.
Labeling the previously known solution at time $t_n$ as $y_n$ and the new solution at time $t_{n+1} = t_n + \Delta t$ as $y_{n+1}$,
we write a scheme of $s$ stages as
\begin{equation}
y_{n+1} = y_n + \sum_{i=1}^s b_i k_i. \label{eq:RungeKutta}
\end{equation}
Here the constants $b_i$ are `weights,' and the increment $k_i$ at stage $i$ is given by 
\begin{equation}
k_i = \Delta t \, f \left( y_n  +  \sum_{j=1}^{i-1} a_{ij} k_j \right), \label{eq:Increment_RK}
\end{equation}
that is, by $\Delta t$ times the right-hand side of Eq.~\ref{eq:ODE} evaluated at the previous solution plus some combination of the increments of previous stages.
The constants $a_{ij}$ can be regarded as elements of a lower-diagonal `Runge-Kutta matrix' having non-zero values only for $j < i$; in particular, at the first stage $a_{10} = 0$ and the first increment is $k_1 = \Delta t \, f \left( y_n \right)$.

The classes included so far in \texttt{Steps} implement this Runge-Kutta scheme in general, and the evolution of conservation laws in particular.
The class \texttt{Step\_RK\_Template} is initialized with arbitrary $a_{ij}$ and $b_i$, keeps track of the increments $k_i$ and computes the argument going in the right-hand side of Eq.~\ref{eq:Increment_RK}, and computes the final updated variables according to Eq.~\ref{eq:RungeKutta}. 
This class is \texttt{abstract} because it has a \texttt{deferred} method \texttt{ComputeIncrement} to specify the function $f$ (the right-hand side of the original Eq.~\ref{eq:ODE}) needed in Eq.~\ref{eq:Increment_RK}.
The extension \texttt{Step\_RK\_C\_Template}---where \texttt{C} stands for `current,' as in `conserved current'---gives a concrete implementation of the method \texttt{ComputeIncrement} for conservation laws, having among its members an instance of \texttt{IncrementDivergence\_C\_Form} to compute the contribution of Eq.~\ref{eq:Increment}.
And as we saw in the example problems, \texttt{Step\_RK\_C\_Template} has a procedure pointer member \texttt{ApplySources} that, when associated, adds source terms to the Runge-Kutta increment.
Finally, the class \texttt{Step\_RK2\_C\_Template} specifies the constants $b_1 = b_2 = 1/2$ and $a_{21} = 1$ for a two-stage second-order scheme.

%%%%%%%%%%%
\textit{\textbf{Integrators.}}
The main purpose of the classes in this division is to evolve a system forward in time with a loop over time steps.
The \texttt{abstract} class \texttt{Integrator\_Template}, with its \texttt{Evolve} method, is a generic foundation for such time integration.
Because different types of systems involve different solvers---indeed, in multiphysics problems, different sequences or `cycles' of solvers---a deferred method \texttt{ComputeCycle} must be given a concrete implementation by an extension of this class.
A method \texttt{ComputeTimeStep} contains some basic functionality for determining a time step, including a reduction over parallel processes; but it relies on a deferred method \texttt{ComputeTimeStepLocal} to determine limits on the time step from the data assigned to a particular process according to the needs of a particular type of system.
The member \texttt{WriteTimeInterval} controls how often output to disk takes place.
%The time step immediately before a write is shortened to land on the target time, and a method \texttt{AdministerCheckpoint} is called to execute the write and perform other periodically needed tasks, such as displaying memory and timing information and computing tallies of global quantities. 
A method \texttt{AdministerCheckpoint} is called to execute the write after completion of the time step closest to the target time\footnote{The command line or parameter file option \texttt{WriteTimeExact=.true.} modifies this default behavior to shorten the time step immediately before a write to land exactly on the target time.} and perform other periodically needed tasks, such as displaying memory and timing information and computing tallies of global quantities. 

An extension provided in this release is \texttt{Integrator\_C\_Template}, where \texttt{C} stands for `current,' as in `conserved current' (see Section~\ref{sec:Fields}).
It provides concrete implementations of the \texttt{ComputeCycle} and \texttt{ComputeTimeStepLocal} methods appropriate for conservation laws, but we nevertheless label it \texttt{abstract} because it adds instances of the \texttt{abstract} classes \texttt{Current\_Template} and \texttt{Step\_RK\_Template} as members that must be allocated as concrete extensions by driver programs, as we saw in the examples in Section~\ref{sec:ExampleProblems}.
Indeed all the problems in bold outline across the top of Fig.~\ref{fig:FluidDynamics} are extensions of \texttt{Integrator\_C\_Template} (located towards the lower left of the figure).

This division also includes classes that produce time series, recording data at checkpoint times and writing a file at the end of a run.
The basic class \texttt{TimeSeriesForm} records memory and timing information.
An extension \texttt{TimeSeries\_C\_Form} (where, again, \texttt{C} stands for `current') adds additional time series relevant to conservation laws, specifically the tallies---that is, volume and surface integrals, and relevant combinations thereof---included in \texttt{Current\_ASC\_Template} (see Section~\ref{sec:Fields}).

%%%%%%%%%%%%%%%%%%%%%%%%%%%%%%%%%%%%%%%%%%%%%%%%%%
\section{Building Examples and Unit Tests}
\label{sec:Building}

\genasis\ \texttt{Mathematics} is distributed as a gzip-compressed tarball (\texttt{.tar.gz}) file. 
Upon uncompression and extraction, the top-level directory has \texttt{README} and {\tt LICENSE} files and three subdirectories: \texttt{Build}, \texttt{Modules}, and \texttt{Programs}. 
The directory structures inside \texttt{Modules} and \texttt{Programs} are as described in Sec.~\ref{sec:Introduction}. 
In particular, the example programs described in Sec.~\ref{sec:ExampleProblems} and the unit test programs are available under the \texttt{Programs/Examples/Mathematics} and \texttt{Programs/UnitTests/Mathematics} subdirectories, respectively.

A machine-specific \texttt{Makefile} is needed to build \genasis\ programs. 
Several sample \texttt{Makefile}s are provided under the subdirectory \texttt{Build/Machines}. 
Minor modifications of one of the provided \texttt{Makefile}s that most approximates one's computing environment is often sufficient to get started. 
The essential information needed includes the name of the compiler wrapper to compile MPI-based code (e.g. commonly \texttt{mpif90} for Fortran), compiler-specific flags for various debugging and optimization options, and the flags and locations to include and link with the required third-party I/O library Silo.\footnote{https://wci.llnl.gov/simulation/computer-codes/silo} 

The version accompanying this paper is compatible with Silo-4.10.x. To use Silo with GenASiS, the Fortran interface to Silo should be enabled (which it is by default). In the simplest case, one can build Silo to be used with GenASiS
with the following commands inside the Silo distribution: 
\begin{verbatim}
> ./configure --enable-fortran
> make
> make install
\end{verbatim}
Silo's documentation (e.g. its INSTALL file) provides more complete information on building Silo.

Once the machine-specific \texttt{Makefile} is set up, the environment variable \texttt{GENASIS\_MACHINE} has to be set to tell the \genasis\ build system to use the corresponding \texttt{Makefile}. For example, to use the \texttt{Makefile} for the GNU compiler on a typical Linux cluster (i.e. \texttt{Makefile\_Linux\_GNU}), in a Bash Unix shell one can type
\begin{verbatim}
> export GENASIS_MACHINE=Linux_GNU
\end{verbatim}

In most common computing environments with a generic MPI library, the fluid dynamics programs described in Sec.~\ref{sec:ExampleProblems} can then be built and executed.
For instance, the commands
\begin{verbatim}
> cd Programs/Examples/Mathematics/FluidDynamics/Executables
> make PURPOSE=OPTIMIZE all
\end{verbatim} 
build all the examples.
The first few of these are run (here with 8 MPI processes) with the commands 
\begin{verbatim}
> export OMP_NUM_THREADS=1
> mpirun -np 8 ./SineWave_Linux_GNU nCells=128,128,128
> mpirun -np 8 ./SawtoothWave_Linux_GNU nCells=128,128,128 \
nWavelengths=2,2,2
> mpirun -np 8 ./RiemannProblem_Linux_GNU nCells=128,128,128 \
FinishTime=0.25
\end{verbatim} 
(To compile in a manner that is unoptimized but useful for debuggers, replace {\tt PURPOSE=OPTIMIZE} with {\tt PURPOSE=DEBUG}.
Or omit it altogether; in the absence of a specification of {\tt PURPOSE}, the {\tt Makefile} in {\tt FluidDynamics/Executables} sets {\tt PURPOSE=DEBUG} as a default.)
Note that in these examples, the optional non-default parameter values for {\tt nCells}, {\tt nWavelengths}, and {\tt FinishTime}---which were used in generating the lower panels of Figures~\ref{fig:SineWave}, \ref{fig:SawtoothWave}, and \ref{fig:RiemannProblem}---are passed to the programs in this case via command-line options. The 1D and 2D cases of these programs---which were used in generating the upper and middle panels of Figures~\ref{fig:SineWave}, \ref{fig:SawtoothWave}, and \ref{fig:RiemannProblem}---can also be executed by specifying fewer elements for {\tt nCells} and the {\tt Dimensionality} option, for example
\begin{verbatim}
> mpirun -np 2 ./RiemannProblem_Linux_GNU \
nCells=128 Dimensionality=1D FinishTime=0.25
> mpirun -np 4 ./RiemannProblem_Linux_GNU \
nCells=128,128 Dimensionality=2D FinishTime=0.25
\end{verbatim}
The {\tt Dimensionality} option is also used as an appendix to the name of the output file and it should be consistent with the number of elements given to {\tt nCells} to determine the desired dimensionality of the mesh.

In the above examples we explicitly set the number of OpenMP threads with the environment variable \texttt{OMP\_NUM\_THREADS}. It is imperative to do so since the default number of threads varies among different compilers if this environmental variable is not set. When running with more than one OpenMP thread per MPI task, one must take care so that thread placement on the processors is set correctly to avoid unintended resource contention.

By default the output files are written in the directory \texttt{Output} that resides on the same level as the \texttt{Executables} directory, but this can be changed with an optional {\tt OutputDirectory} command line option. 

If the VisIt visualization package is available, animated versions of plots similar to Figs.~\ref{fig:SineWave}, \ref{fig:RiemannProblem}, and \ref{fig:RayleighTaylor} can be generated using the supplied visualization script called from the \texttt{Output} directory. The script takes one argument, which is the program name appended with the {\tt Dimensionality} string. Assuming the executable {\tt visit} is available, the visualization script can be called, for example,  as follows:

\begin{verbatim}
> cd Programs/Examples/Mathematics/FluidDynamics/Output
> visit -cli -s ../FluidDynamics.visit.py SineWave_3D
> visit -cli -s ../FluidDynamics.visit.py RiemannProblem_2D
> visit -cli -s ../FluidDynamics.visit.py RayleighTaylor_2D
\end{verbatim}
(The option \texttt{-nowin} can be added to above VisIt invocation to prevent VisIt from trying to display its visualization windows, resulting in the image files being drawn off-screen instead.)

Unit test programs exercising individual \genasis\ classes can similarly be built and executed inside the \texttt{Executables} directory of each leaf division of the code under {\tt Programs/UnitTests}.
For example, the following commands build and execute the unit test programs for classes in the \texttt{Fields} division (see Section~\ref{sec:Fields}):
\begin{verbatim}
> cd Programs/UnitTests/Mathematics/Solvers/Fields/Executables
> make all
> mpirun -np 1 [program_name]
\end{verbatim}
This blanket {\tt make all} builds all the unit test targets in the {\tt Makefile} fragment {\tt Programs/UnitTests/Mathematics/Solvers/Fields/Makefile\_Fields}.
Individual targets of course also can be built.

\genasis\ \texttt{Mathematics} has been tested with the following compilers: GNU Fortran (gfortran, part of GCC) 6.2.0, Intel Fortran 16, and Cray Compiler Environment 8.5.3. 
Newer versions of these compilers are likely to work as well. 
\genasis\ \texttt{Mathematics} is written in full compliance with the Fortran 2003 standard to enhance portability.

%%%%%%%%%%%%%%%%%%%%%%%%%%%%%%%%%%%%%%%%%%%%%%%%%%
\section{Conclusion}
\label{sec:Conclusion}

In this paper we describe, make available, and illustrate with examples the \texttt{Mathematics} division of \textsc{GenASiS}, which for our purposes contemplates systems of physical fields governed by partial differential equations. 
The content of \texttt{Mathematics} is outlined in Fig.~\ref{fig:Mathematics_Structure}.
In implementing this functionality we continue with the object-oriented philosophy described in some detail in connection with \textsc{GenASiS} \texttt{Basics} \cite{Cardall2015GenASiS-Basics:}, using the features of Fortran 2003 that support this programming paradigm.
In our context, an object-oriented approach enables the flexibility connoted by the `General' in \textsc{GenASiS}---the capacity of the code to include and refer to multiple algorithms, solvers, and physics and numerics choices with the same abstracted names and/or interfaces.
The object-oriented principles of inheritance and polymorphism---embodied in the mechanisms of \texttt{type} extension and method overriding---also make it much easier to allow lower-level code to access higher-level code.
This facilitates a primary purpose of \textsc{GenASiS} \texttt{Mathematics}, namely, that it provide versatile and widely-applicable solvers that can be invoked by a range of different systems whose details are specified in the (future) \texttt{Physics} division of \textsc{GenASiS} and in driver programs.

The first division of \textsc{GenASiS} \texttt{Mathematics}, \texttt{Manifolds}, contains meshing infrastructure; as shown in the right diagram of Fig.~\ref{fig:Mathematics_Structure}, this release includes \texttt{Atlases} suitable (for instance) for representing position space, while \texttt{Bundles} in future releases will allow treatment of momentum space as well.
An `atlas' is a collection of `charts' or coordinate patches glued together to cover the manifold as a whole. 
As shown in Fig.~\ref{fig:Atlases_Structure}, \texttt{Atlases} includes subdivisions \texttt{AtlasBasics}, containing classes handling some basic information about an atlas; \texttt{Charts}, whose classes embody single coordinate patches; and \texttt{Intercharts}, whose functionality includes the handling of manifold boundary conditions and the overlaps of separate charts.
This release only includes atlases with a single chart, and charts represented by a single level of meshing. 

After \texttt{Manifolds} in the left diagram of Fig.~\ref{fig:Mathematics_Structure} comes \texttt{Operations}.
The middle right diagram of Fig.~\ref{fig:Mathematics_Structure} shows its subdivisions \texttt{Algebra}, which includes matrix operations, and \texttt{Calculus}, which---as seen in Fig.~\ref{fig:Calculus_Structure}---implements \texttt{Derivatives} and \texttt{Integrals} on \texttt{Manifolds}. 

In this release, the classes in \texttt{Solvers}---the last division in the left diagram of Fig.~\ref{fig:Mathematics_Structure}---are aimed at time-explicit evolution of hyperbolic conservation laws (continuity equations).
The middle left diagram of Fig.~\ref{fig:Mathematics_Structure} shows subdivisions \texttt{Fields}, \texttt{Constraints}, and \texttt{Evolutions}, but \texttt{Constraints}---which will contain solvers for equations expressing constraints on the position space dependence of fields within a given time slice---is not included in this release.
\texttt{Fields} contains classes that describe field types acted on by the solver classes.
As shown in Fig.~\ref{fig:Evolutions_Structure}, \texttt{Evolutions} is divided into \texttt{Increments}, which implement first-order (in time) changes to dependent variables;
 \texttt{Steps}, which assemble multiple increments into a higher-order time step; and \texttt{Integrators}, which evolve a system forward in time with a loop over time steps. 

We provide and discuss several nontrivial example programs in order to illustrate usage of the functionality available in \texttt{Mathematics}.
In this release these are fluid dynamics problems illustrating the solution of conservation laws.
The advection of a periodic plane wave in density is discussed in greatest detail, both in terms of explicit coding and performance outcomes (including convergence, scaling, and threading), as it is a simple case that illustrates the main aspects of setting up and running a physics problem built upon \textsc{GenASiS} \texttt{Mathematics} classes.
The Riemann problem example shows how to specify boundary conditions.
A demonstration of the Rayleigh-Taylor instability illustrates the use of non-default parameters for atlases and charts and the introduction of source terms.
The Sedov-Taylor blast wave shows how to use different coordinate systems, allowing comparison of the results using spherical coordinates in 1D, cylindrical coordinates in 2D, and Cartesian coordinates in 3D.
Finally, a 2D Fishbone-Moncrief equilibrium torus exemplifies the use of physical units and non-uniform grid spacing, providing also an additional example and test of source terms and spherical coordinates.

Our example programs illustrate the utility of \texttt{Basics} and \texttt{Mathematics} for nontrivial problems, and also foreshadow the utility of future releases developing the \texttt{Physics} portion of \textsc{GenASiS}.
Figure~\ref{fig:FluidDynamics} shows the classes that compose our fluid dynamics examples, illustrating how computational physics problems can be broken down into separate concepts, many of which can be abstracted for reuse.
Line counts are by no means a foolproof measure of code significance or human productivity, but they can provide rough initial indications.
\texttt{Basics} and \texttt{Mathematics} currently comprise 19241 and 10786 lines of code respectively.
The classes labeled `Physics' by their green coloration in Fig.~\ref{fig:FluidDynamics} occupy 3581 lines of code.
The line counts for the classes labeled `Problem' by their blue coloration in Fig.~\ref{fig:FluidDynamics} range from 284 (\texttt{RiemannProblem}) to 673 (\texttt{FishboneMoncrief}).
Thus in our example problems, \texttt{Basics} takes 56-57\% of the line count, \texttt{Mathematics} takes 31-32\%, the `Physics' category takes 10-11\%, and the specific `Problem' definition takes 1-2\%.
The line counts of the \texttt{Physics} divisions of \textsc{GenASiS} ultimately will be significantly larger than the count taken here from our example classes, and the \texttt{Mathematics} division will grow as well. 
But already it can be glimpsed that well-designed, reusable code can enable researchers to focus more on their problems and less on computational details.

Subsequent papers will describe and make available additions to the \texttt{Mathematics} division of \textsc{GenASiS} and introduce the \texttt{Physics} division as well.
Some of this functionality will be more specialized towards our focus on an astrophysics application (core-collapse supernovae), but many capabilities will be applicable to other physical problems as well.
Atlases with multiple charts will allow the use of curvilinear coordinates without coordinate singularities.
We also intend to include adaptive mesh refinement, in which we approximate the ideal of continuity with a finite sequence of meshes which provide, as necessary, increasing refinements of the coarsest (top-level) mesh.
More sophisticated Riemann solvers and the use of discontinuous Galerkin methods would be valuable technical improvements.
Self-gravitation, nuclear species and a microphysical equation of state, and magnetic fields are important aspects of many astrophysics problems.
Finally, a more significant challenge for our own target application of core-collapse supernovae is neutrino transport.

\section*{Acknowledgements}

This material is based upon work supported by the U.S. Department of Energy, Office of Science, Office of Nuclear Physics under contract number DE-AC05-00OR22725 and the National Science Foundation under Grant No. 1535130. 
This research used resources of the Joint Institute for Computational Sciences at the University of Tennessee; the Extreme Science and Engineering Discovery Environment (XSEDE), which is supported by National Science Foundation grant number ACI-1053575; and the resources of the Oak Ridge Leadership Computing Facility, which is a DOE Office of Science User Facility supported under Contract DE-AC05-00OR22725.

%% The Appendices part is started with the command \appendix;
%% appendix sections are then done as normal sections
%% \appendix

%% \section{}
%% \label{}

%% References
%%
%% Following citation commands can be used in the body text:
%% Usage of \cite is as follows:
%%   \cite{key}         ==>>  [#]
%%   \cite[chap. 2]{key} ==>> [#, chap. 2]
%%

%% References with bibTeX database:

\def\cpc{Comp. Phys. Commun.}
\def\apjs{Astrophys. J. Suppl. Ser. }
\def\apjl{Astrophys. J. Lett.}
\def\physscr{Phys. Scr.}
\def\apj{Astrophys. J.}

%\bibliographystyle{elsarticle-num}
%\bibliography{bibliography18}

%% Authors are advised to submit their bibtex database files. They are
%% requested to list a bibtex style file in the manuscript if they do
%% not want to use elsarticle-num.bst.

%% References without bibTeX database:

% \begin{thebibliography}{00}

%% \bibitem must have the following form:
%%   \bibitem{key}...
%%

% \bibitem{}

% \end{thebibliography}

\end{document}